\documentclass[twocolumn]{aastex63}

\usepackage{ulem}
\usepackage{hyperref}
\usepackage{amsmath}

\defcitealias{2009ApJS..183...67D}{D09}
\defcitealias{2009AJ....138..332J}{J09}
\defcitealias{tamann}{T11}
\defcitealias{2021arXiv210502120Z}{Z21}
\defcitealias{2020arXiv200510793F}{FM20}
\defcitealias{2021MNRAS.501..933P}{P21}
\defcitealias{2023MNRAS.tmp..926P}{P23}

\shorttitle{JAGB Distances}
\shortauthors{Lee et al.}
\graphicspath{{./}{}}

\begin{document}

\title{Resolved Near-infrared Stellar Photometry from the Magellan Telescope for 13 Nearby Galaxies:\\ JAGB Method Distances}

\author{Abigail~J.~Lee}\affil{Department of Astronomy \& Astrophysics, University of Chicago, 5640 South Ellis Avenue, Chicago, IL 60637}\affiliation{Kavli Institute for Cosmological Physics, University of Chicago,  5640 South Ellis Avenue, Chicago, IL 60637}

\author{Andrew~J.~Monson}\affil{Steward Observatory, The University of Arizona, 933 N. Cherry Avenue, Tucson, AZ 85721, USA}

\author{Wendy~L.~Freedman}\affil{Department of Astronomy \& Astrophysics, University of Chicago, 5640 South Ellis Avenue, Chicago, IL 60637}\affiliation{Kavli Institute for Cosmological Physics, University of Chicago,  5640 South Ellis Avenue, Chicago, IL 60637}

\author{Barry~F.~Madore}\affil{Observatories of the Carnegie Institution for Science 813 Santa Barbara St., Pasadena, CA~91101}\affil{Department of Astronomy \& Astrophysics, University of Chicago, 5640 South Ellis Avenue, Chicago, IL 60637}

\author{Kayla~A.~Owens}\affil{Department of Astronomy \& Astrophysics, University of Chicago, 5640 South Ellis Avenue, Chicago, IL 60637}\affiliation{Kavli Institute for Cosmological Physics, University of Chicago,  5640 South Ellis Avenue, Chicago, IL 60637}

\author{Rachael~L.~Beaton}\affil{Space Telescope Science Institute, Baltimore, MD, 21218, USA}\affiliation{Department of Physics and Astronomy, Johns Hopkins University, Baltimore, MD 21218, USA}

\author{Coral~Espinoza}\affil{Lake Forest College Physics Department, Lake Forest, IL 60045}\affil{Department of Astronomy \& Astrophysics, University of Chicago, 5640 South Ellis Avenue, Chicago, IL 60637}

\author{Tongtian Ren}\affil{Jodrell Bank Centre for Astrophysics, University of Manchester, Oxford Road, Manchester, M13 9PL, UK}

\author{Yi Ren}\affil{College of Physics and Electronic Engineering, Qilu Normal University, Jinan 250200, China}

\correspondingauthor{Abigail J. Lee}\email{abbyl@uchicago.edu}

\begin{abstract}
We present near-infrared $JHK$ photometry for the resolved stellar populations in 13 nearby galaxies: NGC~6822, IC~1613, NGC~3109, Sextans~B, Sextans~A, NGC 300, NGC~55, NGC~7793, NGC~247, NGC~5253, Cen~A, NGC~1313, and M83, acquired from the 6.5m Baade-Magellan telescope.
We measure distances to each galaxy using the J-region asymptotic giant branch (JAGB) method, a new standard candle that leverages the constant luminosities of color-selected, carbon-rich AGB stars.
While only single-epoch, random-phase photometry is necessary to derive JAGB distances, our photometry is time-averaged over multiple epochs, thereby decreasing the contribution of the JAGB stars' intrinsic variability to the measured dispersions in their observed luminosity functions.
To cross-validate these distances, we also measure near-infrared tip of the red giant branch (TRGB) distances to these galaxies.
The residuals obtained from subtracting the distance moduli from the two methods yield an RMS scatter of $\sigma_{JAGB - TRGB}= \pm 0.07$~mag. 
Therefore, all systematics in either the JAGB method and TRGB method (e.g., crowding, differential reddening, star formation histories) must be contained within these  $\pm0.07$~mag bounds for this sample of galaxies because the JAGB and TRGB distance indicators are drawn from entirely distinct stellar populations, and are thus affected by these systematics independently.
Finally, the composite JAGB star luminosity function formed from this diverse sample of galaxies is well-described by a Gaussian function with a modal value of $M_J = -6.20 \pm 0.003$~~mag (stat), indicating the underlying JAGB star luminosity function of a well-sampled full star formation history is highly symmetric and Gaussian, based on over 6,700 JAGB stars in the composite sample.
\end{abstract}


\section{Introduction}
Carbon stars were first serendipitously discovered more than 150 years ago by Father Secchi at the Vatican Observatory, when he noticed the similarity in the spectra between a small group of peculiarly red stars and the light in carbon arc lights\footnote{The first practical electrical lights.} \citep{secchi}. 
More than 100 years later, \cite{1981ApJ...243..744R} laid the path for carbon stars as standard candles when he realized that the carbon star bolometric luminosity function was bright and fairly symmetric. 
Soon after, \cite{1986ApJ...305..634C} speculated the carbon star luminosity function could be used as a distance indicator after observing the similarities in the I-band luminosity function (LF) of carbon stars in M31, M33, NGC 6822, IC 1613, and M31.  
15 years later, \cite{2000ApJ...542..804N, 2001ApJ...548..712W} identified a region of the NIR CMD populated nearly exclusively by  carbon-rich AGB stars (identified photometrically) and successfully used them as standard candles to probe the 3D structure of the LMC. Twenty years later, \cite{2020ApJ...899...66M, 2020arXiv200510793F, 2020MNRAS.495.2858R} showed for the first time the promise of  carbon stars in the J band for measuring extragalactic distances, where they have a constant average magnitude. 

Carbon stars are theoretically predicted to be robust standard candles.
Intermediate-age and -mass stars inevitably evolve onto the Thermally-Pulsating Asymptotic Giant Branch (TP-AGB), a short final phase of a star's life which lasts only $\sim 10^6$ years \citep{2003agbs.conf.....H}. 
During this phase, the star alternates between burning helium and hydrogen in shells around a degenerate carbon-oxygen core. For most ($\sim90\%$) of the TP-AGB stage, the helium-burning shell is essentially dormant, while the hydrogen shell burns and rains helium ash onto the helium shell, increasing its pressure and temperature until helium fusion is eventually triggered. The helium shell burning phase is called a ``thermal pulse,'' and lasts a few hundred years.  As the helium shell fuses helium into carbon (via the triple-$\alpha$ reaction), it expands, forcing the hydrogen shell layer above it to expand as well, lowering its temperature and density and thereby terminating the hydrogen shell fusion. 
Furthermore, the energy from the helium shell burning is now too great to be transported through the star by radiation alone, so a convective shell develops above the helium burning region.
These convective cells may also transport carbon from the inner layers of the star up to the stellar surface and enrich it. This is called the ``third dredge-up event.'' The thermal pulse phase ends when the helium shell depletes its fuel and then contracts, which in turn increases the pressure and temperature in the hydrogen shell zone above it, re-triggering hydrogen fusion.
Between the thermal pulses, the star solely burns via the hydrogen shell for 10,000 to 100,000 years \citep{2003agbs.conf.....H}. The TP-AGB phase continues until the convective envelope dissipates via mass loss and the star becomes a planetary nebula and then eventually a white dwarf.

In some stars, a sufficient number of dredge-up events occur for the abundance of carbon to exceed that of oxygen on the stellar surface ($\rm{C/O}>1$),\footnote{CO is the most strongly bonded molecule in the atmospheres of cool stars. Therefore, if all the O is already bound in CO, surplus C atoms are then available to form $\text{C}_{\text{2}}$ and CN. The high dissociation energy (11.1 eV) of carbon monoxide therefore profoundly affects the distinction in the spectra and observational properties between O-rich and C-rich AGB stars.} and the star becomes known as ``carbon-rich.'' Conversely, the ``oxygen-rich'', or M-type AGB star predecessors to C-rich AGB stars, have ratios of $\rm{C/O}<1$ in their atmospheres. 
C stars' spectra are dominated by bands of the $\text{C}_{\text{2}}$ and CN molecules, which increase the opacity of the AGB star's stellar atmosphere in typical photometric bandpasses and are therefore responsible for the observed cooling of the star's effective temperature and carbon stars' distinctive red color \citep{2003A&A...403..225M}. A notable dichotomy exists in the effective temperatures and near-infrared colors between oxygen-rich and carbon-rich AGB stars, where carbon stars are distinctively cooler and redder.

Carbon stars are also well constrained in luminosity because of the small range of masses for which the third dredge-up event is effective at forming carbon stars ($\approx2-4.5 M_{\odot}$).\footnote{Theoretical models predict that increased metallicity decreases the third dredge-up's efficiency, and therefore the mass range of carbon star formation. For example, \cite{2020MNRAS.498.3283P} found that at an initial metallicity of $Z_i=0.004$ (the metallicity of the SMC), carbon stars formed at initial masses of $M_i\approx1.4M_{\odot}$ to $M_i\approx2.8M_{\odot}$, whereas at a higher metallicity of $Z_i=0.008$ (the metallicity of the LMC), carbon stars formed at initial masses of $M_i\approx1.7M_{\odot}$ to $M_i\approx3.0M_{\odot}$. Because the initial mass of the star translates to its luminosity, the metallicity of the stellar environment would be theoretically expected to affect a population's average carbon star luminosity. On the other hand, \textit{empirically}, \cite{2023arXiv230502453L} found zero metallicity dependence of the JAGB magnitude in the high-metallicity galaxy M31. However, further observational tests in a broader range of galaxy environments are needed to definitively constrain the JAGB method's dependence on metallicity.} 
Hot-bottom burning occurs in stars with initial masses of $M>4.5 M_{\odot}$, where the carbon is burned into nitrogen before it can reach the stellar surface, because the star's interior is so massive and therefore too hot.  Stars with initial masses of $M<2 M_{\odot}$ eject all the matter between the core and stellar surface after a few thermal pulses because the mass of the envelope was so small to begin with, and all evolution is terminated before the conversion to a carbon star can take place \citep{2003A&A...403..225M, 2021arXiv211205535K}.

These theoretical predictions are reflected observationally in the low dispersion of carbon star luminosities, particularly in the near infrared (0.3~mag, \citealt{2001ApJ...548..712W}), making these stars excellent candidates for standard candles.
Aptly, carbon stars with colors of $1.5<(J-K)<2.0$~mag, denominated J-region Asymptotic Giant Branch (JAGB) stars by \cite{2020ApJ...899...66M},  have been found to to be precise and accurate standard candles in the near infrared by several different groups and papers \citep{2001ApJ...548..712W, 2020arXiv200510793F, 2020MNRAS.495.2858R,2021ApJ...907..112, 2021MNRAS.501..933P, 2021arXiv210502120Z, 2022ApJ...933..201L, 2023MNRAS.tmp..926P}. 

The JAGB method has several strengths when compared with the tip of the red giant branch (TRGB) and Leavitt law (Cepheid) distance indicators.
First, with an intrinsic average magnitude of $M_J=-6.2$~mag, carbon stars are at least one magnitude brighter, in the near infrared, than TRGB stars (which have an average intrinsic magnitude of $M_J\approx-5.1$~mag), thereby allowing farther distances to be probed.
% Carbon stars are approximately the same brightness in the near infrared as a Cepheid with a period of 25 days; however, Cepheids must first be identified in the optical wavelengths where their amplitudes are large enough to be detected (due to the greater sensitivity of the surface brightness to temperature in optical wavelengths). 
% Therefore, JAGB stars overall require the least amount of observing time of the three distance indicators.
Second, only one epoch of observations is required to obtain a JAGB distance. Adding two to three epochs of data will significantly improve the precision of the JAGB measurement as discussed at the end of the section, but only one is necessary. On the other hand, more than a dozen epochs are needed to adequately measure the light curves (amplitudes, phases) and periods of Cepheids. 
Third, JAGB stars are found in all galaxies with stellar populations having ages between 200 Myr and 1~Gyr, whereas Cepheids can only be found in the young star-forming disks of dusty, gas-rich spiral and irregular galaxies. Fourth, JAGB stars are easily identified solely on the basis of their colors and magnitudes in the near infrared. Fifth, JAGB stars are best observed in near-infrared observations where effects of reddening are significantly decreased compared with optical wavelengths.

A significant limitation of the JAGB method is, however, the small number of galaxies to which it has currently been applied, historically largely due to limits of NIR facilities.
Now, in this paper, we present high-precision $JHK$ imaging of nearby galaxies to test and apply the JAGB method across a diverse sample of galaxies, using imaging optimized for the carbon stars and analyzed homogeneously. 

Furthermore, we measure NIR TRGB distances, a useful cross-checking tool for our JAGB distances, because any systematics affecting the JAGB method will be largely independent from those affecting the TRGB method in measuring distances. First, the JAGB method is based on a intermediate age and mass population of carbon-rich AGB stars (200 Myr - 1 Gyr, 2 - 4.5$M_{\odot}$), whereas RGB stars are significantly older and of lower mass ($>4$~Gyr, $<2M_{\odot}$). The JAGB method (optimally) uses stars in the extended disks of galaxies, whereas the TRGB targets stars only in the less-crowded and low-reddening (dust-free) stellar halos. Finally, the astrophysical mechanism by which these two methods are standard candles are completely independent; the JAGB method is based on the third dredge-up event for TP-AGB stars and the TRGB is controlled by the explosive onset of helium burning in the cores of red giants. Therefore, any systematic related to any of these aforementioned characteristics (e.g., crowding, star formation history, metallicity) will affect the methods in independent ways.  By comparing the two distances pairwise in the same galaxies we can constrain these effects by sampling galaxies having a wide range a diverse star-formation histories, metallicities, and dust environments.

This is the first study designed to provide a homogeneous sample for the purpose of measuring JAGB distances, where all galaxies have been observed with the same telescope/instrument and the data analyzed in a self-consistent manner. Previous studies that have calculated JAGB distances (which were then compared with Cepheid or TRGB distances) to multiple galaxies were based on available published data (e.g., \citealt{2020arXiv200510793F, 2021arXiv210502120Z, 2023MNRAS.tmp..926P}). Differences in the photometry software, telescope, and sample selection of JAGB stars could have introduced additional scatter in comparisons between different distance indicators.

\begin{figure}\figurenum{1}
\centering
\includegraphics[width=\columnwidth]{"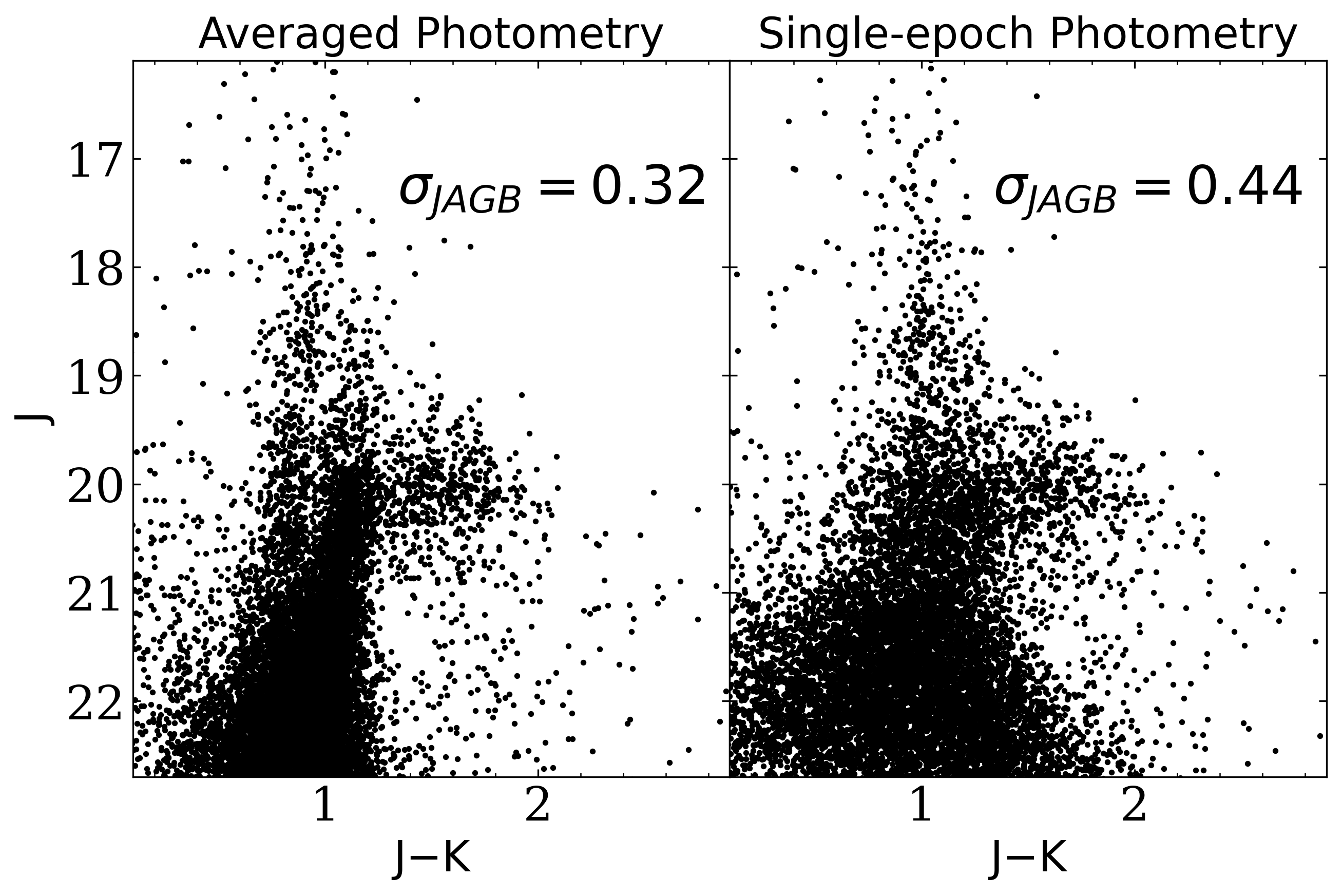"}
\caption{Example of how using time-averaged photometry decreases the measured scatter in the JAGB star LF. These two color-magnitude diagrams are of the galaxy NGC 300. On the left we show photometry averaged from five nights of observing, and on the right, we show photometry from only one night of observing. The scatter shown in the upper right-hand corner was measured from the JAGB stars with magnitudes $m_J\pm0.75$~mag, where $m_J$ is the measured mode. The CMDs have the same stars.}
\label{fig:scatter}
\end{figure}

Furthermore, this study is idealized because we use multi-epoch photometry to increase the precision of our measurements.
Most JAGB stars are variable \citep{2001ApJ...548..712W}. 
The intrinsic variability of JAGB stars will thus contribute to the observed dispersion in the JAGB luminosity function.
The intrinsic scatter in the JAGB LF resulting from the variability of JAGB stars is $\pm0.2$~mag \citep{2020ApJ...899...66M}.
Simply averaging two or more randomly observed epochs will bring down this scatter as $1/\sqrt{N_e}$, where $N_e$ is the number of widely spaced epochs. For example, in Figure \ref{fig:scatter}, we show two color-magnitude diagrams of NGC 300, one derived from one epoch of observing and one derived from three temporally-averaged widely-spaced epochs of observing (although these data encompasses five total days, September 9 2011, October 5 2011, November 3 2011, September 17 2021, September 17 2022), the data only span three total years). The observed scatter (which includes the intrinsic scatter from the AGB stars' variability plus random scatter like photometric errors) measured from single-epoch data was 0.44~mag, whereas the scatter determined from temporally-averaged photometry was measured to be 0.32~mag. This example demonstrates that JAGB measurements are more precise when derived from multiple epochs of data. 
Therefore, the data presented in this study are optimal for measuring high-precision JAGB distances.

The outline of this paper is as follows. In Section \ref{sec:data}, we describe our Magellan FourStar observations and data reduction procedure. In Section \ref{sec:jagb}, we describe the JAGB method and measured distances. In Section \ref{sec:trgb}, we measure NIR TRGB distances to the galaxies in our sample. In Section \ref{sec:literature}, we compare our measured JAGB and TRGB distances with those from the literature. And finally, in Section \ref{sec:summary}, we summarize and conclude this paper. 

\section {Data}\label{sec:data}

\begin{deluxetable*}{cccccccc}
\tablecaption{Galaxy Sample Properties (Ordered by Distance)}\label{tab:observ}
\tablehead{
\colhead{Galaxy} & 
\colhead{R.A. (J2000)} & 
\colhead{Dec. (J2000)} & 
\colhead{Morphological Type} & 
\colhead{$A_J$\tablenotemark{a}}  & 
\colhead{$d$ (Mpc)\tablenotemark{b}} &
\colhead{Galaxy Group} & 
\colhead{JAGB LF Width (mag)\tablenotemark{c}}}
\startdata
NGC 6822 & 19 44 55 & $-$14 48 00 &Irregular & 0.17 & 0.5 & Local Group & 0.30 \\
IC 1613& 	01 04 48 & +02 07 04  & Irregular  & 0.02 & 0.8 & Local Group & 0.32 \\ 
NGC 3109  &10 03 09 & $-$26 09 25 & 	Spiral & 0.05  & 1.2 & Local Group & 0.33\\ 
Sextans B   &  09 59 59& +05 20 20 & 	Irregular & 0.02  & 1.3 & Local Group & 0.30\\ 
Sextans A &  10 11 01 & $-$04 42 00 & Irregular  &  0.03 & 1.4 & Local Group & 0.30\\ 
NGC 0300 &  00 54 54 & $-$37 41 04 &  	Spiral & 0.01  & 1.8 & Sculptor Group & 0.32\\
NGC 0055 & 00 14 57 & $-$39 11 48 &  	Spiral & 0.01 & 1.9 & Sculptor Group & 0.31 \\
NGC 7793 &  23 57 50 & $-$32 35 28 & Spiral & 0.01 & 3.1 & Sculptor Group & 0.35 \\
NGC 247 & 00 47 09 & $-$20 45 37 & Spiral & 0.01 & 3.2 & Sculptor Group & 0.32\\
NGC 5253 & 13 39 56 & $-$31 38 24  & Amorphous & 0.04  & 3.3 & Cen A/M83 Group & 0.33\\
Cen A & 13 25 28 & $-$43 01 09 & E/S0 (pec) & 0.08 & 3.7 & Cen A/M83 Group & \nodata \\
NGC 1313 & 03 18 15 & $-$66 29 50 & Spiral  & 0.08 &  3.9 & Isolated  & 0.36\\
M83 & 13 37 01 & $-$29 51 57 & Spiral & 0.05 & 4.9 & Cen A/M83 Group & \nodata \\
\enddata
\tablenotetext{a}{Tabulated from NED: \url{https://irsa.ipac.caltech.edu/applications/DUST/}}
\tablenotetext{b}{The JAGB distances measured in this paper. For Cen A and M83, distances are from the Extragalactic Distance Database }
\tablenotetext{c}{Scatter on the mode $m_{JAGB}$ for stars between $m_{JAGB}\pm0.75$~mag}
\end{deluxetable*}

\begin{figure*}
\centering\figurenum{2}\label{fig:imaging}
\includegraphics[width=.85\textwidth] {"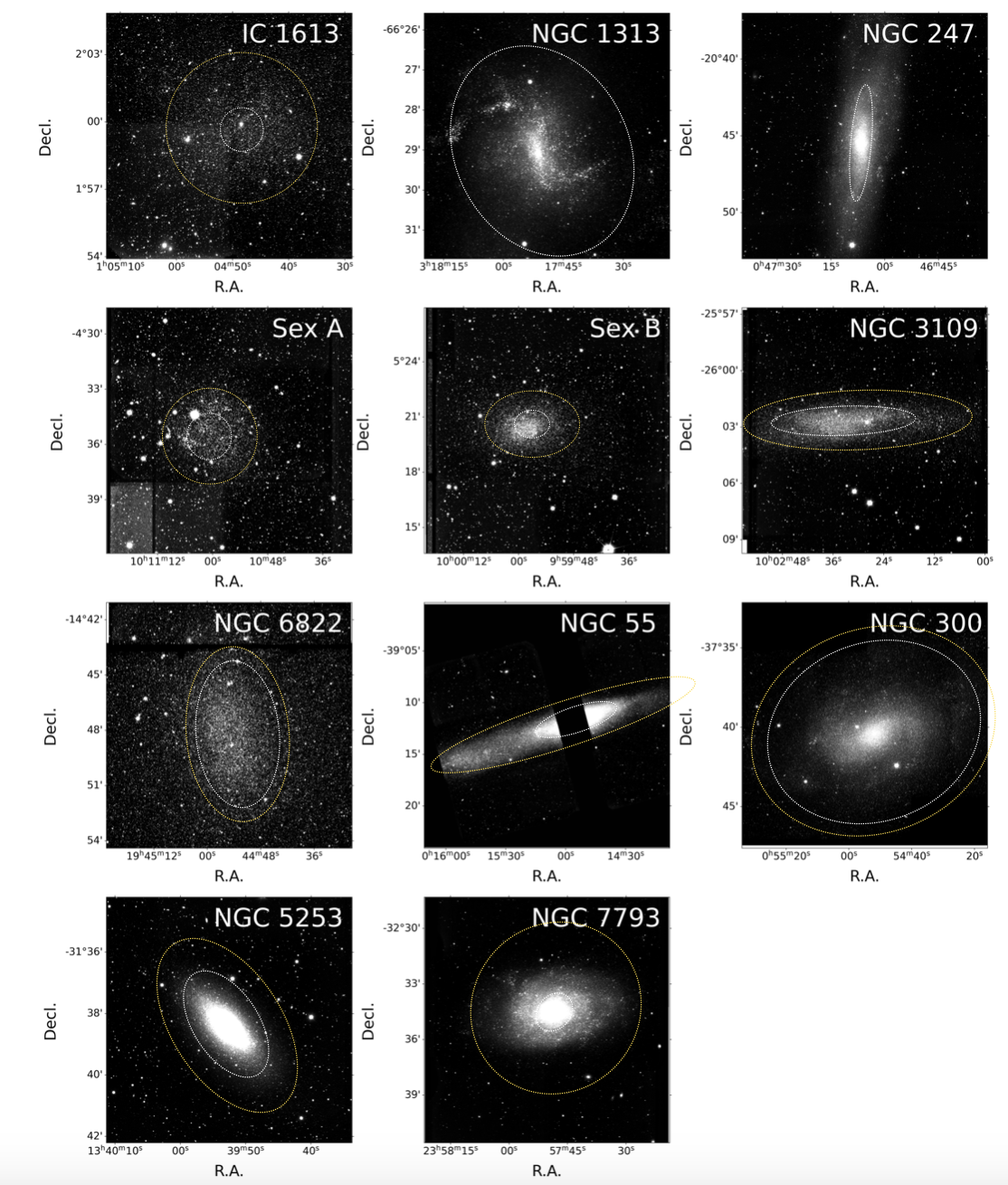"}
\caption{Images of the 11 galaxies observed with the FourStar camera on the 6.5 m Magellan-Baade telescope. The JAGB measurements were measured from all the data outside the dotted white ellipses and the TRGB measurements were measured from all the data} outside the dotted yellow ellipses.
\end{figure*}

We obtained near-infrared $JHK$ observations of the 13 galaxies at the 6.5 m Magellan-Baade telescope at Las Campanas Observatory with the wide-field FourStar near-infrared camera \citep{2013PASP..125..654P}. 
The FourStar imager has a field of view of $10\arcmin.8 \times 10\arcmin.8$ and a plate scale of $0.\arcsec 159~ \rm{pixel}^{-1}$.
The names and properties of these 13 galaxies are listed in Table \ref{tab:observ}.
The FourStar images of the galaxies are shown in Figure \ref{fig:imaging}. A log of our observations, which were taken from 2011 to 2023, can be found in Table \ref{tab:observloog}. These observations were optimized for a study of JAGB stars in three ways: 
 \begin{enumerate}
     \item For JAGB stars, only single-epoch photometry is needed to straightforwardly derive distances. However, by deriving mean photometry from multiple epochs, the intrinsic variability of the JAGB stars can be averaged over, thereby decreasing the dispersion in the observed luminosity function. All of our galaxies have at least two observations in the J band. 
     \item Our observations targeted the outer disks of galaxies, where reddening, crowding, and blending effects are minimized for the carbon stars.
     \item All of these observations were taken with the same telescope and photometry was extracted uniformly using the same suite of photometry software.
 \end{enumerate}

Photometry was extracted from the images using the \textsc{daophot/allframe} suite of photometry software \citep{1987PASP...99..191S}. For a given galaxy, empirical PSFs were fit individually for each frame. A master source list was then constructed from an aligned and co-added image built from all of the individual images using \textsc{montage2}. We then photometered each of the individual frames simultaneously using the master source list. 

Our final catalogs were cleaned using photometric quality cuts based on the J-band $\chi$, sharp, and $\sigma$ parameters of each star, following the procedure performed in \cite{2019ApJ...885..141B} for the Carnegie-Chicago Hubble Program (CCHP).
Photometric zero-points with respect to the Two Micron All-Sky Survey (2MASS) \citep{2006AJ....131.1163S} were derived by matching bright, unsaturated stars in our catalogs to 2MASS. We considered the following uncertainties associated with the photometric calibration in our error budget for both the JAGB method and TRGB (Tables \ref{tab:jagb_error} and \ref{tab:trgb_error}, respectively): the 2MASS photometric zeropoints (0.01 mag; \citealt{2006AJ....131.1163S}) and the statistical uncertainty on the (2MASS - Fourstar) transformation.

The deprojected galactocentric distance for every object in the catalogs was also calculated using the galaxy's position angle, inclination, and galactic center. These radial distances were then used to perform spatial cuts for the JAGB method (which optimally uses stars in the outer disk and halo) and TRGB method (which optimally uses stars in the halo). The spatial cuts used are shown in Figure \ref{fig:imaging}. 

All the final cleaned catalogs can be found at doi:\href{https://zenodo.org/records/10606945}{10.5281/zenodo.10606944}.
Although the primary purpose of these observations was to measure JAGB distances, these catalogs could be used for other purposes. For example, these data could enable further studies on the near-infrared flux contribution of TP-AGB stars to the integrated luminosities of galaxies (e.g., \citealt{2013ApJ...764...30M}), reconstructing the star formation history of a galaxy from NIR data (e.g., \citealt{2021MNRAS.508..245M}), tracing the age and metallicity gradients of galaxies using the C/M ratio of AGB stars (e.g., \citealt{2008A&A...487..131C}), and studying the red supergiant content of galaxies (e.g., \citealt{2021AJ....161...79M}).

\begin{deluxetable}{cccr}
\tablecaption{Observation Log}\label{tab:observloog}
\tablehead{
\colhead{Galaxy} & 
\colhead{Date (y m d)} & 
\colhead{Filter(s)}}
\startdata
NGC 6822 & 2012 05 03 & JHK\\
NGC 6822 & 2012 05 07 & JHK\\
NGC 6822 & 2012 05 11 & JHK\\
NGC 6822 & 2012 05 28 & JHK\\
NGC 6822 & 2012 06 01 & K \\
\enddata
\tablecomments{Table 2 is published in its entirety in the machine-readable format.
      A portion is shown here for guidance regarding its form and content.}
\end{deluxetable}

\section{The J-region Asymptotic Giant Branch method}\label{sec:jagb}

In this section, we determine JAGB method distances to the galaxies in our sample.

\subsection{Measurement Procedure}\label{subsec:jagb_measure}

JAGB stars are selected via their color to delineate them from oxygen-rich AGB stars on the blue side and extreme carbon stars on the red side. However, the exact colors used to select JAGB stars has varied inconsistently throughout the literature; in particular, the blue cutoff has ranged from $(J-K)>$ 1.3 to 1.5.
With our current homogeneous sample of galaxies, we are now in a position to standardize the color range for selecting JAGB stars. Moving forward, we chose $1.5<(J-K)<2.0$~mag as the standard JAGB color cuts for every galaxy. The colors of carbon stars depend on the metallicity of the galaxy, where carbon stars have a bluer cutoff in low-metallicity environments. For example, \cite{2024arXiv240114889B} simulated carbon stars at two metallicities, Z = 0.004 and Z = 0.014, and found the carbon stars at Z = 0.004 spanned approximately $1.3<(F090W-F150W)<4.5$~mag in color, whereas the carbon stars at Z = 0.014 spanned approximately $1.8<(F090W-F150W)<4.5$~mag in color. Therefore, by selecting carbon stars in a more conservative color range, we are prioritizing  carbon star sample purity over completeness, particularly in the higher-metallicity galaxies.  
Furthermore, in section \ref{sec:literature}, we found excellent agreement between the independent TRGB and JAGB distances, demonstrating that these chosen color cuts are effective for measuring accurate JAGB distances. 

 JAGB measurements are optimally performed in the outer disks of galaxies, where the JAGB stars are minimally affected by reddening, crowding, and blending \citep{2022ApJ...933..201L}.
Therefore, first, we performed spatial cuts on our photometry, only using data in the outer disks. The spatial cuts are shown by the dotted white ellipses in Figure \ref{fig:imaging}. Data inside these ellipses were excluded from the analysis.

\begin{figure*}\figurenum{3}\label{fig:jagb_cmd}
\centering
\includegraphics[width=\textwidth]{"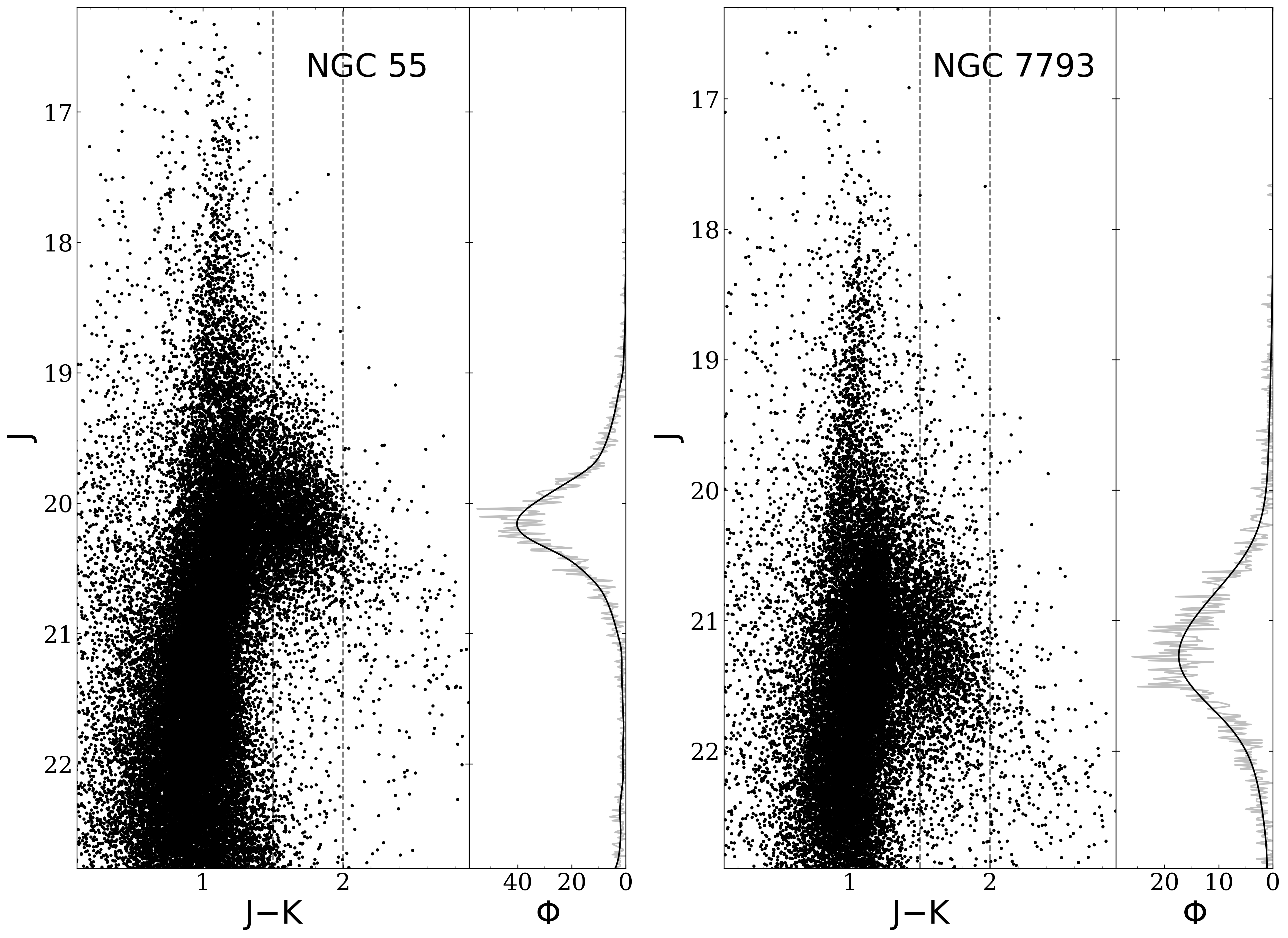"}
\caption{J vs. (J-K) color-magnitude diagrams (left panels) and GLOESS-smoothed luminosity functions in black overplotted on the binned LF in grey (right panels) for two example galaxies, NGC 55 and NGC 7793. The JAGB stars are located in the color range $1.5<(J-K)<2.0$~mag, between the dotted grey lines. The CMDs for the other galaxies can be found in Appendix \ref{sec:jagb_cmd_cont}. The range of the y-axis is 6.6~mag for all the CMDs.}
\end{figure*}

Then, the JAGB stars were selected solely on the basis of their color, using color cuts of $1.5<(J-K)<2.0$~mag. 
Next, we generated the JAGB star luminosity function by binning the J-band magnitudes of these stars using bins of 0.01~mag. We then smoothed the binned luminosity function using a Gaussian-windowed, locally weighted scatterplot smoothing (GLOESS) algorithm, a data-smoothing interpolating technique that has been found to be effective at suppressing false, Poisson noise-induced edges and peaks in luminosity functions \citep{loess_ref, loader, 2004AJ....128.2239P, 2017AJ....153...96M}. The peak location of this luminosity function marks the apparent JAGB magnitude. 
The only user input in this procedure is the smoothing parameter $\sigma_s$. We explored the systematics of the choice of $\sigma_s$ later in Section \ref{subsec:smoothing}. 
The color magnitude diagrams and smoothed luminosity functions for 11 of the galaxies are shown in Figure \ref{fig:jagb_cmd}. Two galaxies, Cen A and M83 (the third farthest and farthest galaxies in this sample, respectively), had significantly lower-quality CMDs and JAGB star luminosity functions than the other 11 galaxies in our main sample. We discuss these two galaxies in Appendices \ref{subsec:cena} and \ref{subsec:m83}, and excluded them from the JAGB and TRGB analyses.

To measure the distance modulus to each galaxy, we used the JAGB zeropoint from \cite{2022ApJ...938..125M} of $M_J=-6.20\pm0.01 ~\rm{(stat)} \pm0.04$ (sys)~mag, which combines the geometric calibrations of the JAGB method from the LMC/SMC detached-eclipsing binaries \citep{2020ApJ...899...66M}, Milky Way Gaia DR3 parallaxes \citep{2021ApJ...923..157L}, and Milky Way open clusters \citep{2022ApJ...938..125M}.

All of the apparent magnitudes were corrected for their line-of-sight foreground Galactic extinction, adopted from the \cite{1998ApJ...500..525S} full-sky Galactic dust map recalibrated by \cite{2011ApJ...737..103S}.\footnote{Determined from \url{https://irsa.ipac.caltech.edu/applications/DUST/}} These corrections are tabulated in Table \ref{tab:observ}. We adopted half of the galactic extinction value as its systematic uncertainty, per CCHP procedure (e.g., \citealt{2019ApJ...885..141B}).
Error budgets for each JAGB method measurement are shown in Table \ref{tab:jagb_error}. The error on the mode of the smoothed luminosity function was measured as $\sqrt{\Sigma({m_i-m_{JAGB}})^2}/N$, where $m_{JAGB}$ is the mode, for stars with magnitudes $m_{JAGB}\pm0.5$~mag.

\subsection{Effect of Smoothing Parameter Choice}\label{subsec:smoothing}

To measure the effect of the choice of smoothing parameter $\sigma_s$, we varied the smoothing parameter from 0.15~mag to 0.40~mag in steps of 0.05~mag. We then adopted the largest change from the fiducial values listed in Table \ref{tab:jagb_error} as the statistical error due to the smoothing parameter choice, a procedure that was also utilized in \cite{2022ApJ...933..201L}. For example, in NGC 300, the faintest mode was measured using a smoothing parameter of $\sigma_s=0.15$~mag: $m_{JAGB}=20.04$~mag. The brightest mode was measured using a smoothing parameter of $\sigma_s=0.40$~mag: $m_{JAGB}=20.00$~mag. For this galaxy, the fiducial mode of $m_{JAGB}=20.01$~mag was measured using a smoothing parameter of $\sigma_s=0.25$~mag. The adopted statistical error due to the smoothing parameter for NGC 300 was then 0.02~mag.

However, robust statistical errors due to the effect of smoothing parameter would be most accurately measured with a large sample of artificial stars.
In the second paper of this series, we plan to show results for the effects of the smoothing parameter on the JAGB measurement error via artificial star tests. For now, however, we adopted these conservative statistical errors.

% \subsection{Internal Reddening}
% Therefore, because we have only used outer disk data for our sample of galaxies, we assume negligible error due to the internal reddening. 

\begin{deluxetable*}{c|ccccc|cccc}\tablenum{3}
\tablecaption{JAGB Method Systematic (left) and Statistical (right) Error Budget}\label{tab:jagb_error}
\tablehead{
\colhead{Galaxy} & 
\colhead{Galactic} & 
\colhead{2MASS} &
\colhead{2MASS - Fourstar} & 
\colhead{Zeropoint} & 
\colhead{Total} & 
\colhead{Error on} & 
\colhead{Choice} & 
\colhead{Zeropoint} & 
\colhead{Total} \\
\colhead{} & 
\colhead{Extinction} &
\colhead{Phot. ZP} &
\colhead{Correction}  & 
\colhead{} & 
\colhead{$\sigma_{sys}$} & 
\colhead{the mode} &
\colhead{of $\sigma_s$} & 
\colhead{} & 
\colhead{$\sigma_{stat}$}\\
\colhead{} & 
\colhead{(mag)} &
\colhead{(mag)} & 
\colhead{(mag)} & 
\colhead{(mag)} & 
\colhead{(mag)} & 
\colhead{(mag)} & 
\colhead{(mag)} & 
\colhead{(mag)} & 
\colhead{(mag)}
}
\startdata
NGC 6822 & 0.09 & 0.01 & 0.01 &0.04 & 0.10 &0.02 &0.05&0.01 & 0.05\\
IC 1613&  0.01  & 0.01 &0.01& 0.04&0.04&0.04 &0.09&0.01 & 0.10\\
NGC 3109 & 0.03 & 0.01 &0.01 &0.04&0.05&0.03 &0.02&0.01 & 0.04 \\
Sextans B& 0.01  & 0.01 & 0.01& 0.04&0.05& 0.03&0.09&0.01 & 0.09\\
Sextans A& 0.02  & 0.01 & 0.01&0.04&0.05&0.05 &0.13&0.01  & 0.14\\
NGC 0300& 0.01  & 0.01 & 0.02&0.04&0.05&0.02&0.02 &0.01 & 0.03 \\
NGC 0055& 0.01 & 0.01 &0.01& 0.04&0.04&0.01 &0.02 &0.01 & 0.02\\ 
NGC 7793& 0.01 & 0.01 &  0.02&0.04&0.05& 0.01 & 0.10 &  0.01 & 0.10\\
NGC 247& 0.01  & 0.01 &0.01&0.04 &0.04 &0.01 &0.02 &0.01 &0.02 \\ 
NGC 5253& 0.02 & 0.01 & 0.01& 0.04&0.05 & 0.02&0.03&0.01 & 0.04\\
NGC 1313& 0.04  & 0.01 & 0.01&0.04& 0.06& 0.01 &0.02&0.01 & 0.03\\
\enddata
\end{deluxetable*}

\subsection{A Composite JAGB Luminosity Function}

\begin{figure}\figurenum{4}
\centering
\includegraphics[width=\columnwidth]{"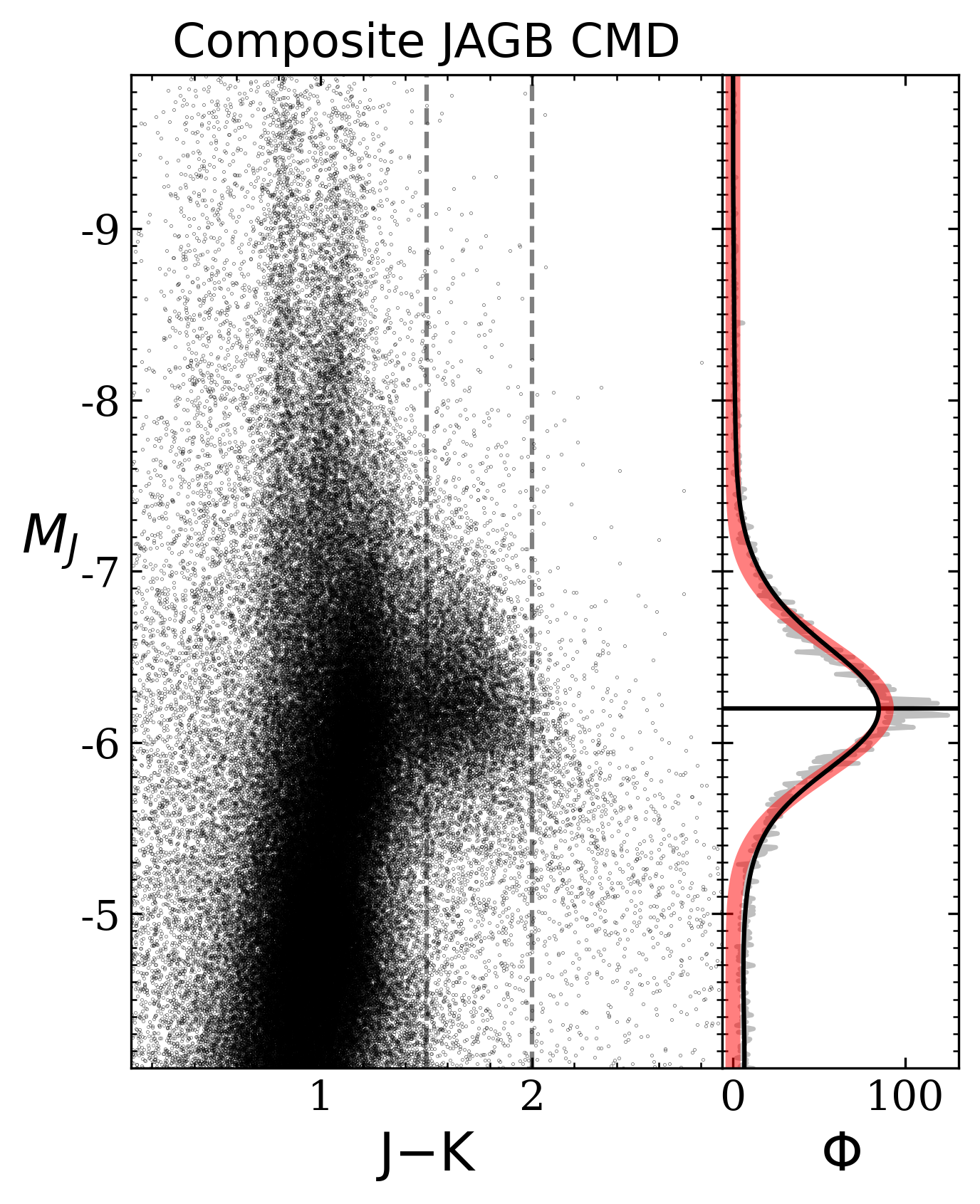"}
\caption{Composite distance-corrected CMD for the 11 galaxies with JAGB distances in this paper. The smoothed JAGB star luminosity function is shown in black in the right panel, with a Gaussian fit in red over-plotted. The black line marks the modal value at $M_J = $ -6.20 $\pm$
0.003~mag. }
\label{fig:composite}
\end{figure}

Figure \ref{fig:composite} shows a distance-corrected composite CMD of the 11 galaxies for which we measured JAGB distances. In this CMD, 
there are a total of $\sim$6700 JAGB stars plotted $M_J\pm0.5$~mag, where $M_J$ was the measured mode of the composite JAGB star luminosity function.
The measured modal value was $M_J =  -6.20 \pm$ 0.003 (stat)~mag.
This JAGB star luminosity function has a measured dispersion of 0.24~mag for stars with magnitudes of $M_J\pm0.5$~mag. 
 In Figure \ref{fig:composite}, we also show a red Gaussian fit to the binned luminosity function overlaid ontop of the black smoothed luminosity function. 

The composite JAGB star luminosity function is symmetric and closely Gaussian in form. The 11 galaxies plotted here span a wide range of star formation histories, indicating the underlying JAGB star luminosity function of a (well-sampled) full star formation history is nearly perfectly symmetric and Gaussian.

\section{Near-infrared Tip of the Red Giant Branch}\label{sec:trgb}

 The tip of the red giant branch marks the onset of the helium flash for low-mass stars ($<2M_{\odot}$), when the star's degenerate helium core reaches a sufficiently high density and temperature of about $10^8$~K via hydrogen shell fusion. At that point, the degeneracy of the core is lifted and the core reaches a hot enough temperature to begin helium burning   \citep{1997MNRAS.289..406S}. The star then settles onto the horizontal branch or red clump, depending on the initial mass and metallicity of the star.  

In near-infrared wavelengths, the tip of the red giant branch can be observed empirically as an upward-sloping discontinuity in a color magnitude diagram of red giant branch stars. The TRGB is brighter in the NIR than in the optical, and therefore can probe farther distances than the more conventionally-used I-band TRGB, which is flat with color \citep{2018ApJ...858...11M, 2018ApJ...858...12H, 2019ApJ...880...63M, 2020ApJ...891...57F}. 

With our high-precision and well-sampled near-infrared data, we can also measure tip of the red giant branch distances. To measure the NIR TRGB, we first performed spatial cuts on our data to exclude the disks of the galaxies, which have been shown to be problematic for TRGB measurements (e.g., \citealt{2020arXiv200804181J}).
It is imperative that the TRGB be measured in the stellar halos of galaxies, where there is less contamination from the younger AGB stars which may populate the region of color-magnitude space immediately above the TRGB. Furthermore, effects of crowding and reddening are significantly diminished in the stellar halo.
Our spatial cuts are shown in Figure \ref{fig:imaging} by the dotted yellow ellipses. Data inside the ellipse for each galaxy were removed. 

\begin{figure*}\figurenum{5}\label{fig:cmd1_trgb}
\centering
\includegraphics[width=.95\textwidth]{"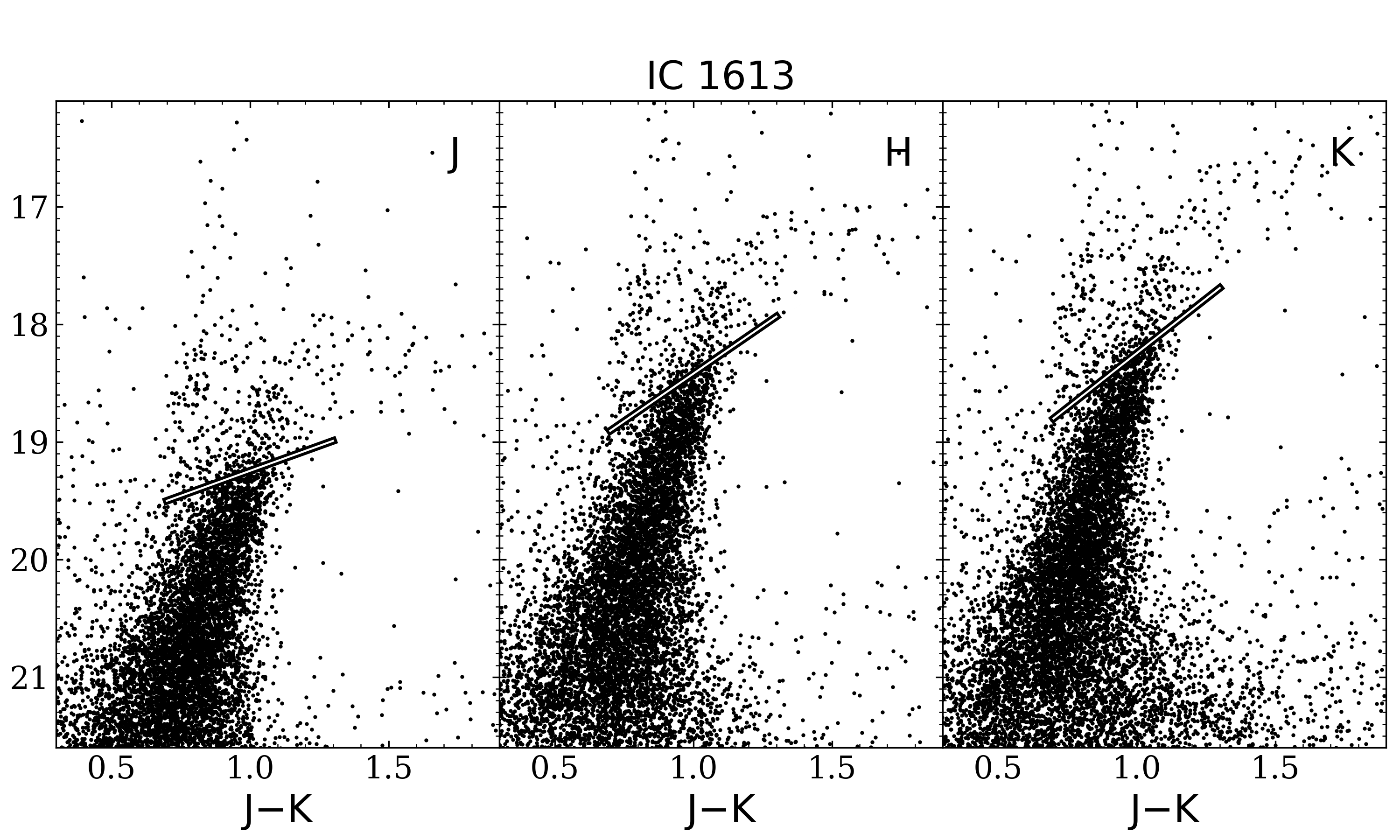"}
\caption{Near-infrared CMDs for one example galaxy, IC 1613.} The solid line shows the measured TRGB detection in the J and H bands. The TRGB was measured for stars with $0.7<(J-K)<1.3$~mag color, which is shown by the width of the solid line. The J-band TRGB was then projected into the K band and shown here in the right panel for each galaxy. The CMDs for the other galaxies can be found in Appendix \ref{sec:jagb_cmd_cont}.
\end{figure*}

Next, we selected the RGB stars using color cuts of $0.7<(J-K)<1.3$. Then, we employed the TRGB detection method first introduced by \cite{2009ApJ...690..389M}. The data were transformed into T-band magnitudes, or `rectified,' using slopes from \citealt{2020ApJ...891...57F} (determined using LMC TRGB stars). The T-band magnitudes are designed to be insensitive to metallicity, thus leading 
the TRGB to be flat with color. For each galaxy, we constructed $T[J, (J-K)]$ and $T[H, (J-K)]$ magnitudes, excluding the K-band because $m_J^{TRGB}$ and  $m_K^{TRGB}$ are mathematically equivalent (see \citealt{2020AJ....160..170M}). Then, the T-band RGB star luminosity function for each band was constructed by finely binning the T-band magnitudes using bins of 0.01~mag. The T-band luminosity functions were then smoothed using the GLOESS smoothing algorithm discussed in Section \ref{subsec:jagb_measure} using a smoothing parameter of $\sigma_s=0.10$~mag.
Smoothing filters have already been widely adopted in measuring the TRGB to suppress false edges in the luminosity functions (due to noise).
To measure the discontinuity in the T-band luminosity function, we employed the method consistently used by the CCHP, introduced in \cite{1993ApJ...417..553L}. First, the smoothed luminosity function was convolved with a Sobel edge detection kernel [-1,0,+1], resulting in a discrete approximation of the first derivative of the luminosity function, i.e. the edge response function. The Sobel filter response was then weighted inversely by the Poisson noise calculated in the adjacent bins in the smoothed luminosity function (described in \citealt{2017ApJ...845..146H}). The maximum value of the edge response then marks the TRGB. Finally, we transformed the T-band TRGB values back into J- and H-band magnitudes to determine the final near-infrared TRGB intercepts. This rectification method has also been employed to measure the NIR TRGB in \cite{2018ApJ...858...11M, 2019ApJ...880...63M, 2020arXiv201209701C, 2022ApJ...933..201L, 2023MNRAS.tmp..926P}. The T-band luminosity functions and edge detection response functions can be found in Appendix \ref{sec:tbandlf}.
The J, H, and K vs. (J$-$K) color magnitude diagrams and TRGB measurements are shown in Figure \ref{fig:cmd1_trgb}.  

\cite{2009ApJ...690..389M} showed that with at least 400 stars below the TRGB, the tip could be measured to within $\sim0.1$~mag at the 85\% confidence level. Then, in a recently updated paper on simulations of TRGB detections across a range of smoothing parameters, photometric errors, and number of RGB stars, \cite{2023AJ....166....2M} found that ``one should not even attempt a tip detection at low signal-to-noise in situations where the [TRGB] population size is only in the hundreds. Spurious signals will be found above and below the true tip." In particular, for a population of 120 RGB stars with average photometric errors of $\pm0.05$~mag and a smoothing parameter of $\sigma_s=0.10$~mag, wild statistical fluctuations in both the LF and edge response function led to false positive detections $\pm1$~mag above and below the TRGB.
Therefore, we have only measured TRGB distances to galaxies in our sample that have ample TRGB halo stars.
NGC 1313 and NGC 247 did not have deep enough data to do any TRGB measurement so were excluded entirely. NGC 7793 did not have deep enough H-band data and NGC 5253 lacked H-band data, so these galaxies only have J-band TRGB measurements. 
We also note that \cite{2023AJ....166....2M} found that for 1200 stars in the RGB which had photometric errors of $\pm0.05$~mag and a LF smoothed with a smoothing parameter of $\sigma_s=0.10$~mag, the tip detection was found to accurate to $\pm0.03$~mag. We confirmed the photometric errors of the RGB stars $\pm0.2$~mag of the TRGB in this sample were typically $\pm0.03$ to $\pm0.05$ mag.

We determined distance moduli based on the absolute calibrations from \cite{2018ApJ...858...12H}, which we repeat below. The errors on these zeropoints are $\pm0.01$ (stat) and $\pm0.06$ (sys)~mag. 
\begin{equation}
M_J^{TRGB}=-5.14-0.85 \times [(J-K)_o - 1.00]
\end{equation}
\begin{equation}
M_H^{TRGB}=-5.94-1.62\times[(J-K)_o - 1.00]
\end{equation}

\begin{figure}
\centering
\includegraphics[width=\columnwidth]{"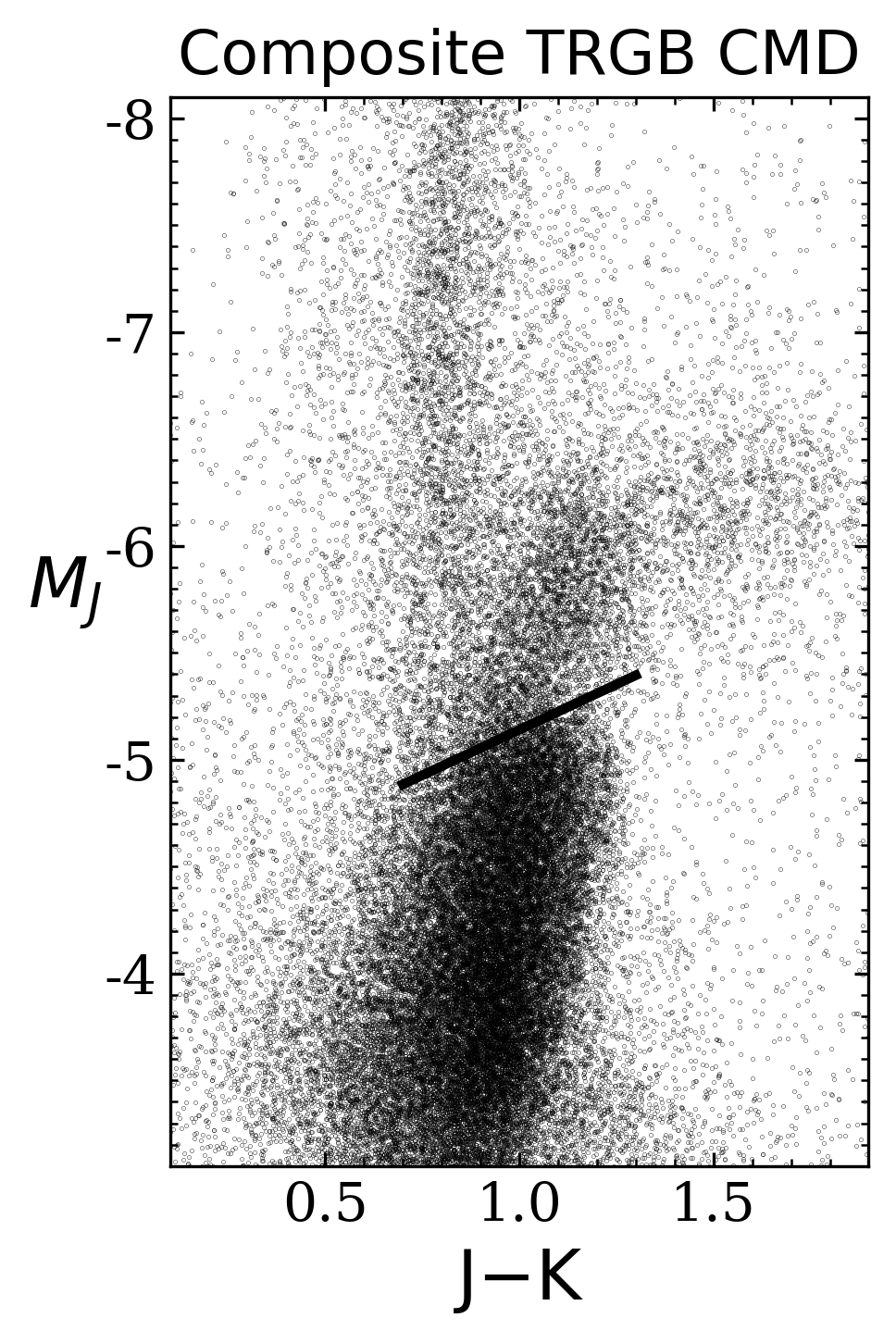"}\figurenum{6}
\caption{Composite distance-corrected CMD for the 9 galaxies with TRGB distances in this paper. The black line denotes the J-band TRGB. }
\label{fig:composite_trgb}
\end{figure}

In Table \ref{tab:trgb_error}, we tabulate all the uncertainties for each TRGB distance. This includes as a statistical uncertainty derived from the width of the T-band LF edge response divided by the square root of the number of stars contributing to the response at that magnitude \citep{2019ApJ...885..141B}. As in Section \ref{sec:jagb}, we also applied galactic extinction corrections to the apparent magnitudes, and adopted half of the extinction value as its systematic uncertainty.  
% All of these TRGB measurements have been performed in the low-reddening stellar halos of galaxies, and in the near infrared band-passes where effects are reddening are significantly diminished. Therefore, we assume negligible internal reddening for our NIR TRGB measurements. 
In Table \ref{tab:distmod}, we also list the final measured TRGB values and extinction corrections for each galaxy. Finally, in Figure \ref{fig:composite_trgb}, we show a composite distance-corrected CMD of the nine galaxies with TRGB distances in this paper.

\begin{deluxetable*}{l|ccccc|ccc}
\tablecaption{NIR TRGB Systematic (left) and Statistical (right) Error Budgets}\tablenum{4}\label{tab:trgb_error}
\tablehead{
\colhead{Galaxy} & 
\colhead{Galactic} & 
\colhead{2MASS} & 
\colhead{2MASS - Fourstar} & 
\colhead{Zeropoint} & 
\colhead{Total} & 
\colhead{Width of} & 
\colhead{Zeropoint} & 
\colhead{Total} \\
\colhead{} & 
\colhead{Extinction} &
\colhead{Phot. ZP} & 
\colhead{Correction} & 
\colhead{} & 
\colhead{$\sigma_{sys}$} & 
\colhead{Edge Response} & 
\colhead{} & 
\colhead{$\sigma_{stat}$}\\
\colhead{} & 
\colhead{(mag)} &
\colhead{(mag)} & 
\colhead{(mag)} & 
\colhead{(mag)} & 
\colhead{(mag)} & 
\colhead{(mag)} & 
\colhead{(mag)} & 
\colhead{(mag)}
}
\startdata
NGC 6822 (J) & 0.09 & 0.01 &  0.01 & 0.06  & 0.11 &   0.002 & 0.01 & 0.01  \\
NGC 6822 (H) & 0.06 & 0.01 & 0.01 & 0.06 & 0.09 &0.002  & 0.01 & 0.01  \\
IC 1613 (J)&  0.01 & 0.01 & 0.01& 0.06 & 0.06 &  0.003& 0.01  & 0.01\\
IC 1613 (H) & 0.01 &  0.01& 0.01 & 0.06 & 0.06 & 0.003 & 0.01 & 0.01 \\
NGC 3109 (J) & 0.05 & 0.01 & 0.01& 0.06 & 0.08  & 0.005& 0.01 & 0.01\\
NGC 3109 (H) & 0.03 & 0.01 & 0.01& 0.06 & 0.07 &  0.004& 0.01  &0.01\\
Sextans B (J) & 0.02 & 0.01 & 0.02& 0.06 & 0.07  & 0.005& 0.01   & 0.01\\
Sextans B (H) & 0.01  & 0.01 & 0.01& 0.06 & 0.06 & 0.004 & 0.01 & 0.01\\
Sextans A (J) & 0.03 & 0.01 & 0.01& 0.06 & 0.07  & 0.007& 0.01  & 0.01\\
Sextans A (H) & 0.02 & 0.01 & 0.02& 0.06 & 0.07  & 0.007 & 0.01 & 0.01\\
NGC 0300 (J) & 0.01  & 0.01 & 0.02& 0.06 & 0.06  & 0.002& 0.01  & 0.01\\
NGC 0300 (H) & 0.01  & 0.01 & 0.01& 0.06 & 0.06  & 0.002& 0.01  & 0.01\\
NGC 0055 (J) & 0.01 & 0.01& 0.01 & 0.06 & 0.06 &  0.003& 0.01  & 0.01\\
NGC 0055 (H) & 0.01 & 0.01& 0.01 & 0.06 & 0.06 & 0.002 & 0.01 & 0.01 \\
NGC 7793 (J) & 0.01 & 0.01& 0.02& 0.06 & 0.07 &  0.004& 0.01  & 0.01\\
% NGC 247 (J) & 0.01 & 0.05 & 0.06 & 0.08 & 0.04 & 0.01 & 0.05 \\ 
% NGC 247 (H) & 0.01  & 0.13 & 0.06 & 0.15 & 0.04 & 0.01 & 0.04\\
NGC 5253 (J)& 0.04 & 0.01 & 0.01 & 0.06 & 0.07& 0.002 & 0.01 & 0.01\\
\enddata
\end{deluxetable*}

\section{Distance Comparisons}\label{sec:literature}

\begin{deluxetable}{cccccc}
\tablecaption{Measured JAGB and TRGB apparent magnitudes and relevant foreground extinctions}\tablenum{5}\label{tab:distmod}
\tablehead{
\colhead{Galaxy} & 
\colhead{$m^{JAGB}$} & 
\colhead{$m^{TRGB}_J$} &
\colhead{$m^{TRGB}_H$} &
\colhead{$A_J$} &
\colhead{$A_H$}
}
\startdata
NGC 6822 & 17.54 & 18.65 & 17.83 & 0.17 & 0.11 \\
IC 1613 & 18.27 & 19.25 & 18.42 & 0.02 & 0.01 \\
NGC 3109 & 19.31 & 20.40 & 19.54 & 0.05 & 0.03 \\
Sextans B & 19.32 & 20.50 & 19.70 & 0.02 & 0.01 \\
Sextans A & 19.54 & 20.66 & 19.83 & 0.03 & 0.02 \\
NGC 0300 & 20.08  & 21.13 & 20.38 & 0.01 & 0.01\\
NGC 0055 & 20.16  & 21.24 & 20.54 & 0.01 & 0.01\\ 
NGC 7793 & 21.27 & 22.31 & \nodata &  0.01  & \nodata\\
% NGC 0247 & 21.35 & 22.74 & 21.43 & 0.01 & 0.01  \\ 
NGC 0247 & 21.35 & \nodata & \nodata & 0.01 & \nodata  \\ 
NGC 5253 & 21.44 & 22.61 & \nodata & 0.04 & \nodata \\
NGC 1313 & 21.83 & \nodata & \nodata & 0.08 & \nodata \\
\enddata
\end{deluxetable}

\begin{figure}
\centering
\includegraphics[width=\columnwidth]{"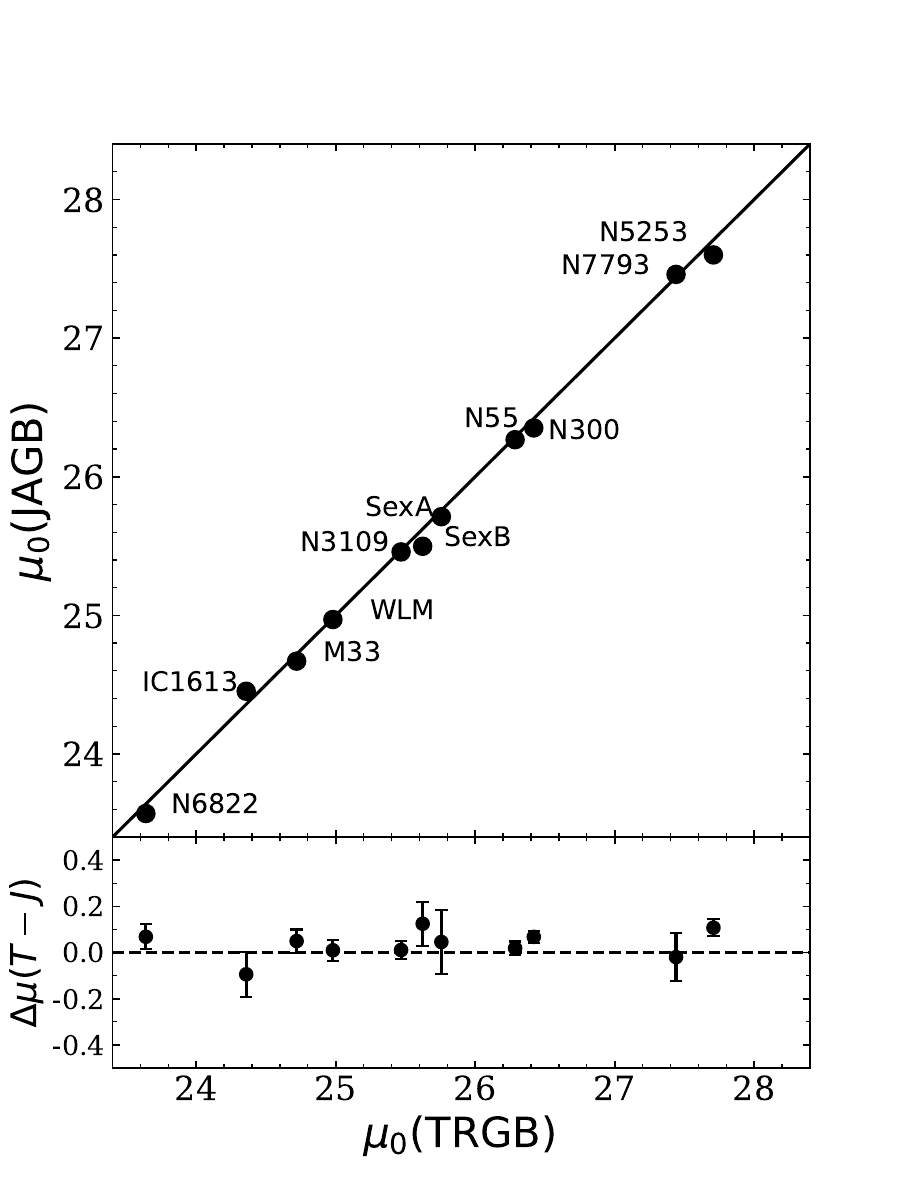"}\figurenum{7}
\caption{Comparison of 11 TRGB and JAGB distance moduli. The diagonal black line is not a fit to the data, but a line of slope 1, showing the excellent agreement between the TRGB and JAGB methods. The one-sigma scatter measured about the unit-slope black line is $\pm0.07$~mag.
The distances span a range from about 0.5 to 3.3 Mpc. The bottom panel shows the difference between the distance moduli (TRGB minus JAGB distance modulus) as a function of the TRGB distance modulus. The error bars represent the statistical TRGB and JAGB errors added in quadrature. }
\label{fig:comp}
\end{figure}

In Figure \ref{fig:comp}, we show a TRGB-JAGB comparison of the distances for the nine galaxies in this sample that have both measured JAGB and TRGB distances, along with our distances measured to WLM \citep{2021ApJ...907..112} and M33 \citep{2022ApJ...933..201L} in companion papers. The mean offset between the all the measured distances was measured to be $<\rm{TRGB-JAGB}>=+0.036\pm0.021$~mag (standard error on the mean). The 11 galaxies in Figure \ref{fig:comp} have an RMS scatter about a unit-slope line of $\pm0.07$~mag. If we attribute this scatter equally to both methods (meaning $0.07^2=\sigma_{JAGB}^2 + \sigma_{TRGB}^2$), this then suggests each of the two methods can individually measure distances with 2\% precision for nearby galaxies. Alternatively, by assuming that the scatter results entirely from the JAGB method, this comparison suggests that JAGB distances would still individually be measured with 3\% precision. 

Two other studies have undertaken similar TRGB-JAGB comparisons; \cite{2020arXiv200510793F} and \cite{2022ApJ...926..153M} both measured an RMS scatter of $\pm0.08$~mag. 
The Araucaria Project \citep{2021arXiv210502120Z} also compared JAGB-Cepheid distances between seven galaxies, finding an RMS scatter of $\pm0.09$~mag. 
However, all these observations originally targeted populations other than the JAGB stars (often Cepheids) and were not analyzed homogeneously using data from the same telescope. In comparison, all our TRGB and JAGB distances were derived from the same imaging and used photometry averaged over multiple epochs, designed to beat down the scatter introduced by the variability of the JAGB stars. Furthermore, we performed spatial cuts to exclude the crowded inner disks of galaxies.

As discussed in the introduction, the TRGB and JAGB are drawn from entirely separate stellar populations, and therefore will likely have completely unrelated astrophysical systematics.  The observed scatter of $\pm0.07$~mag from the TRGB-JAGB comparison includes all the cumulative effects of differences potentially resulting from host galaxy types, metallicities, star formation histories, and uncorrected reddening effects. That is, all systematics are constrained in each distance indicator at the $\pm0.05$~mag (2\% in distance) level or less. 

% \ajm{I very much like the Fig 5 comparison of JAGB vs TRGB.   Can you do this also with the Leavitt PL relation distances?  You don't have to re-derive the distances but between Scowcroft, Rich, and the Aracuaria project there should be cepheids distance to all of these available for a direct comparison?   I guess that could be a bit of scope creep since this is about the JAGB method, and the cepheids comparison was already kinda done in Z21.  On the other hand, it would be cool to be the paper that includes all 3 methods comparisons   }

\subsection{Comparisons with JAGB Distances from the Literature}

In this section, we compare our measured JAGB distance moduli to JAGB distances reported in the literature.

In Table \ref{tab:compare}, we have compiled all available NIR TRGB and JAGB distance moduli to the galaxies in our sample. A visual representation of this table is also shown in Figure \ref{fig:comparison_big}. We also tabulated I-band TRGB distance moduli from the following sources. 
IC 1613 has had its I-band TRGB distance measured by the CCHP, which we listed; for the rest of the galaxies, we recorded I-band TRGB distances from the Extragalactic Distance Database (EDD) \citep{2009AJ....138..323T, 2009AJ....138..332J, 2021AJ....162...80A} and the ACS Nearby Galaxy Survey Treasury (ANGST) survey \citep{2009ApJS..183...67D}. To avoid redundancy, if the EDD distance was based on observations collected for ANGST, we exclusively recorded the ANGST distance. 
We have also tabulated the two Cepheid distances from the Carnegie Hubble Program (CHP), the predecessor to the CCHP. The CHP utilized the FourStar $JHK$ photometry from this paper (taken previous to 2012) and Spitzer [3.6] and [4.5]-band photometry to measure multi-wavelength Leavitt law distances to NGC 6822 \citep{2014ApJ...794..107R} and IC 1613 \citep{2013ApJ...773..106S}.

In the following subsections, we discuss  how our JAGB distances compare with those from the literature. 
For galaxies that have yet to have their distances measured with the JAGB method before this paper (Sextans B, Sextans A, NGC 5253, NGC 1313), we instead cross-checked with literature distances from the I-band TRGB. 
In the cases where our JAGB distances significantly disagreed with the I-band TRGB distances (Sextans B, NGC 1313), we further cross-compared with distances derived from Cepheid stars. We discuss the galaxies in three subsections: Sections \ref{subsubsec:good_agree},  \ref{subsubsec:poor_agree}, and \ref{subsubsec:mixed_agree} discuss galaxies whose JAGB distances agree well with those from the literature, disagree with literature results, and have mixed agreement with the literature, respectively.

 We also hereafter refer to the following papers by the following initialisms: 
\citealt{2020arXiv200510793F}  (\citetalias{2020arXiv200510793F}), 
\citealt{2023MNRAS.tmp..926P} (\citetalias{2023MNRAS.tmp..926P}), 
\citealt{2021arXiv210502120Z} (The Aracucaria Project, \citetalias{2021arXiv210502120Z}), 
\citealt{2009AJ....138..332J} (EDD, \citetalias{2009AJ....138..332J}), and
\citealt{2009ApJS..183...67D} (ANGST, \citetalias{2009ApJS..183...67D}). In particular, \citetalias{2020arXiv200510793F}, \citetalias{2023MNRAS.tmp..926P}, and \citetalias{2021arXiv210502120Z} all measured multiple JAGB distances, and \citetalias{2009AJ....138..332J} and \citetalias{2009ApJS..183...67D} measured I-band TRGB distances for a large sample of galaxies. \citetalias{2023MNRAS.tmp..926P} also measured multiple NIR TRGB distances.

% We note that the I-band TRGB distances appear to be systematically fainter than our NIR TRGB distances (mean offset being $\pm0.14\pm0.04$~mag). We emphasize several of the TRGB distances presented in this paper, particularly the distances for the farthest galaxies suffer from low RGB numbers. 

\subsubsection{Galaxies with good literature agreement: IC 1613, Sextans A, NGC 55, NGC 5253}\label{subsubsec:good_agree}

\textit{IC 1613.} IC 1613's low foreground and internal extinction should make it an easy target for measuring distances, so it is reassuring our JAGB measurement agrees with those from the literature to within $1\sigma$. Our JAGB distance also agrees at the 1.4$\sigma$ level with the CHP multi-wavelength Leavitt law distance from \cite{2013ApJ...773..106S}, which utilized the same $JHK$ imaging from this paper (taken in 2011) as well as Spitzer [3.6] and [4.5]-band photometry.

\textit{Sextans A.} There are no JAGB distances to Sextans A available in the literature, so we instead compared our measured JAGB distance to I-band TRGB distances from ANGST. Our JAGB distance modulus agrees to within $1\sigma$ with the ANGST I-band TRGB measurement.

\textit{NGC 55.} Our measured JAGB distance agrees almost exactly with the JAGB distance modulus measured by \citetalias{2023MNRAS.tmp..926P}.

\textit{NGC 5253.} There are no JAGB distances to NGC 5253 available in the literature, so we instead compared our measured distance to I-band TRGB distances from the EDD.
Our JAGB measurement agrees to within $1\sigma$ with the EDD I-band TRGB measurement.

\subsubsection{Galaxies with poor literature agreement: Sextans B, NGC 7793, NGC 1313}\label{subsubsec:poor_agree}

\textit{Sextans B.} There are no JAGB distances to Sextans B available in the literature, so we instead compared our measured distances to I-band TRGB distances from ANGST and the EDD. Our JAGB measurements is in more than $2\sigma$ tension with the ANGST and EDD I-band TRGB measurements. One potential explanation is that ANGST and EDD both used disk fields for their TRGB measurements, which would lower the precision and accuracy of their detection of the TRGB. 
As an additional cross-check, we also compare our JAGB distance with the only recent Cepheid measurement to Sextans B in the literature from \cite{2011A&A...531A.134T} of $\mu_0=25.53\pm0.10$~mag, which agrees well with our JAGB distance modulus. 

\textit{NGC 7793} Our JAGB distance modulus disagrees at the $2.1\sigma$ level with the \citetalias{2021arXiv210502120Z} JAGB measurement. We note that the field used by \citetalias{2021arXiv210502120Z} for their analysis contains the nucleus of NGC 7793. Reddening effects from the inner disk of this galaxy may explain why the \citetalias{2021arXiv210502120Z} measurement is 0.24~mag fainter than our JAGB measurement. Furthermore, \citetalias{2021arXiv210502120Z} used a larger reddening correction of 0.07~mag (derived from Cepheids), compared with our correction of 0.01~mag (the foreground extinction value from \citealt{1998ApJ...500..525S}). If we used the same reddening correction, this would bring our measurements into $1.6\sigma$ agreement.

% Our TRGB distance modulus agrees well with the I-band TRGB measurement from ANGST. However, 
\textit{NGC 247.} Our JAGB distance disagrees with \citetalias{2021arXiv210502120Z} by $2.4\sigma$.
This difference can be straightforwardly attributed to our differing reddening corrections. \citetalias{2021arXiv210502120Z} used an extinction correction of $A_J=0.16$~mag, derived from Cepheids, compared to our foreground reddening value of $A_J=0.01$~mag. Our measured uncorrected JAGB magnitudes agree to within 0.02~mag.

\textit{NGC 1313.} There are no JAGB distances to NGC 1313 available in the literature, so we instead compared our measured distance to I-band TRGB distances from the EDD.
Our JAGB measurement disagrees with the EDD I-band TRGB measurement by $2.8\sigma$. There are a couple of potential explanations for this discrepancy. First, 
the EDD I-band TRGB measurement was made in the outer disk of NGC 1313, and there seems to be a significant number of AGB stars blurring the tip measurement.\footnote{\url{https://edd.ifa.hawaii.edu/get_cmd.php?pgc=12286}} 
Second, at 3.9 Mpc away, NGC 1313 is 0.6 Mpc farther than NGC 5253, the second farthest galaxy in this sample. As a cross-check, only one Cepheid distance measurement has been made to NGC 1313: $\mu_0=28.31\pm0.10$~mag \citep{2015ApJ...799...19Q}, which disagrees with both our JAGB distance modulus and the EDD I-band TRGB distance modulus significantly. Therefore, it is possible that NGC 1313 is at the distance where it is difficult to measure accurate JAGB method distances with ground-based near-infrared telescopes.

\subsubsection{Galaxies with mixed literature agreement: NGC 6822, NGC 3109, NGC 300}\label{subsubsec:mixed_agree}

\textit{NGC 6822.} Our JAGB  distance modulus agrees to within $\sim 1\sigma$ with the JAGB distance moduli from \citetalias{2020arXiv200510793F} and \citetalias{2023MNRAS.tmp..926P}. 
However, our measurement disagrees at the 2.8$\sigma$ level with the \citetalias{2021arXiv210502120Z} measurement. Some of this disagreement results from the different reddening corrections used; we used a foreground correction of $A_J=0.17$~mag, while \citetalias{2021arXiv210502120Z} used an $A_J=0.31$~mag, which was derived from a reddening law fit for the Cepheid P-L relations in NGC 6822. If \citetalias{2021arXiv210502120Z} used our reddening correction instead, our JAGB distance moduli agree at the $1.6\sigma$ level. 
Secondly, we note that \citetalias{2021arXiv210502120Z} averaged their JAGB measurement from two separate datasets. The first dataset from SOFI camera on the New Technology Telescope, yielded a distance modulus of $\mu_0=23.19\pm0.04$~mag, and the second dataset from the PANIC instrument (the predecessor to the FourStar Camera) on the Magellan-Baade Telescope yielded $\mu_0=23.30\pm0.04$~mag. Therefore,  
the two independent datasets produce JAGB magnitudes that disagree by $0.11$~mag. Thus, the disagreement between our JAGB magnitude vs. the \citetalias{2021arXiv210502120Z} JAGB magnitude may have resulted from a systematic photometric offset difference between our catalogs. As an additional cross-check with, we also compared our JAGB distance with the CHP multi-wavelength Leavitt law distance from \cite{2014ApJ...794..107R}, which utilized the same $JHK$ imaging  from this paper (taken in May 2012) as well as Spitzer [3.6] and [4.5]-band photometry. Our JAGB distances agrees at the $1.6\sigma$ level with this Cepheid distance.
% In conclusion, NGC 6822 has a large foreground extinction as it lies at a low Galactic latitude, making its distance determination more challenging. 

% We note both our JAGB and TRGB distances agree to within the quoted uncertainties with the Mira-based distance from \cite{2013MNRAS.428.2216W} of $\mu_0=23.56\pm0.03$~mag, further adding confidence that our derived distances are accurate. 

% Our TRGB measurement is in $1.8\sigma$ disagreement with the NIR TRGB measurement from \citetalias{2023MNRAS.tmp..926P}. One potential explanation for this discrepancy is that \citetalias{2023MNRAS.tmp..926P} used disk data of NGC 6822, where there is more contamination from AGB stars, as well as more crowding and reddening.

\textit{NGC 3109.} Our JAGB distance to NGC 3109 agrees to within $1.3\sigma$ with \citetalias{2021arXiv210502120Z} and \citetalias{2020arXiv200510793F}.
However, our JAGB measurement disagrees at the $1.9\sigma$ level with the JAGB distance from \citetalias{2023MNRAS.tmp..926P}. This disagreement could potentially be because  \citetalias{2023MNRAS.tmp..926P} performed their JAGB measurement in the inner disk of NGC 3109, where reddening effects could bias their result fainter. 

\textit{NGC 300.} Our JAGB distance agrees well with that from \cite{2022ApJ...926..153M}, which used the HST WFC3/IR F110W filter to measure a distance modulus to NGC 300.
% We note however our TRGB measurement agrees well with the Araucaria project Cepheid measurement from \cite{2005ApJ...628..695G} of $\mu_0=26.37\pm0.06$~mag and the I-band TRGB from \cite{2006ApJ...638..766R} of $\mu_0=26.30\pm0.12$~mag. Interestingly, 
On the other hand, our JAGB measurement disagrees to about $2.5\sigma$ with the \citetalias{2021arXiv210502120Z} Araucaria project JAGB measurement. Some of this disagreement can be straightforwardly attributed to our different reddening corrections. \citetalias{2021arXiv210502120Z} used a reddening correction of $A_J=0.08$~mag derived from Cepheids compared with our correction of $A_J=0.01$~mag; using the same correction brings our measurements into $1.7\sigma$~agreement. Furthermore, the fields used by \citetalias{2021arXiv210502120Z} encompass the inner disk of NGC~300, where reddening could be biasing their result fainter.

\begin{figure*}\figurenum{8}\label{fig:comparison_big}
\gridline{\fig{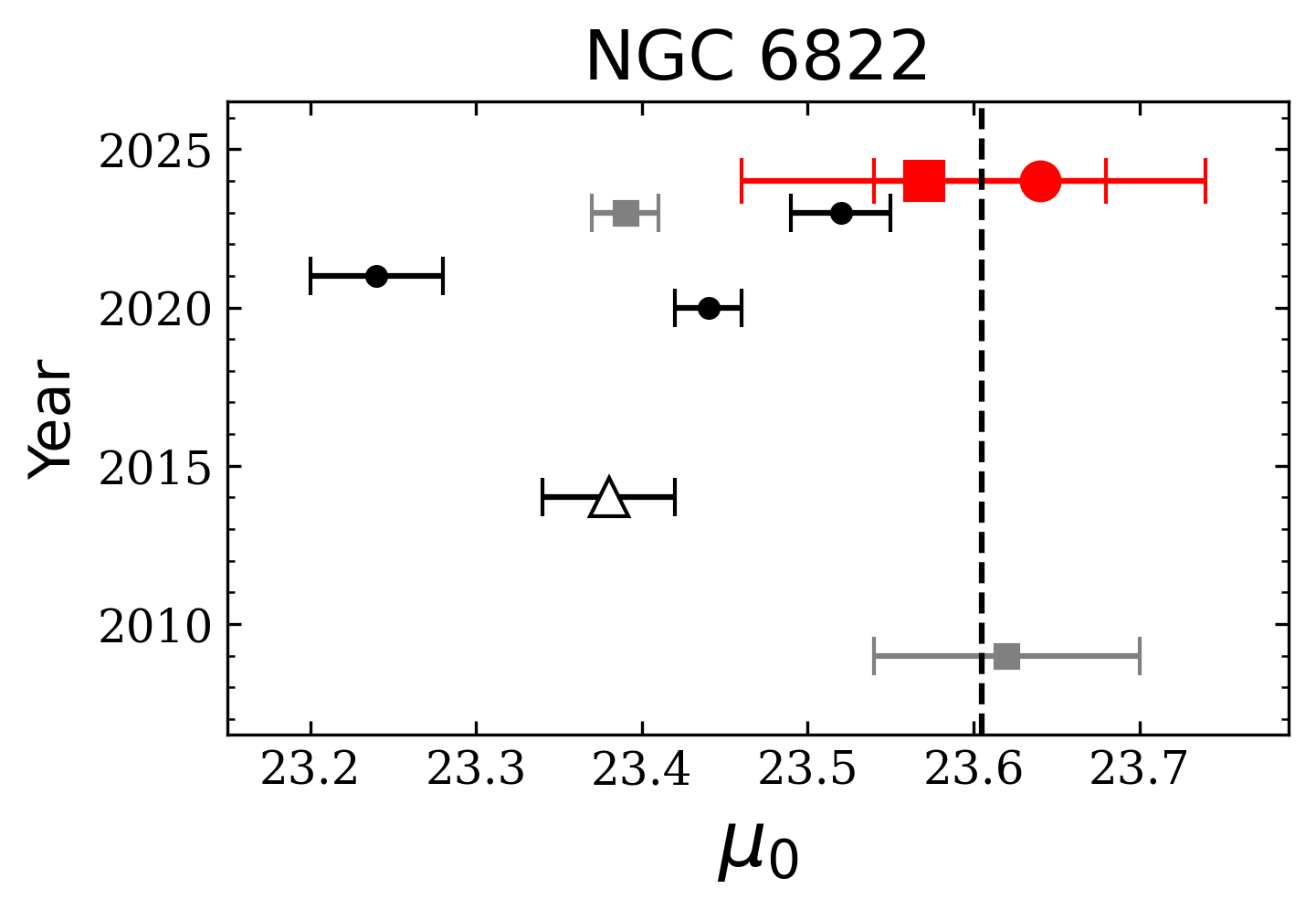}{0.333\textwidth}{}
          \fig{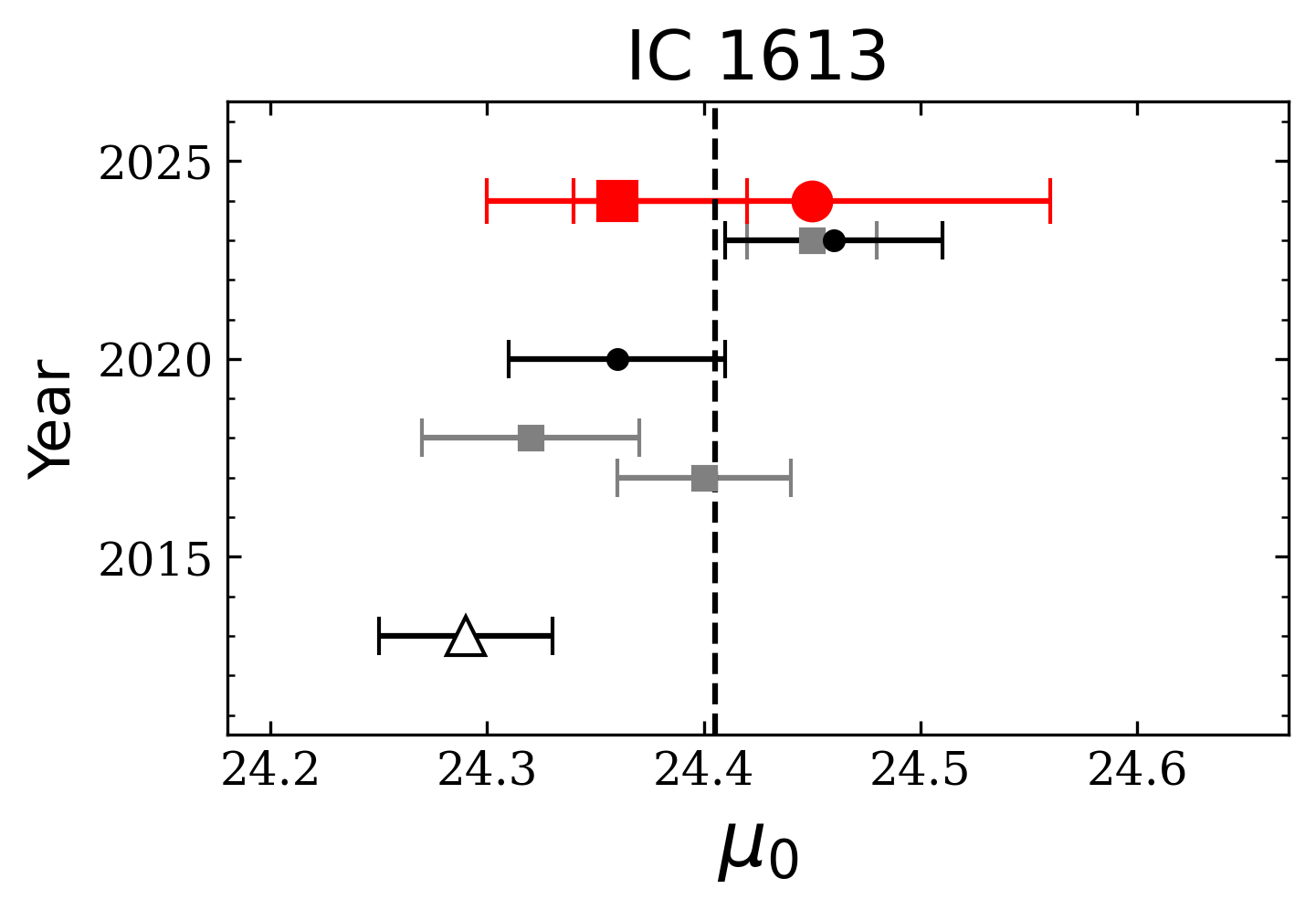}{0.333\textwidth}{}
          \fig{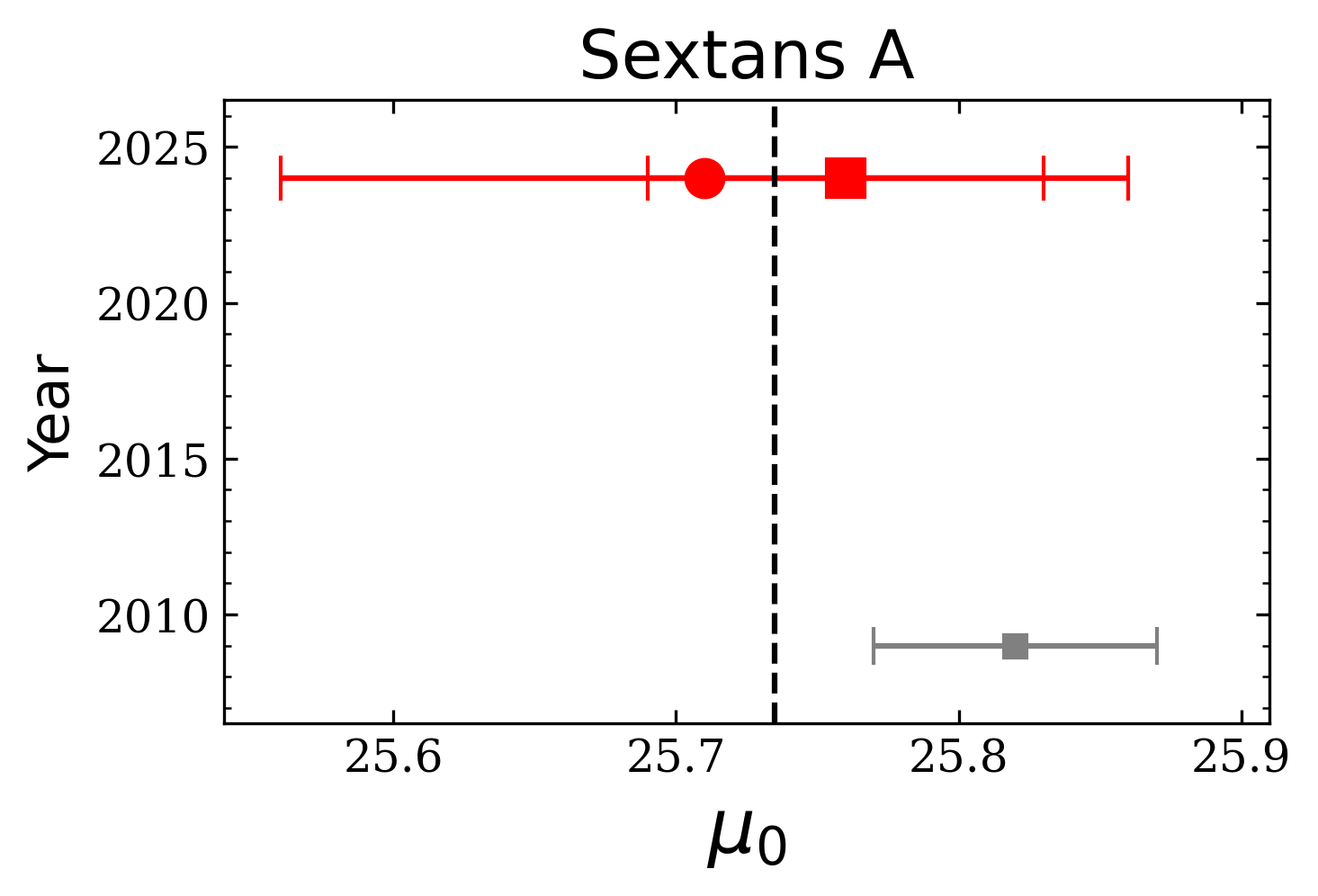}{0.333\textwidth}{}}
\gridline{\fig{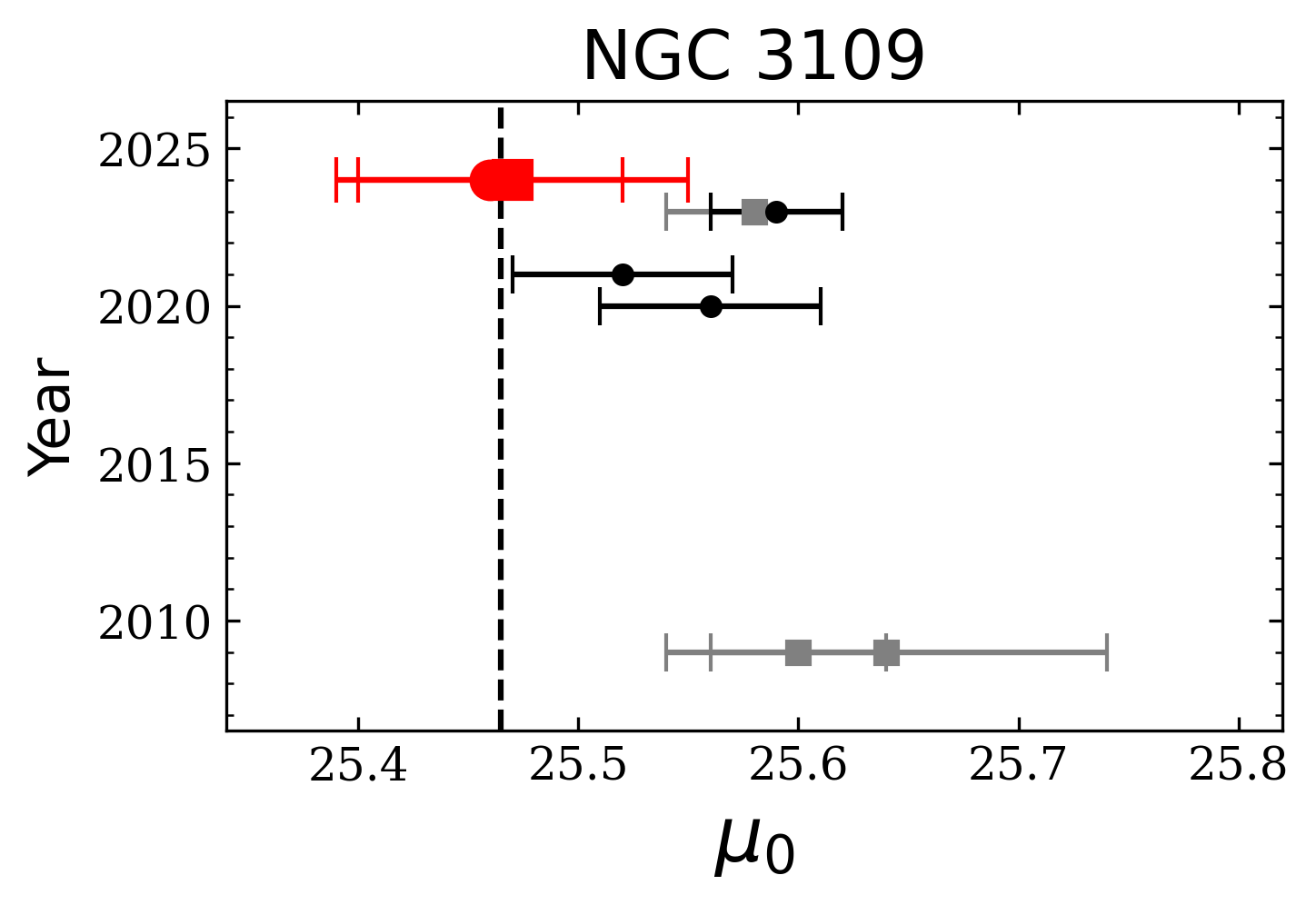}{0.333\textwidth}{}
          \fig{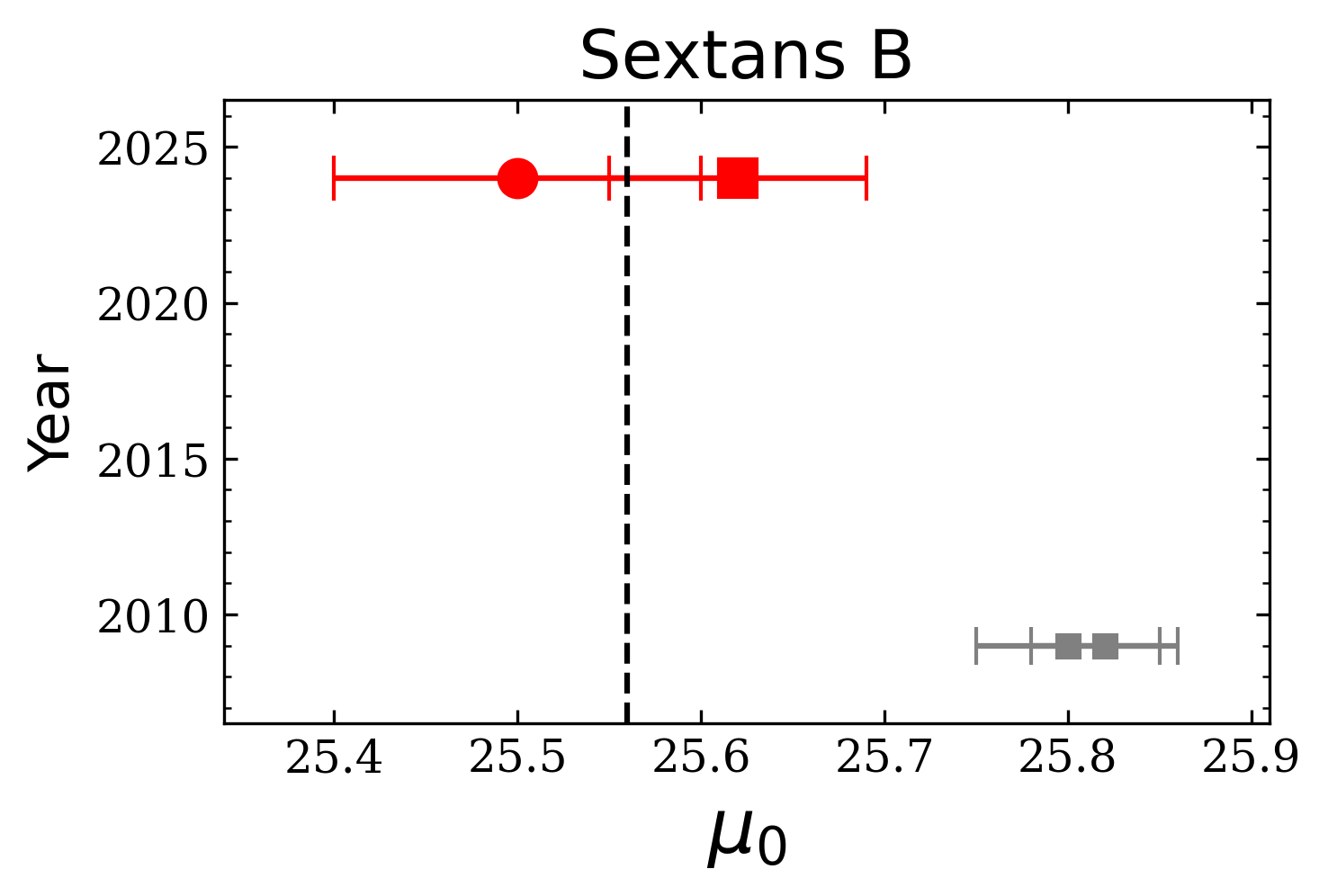}{0.333\textwidth}{}
                    \fig{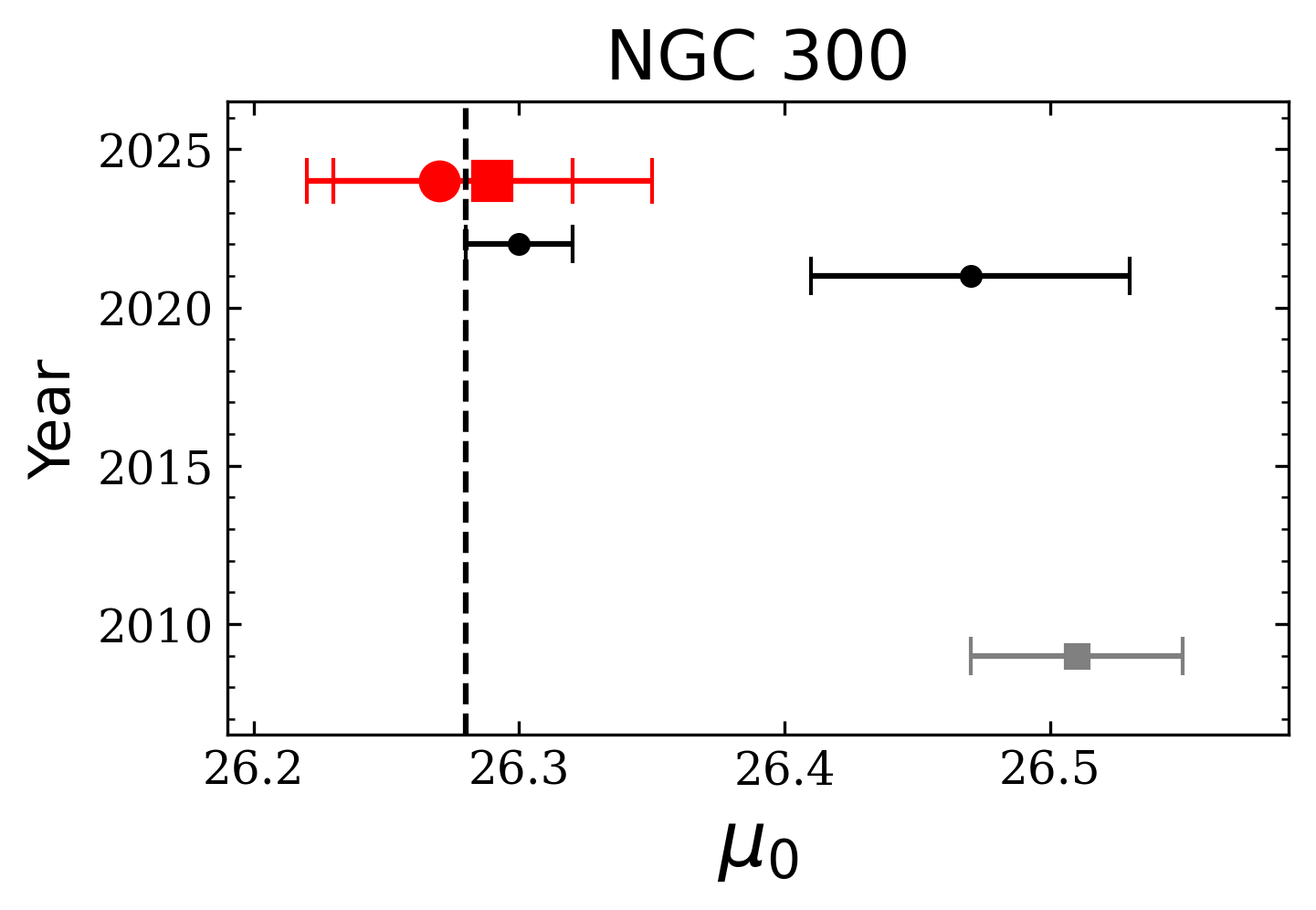}{0.333\textwidth}{}
          }
          \gridline{\fig{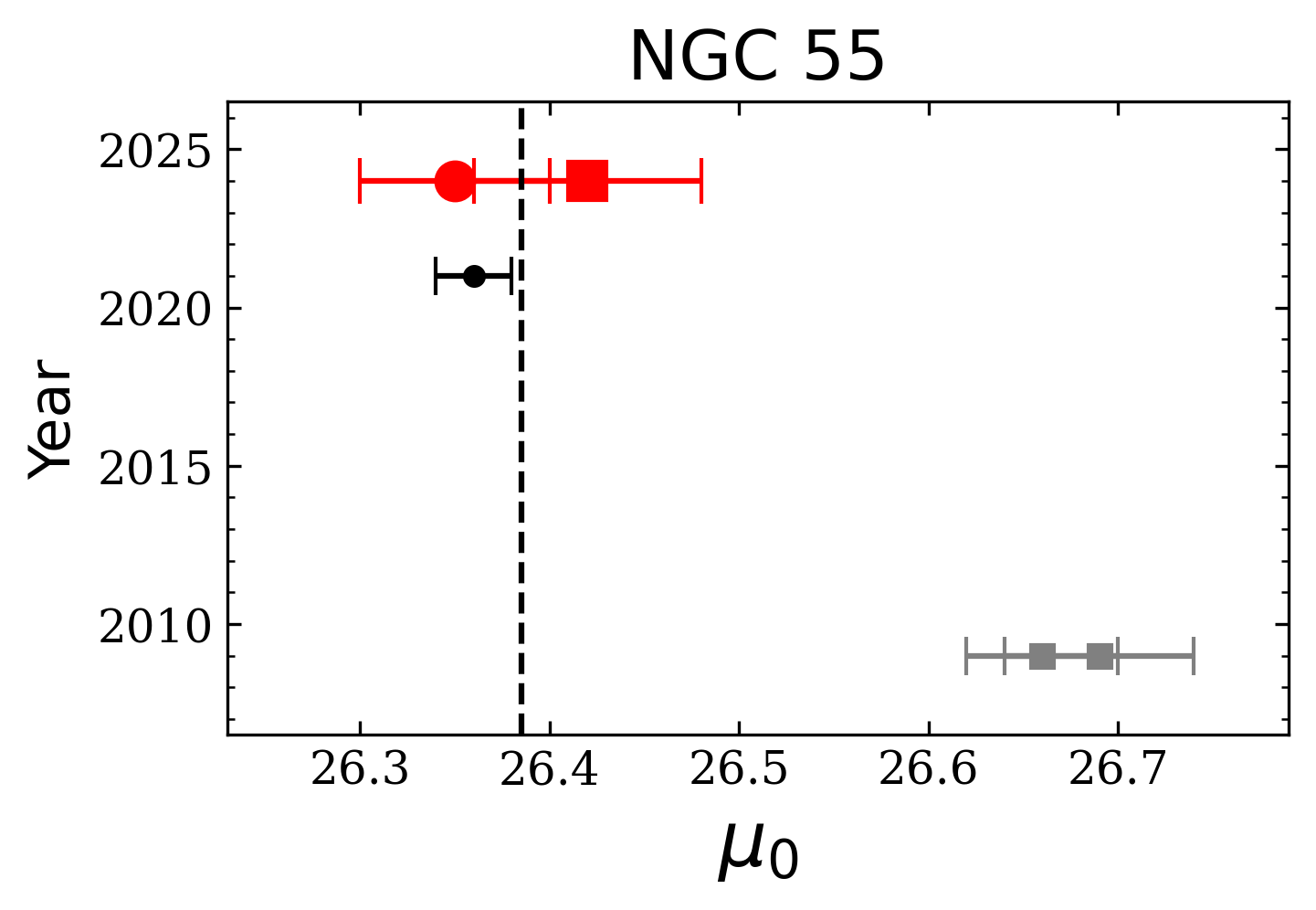}{0.333\textwidth}{}
          \fig{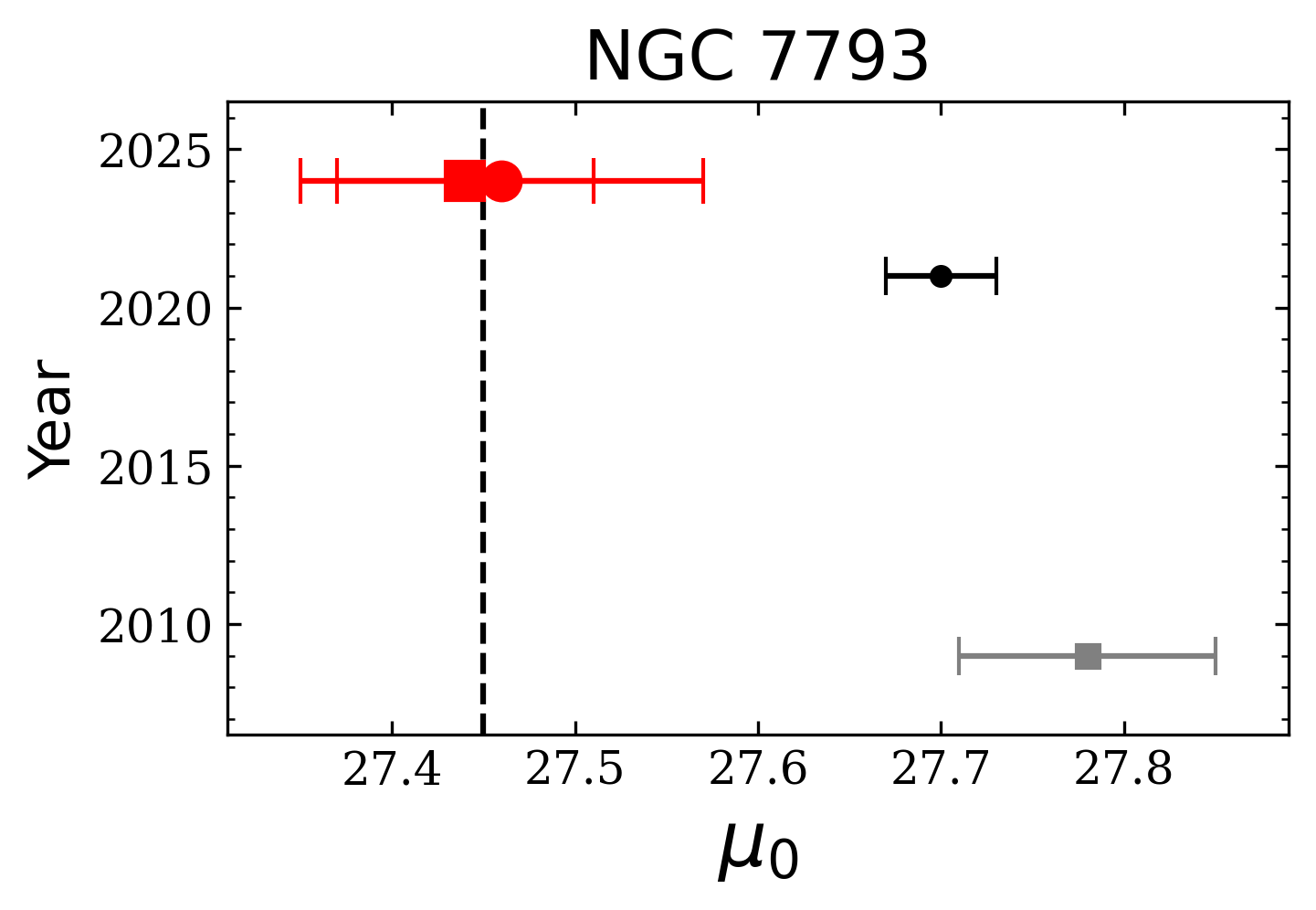}{0.333\textwidth}{}
                    \fig{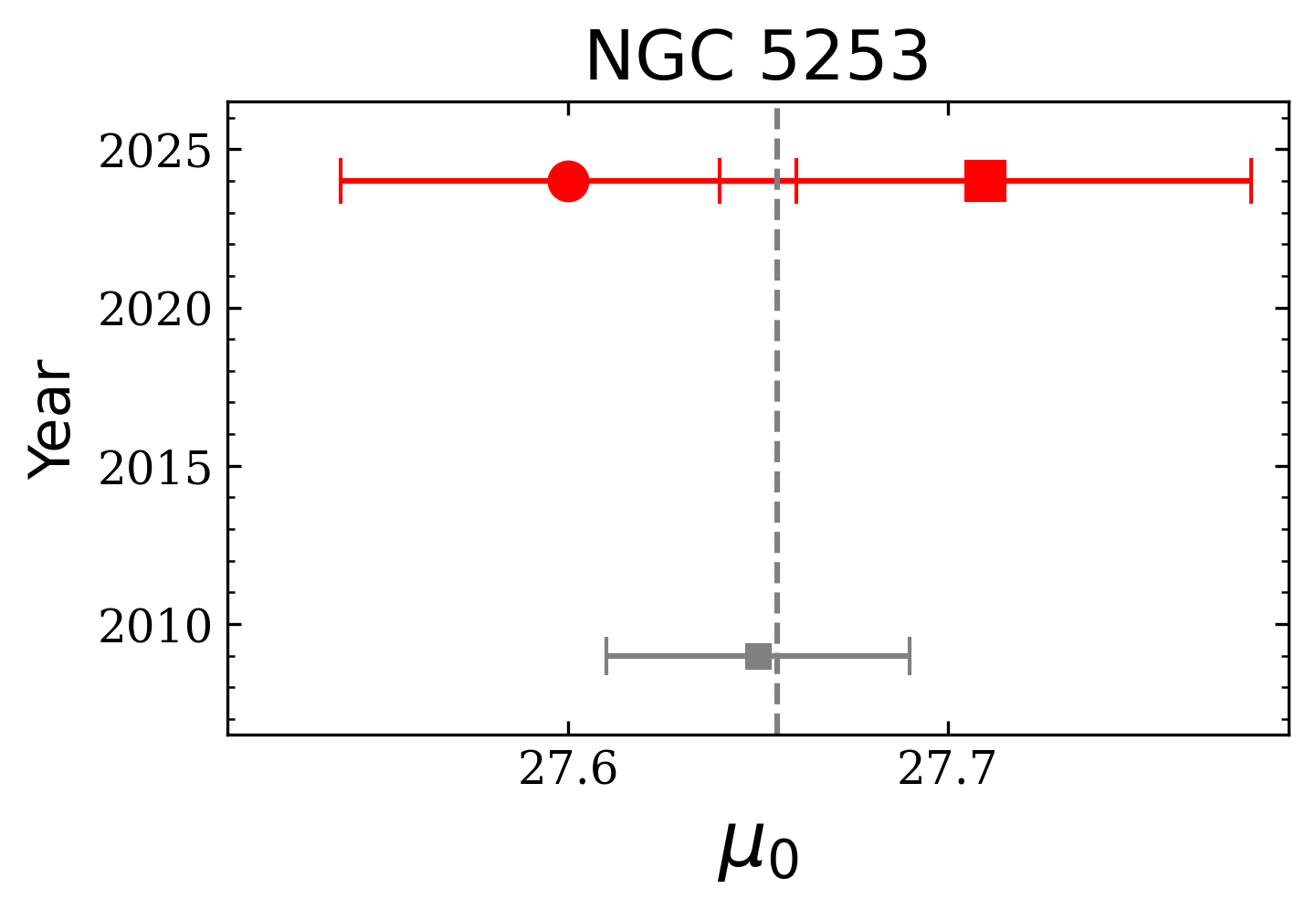}{0.333\textwidth}{}
          }
\gridline{\fig{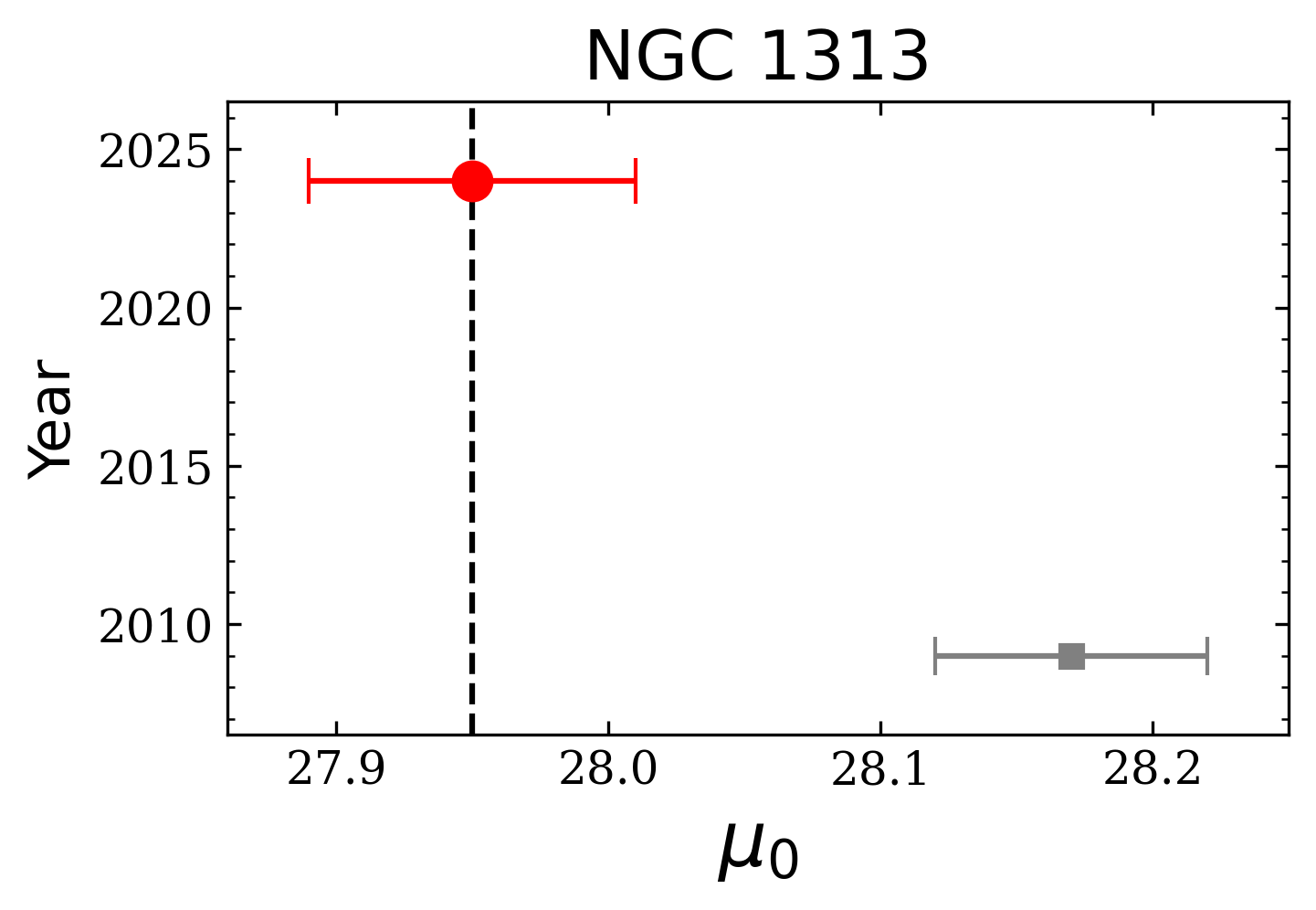}{0.333\textwidth}{}
\fig{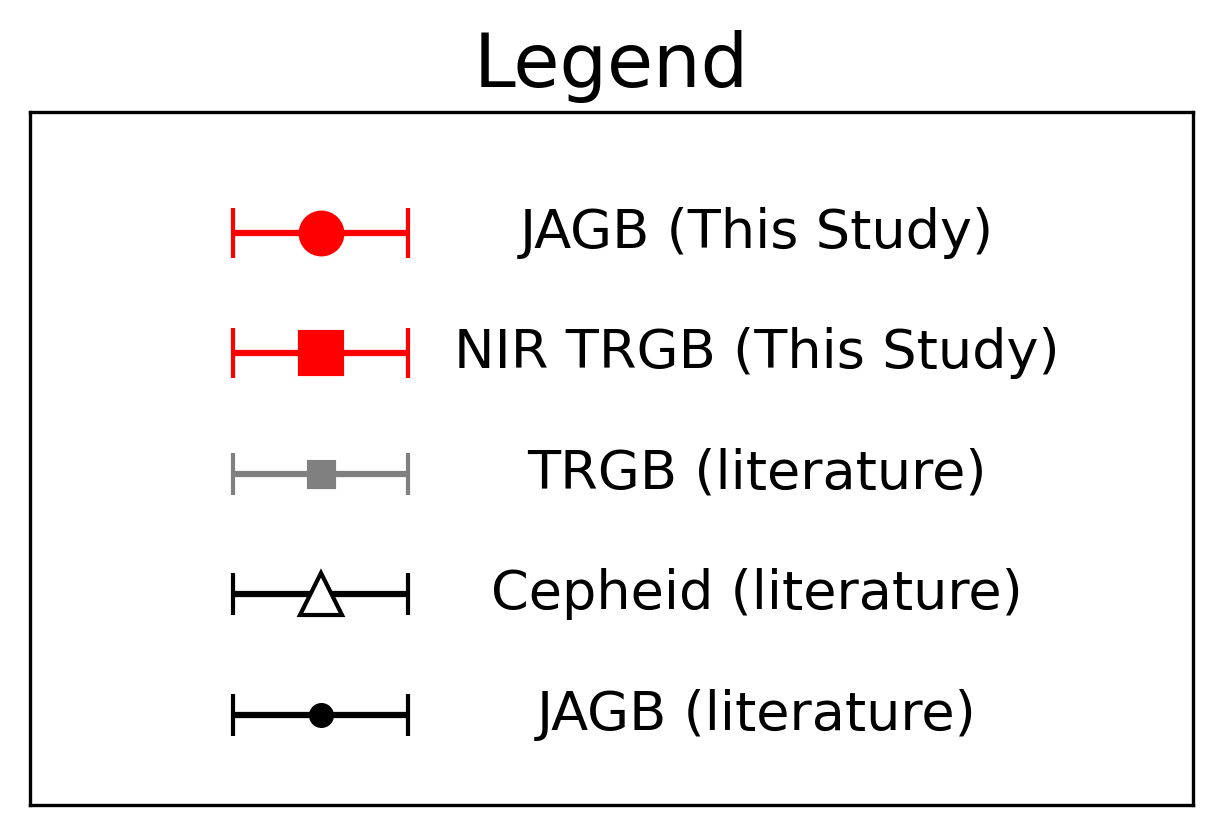}{0.36\textwidth}{}
          }
\caption{A visual representation of Table \ref{tab:compare}, showing comparisons between our measured JAGB and NIR TRGB distances with those from the literature, ordered by year, for each galaxy. For distances from the literature: grey squares represent TRGB measurements, black circles represent JAGB measurements, and white triangles represent CHP Cepheid measurements. Red squares and red circles represent the TRGB and JAGB distances measured in this study, respectively. The dotted black line marks the average of the measured JAGB and TRGB distance moduli from this study.}
\end{figure*}

\subsubsection{Summary}

In summary, most differences between our JAGB measurements with those from the literature can be straightforwardly attributed to differing reddening corrections. Accounting for these differences, we overall find excellent agreement ($\lesssim1.5\sigma$) between our JAGB distances with those from the literature, with the exception of NGC 3109 and NGC 300. However, we speculate the choice of field (inner disk vs. outer disk) and the resulting reddening bias incurred may potentially explain these differences.

% Our NIR TRGB measurements agrees to within about $\sim1\sigma$ with \citetalias{2023MNRAS.tmp..926P}'s TRGB measurement.

 % Second, it appears their photometry is  their modal magnitude was measured to be $J=19.40$~mag, whereas our was $J=19.31$~mag. \citetalias{2023MNRAS.tmp..926P} measured a K-band TRGB of $\mu_0^K (TRGB) = 25.58\pm0.04$~mag, which is 0.11~mag fainter than our TRGB measurement of $\mu_0 (TRGB) = 25.47\pm0.09$~mag. The difference therefore likely results from differences in our photometry, since their distance moduli were measured systematically 0.1~mag fainter than our distance moduli.

% Our TRGB distance modulus disagrees by more than $3\sigma$ with the TRGB EDD distance. However, we note that the EDD distance is based on a sparsely populated CMD from HST program 10523.\footnote{\url{https://edd.ifa.hawaii.edu/get_cmd.php?pgc=73049}} Our NGC 7793 TRGB  distance is also based on sparsely populated data, and is one of the least precise of all our TRGB distances.

% Furthermore, using the multi-wavelength Cepheid P-L relation, \cite{2017ApJ...847...88Z} derived a distance modulus of $\mu_0=27.66\pm0.08$~mag. 

% The widely range of the distance measured by the NIR TRGB, I-band TRGB, JAGB method, and Cepheid P-L relation indicate the distance to NGC 7793 could benefit from further careful scrutinizing and cross-checks that are beyond the scope of this paper.

\clearpage
\startlongtable
\begin{deluxetable*}{ccccc}\tablenum{6}
\centerwidetable
\tablecaption{Literature Distances}\label{tab:compare}
\tablehead{
\colhead{Galaxy} & 
\colhead{Method} & 
\colhead{$\mu_0$} & 
\colhead{Reference} & 
\colhead{Notes}
}
\startdata
NGC 6822  & I-band TRGB & $23.62\pm0.08$& \citetalias{2009AJ....138..332J} & $m_{F814W}=19.97\pm0.07$~mag, $A_{F814W}=0.40$~mag \\
NGC 6822 & NIR TRGB & $23.39\pm0.02$ & \citetalias{2023MNRAS.tmp..926P} & K-band TRGB \\
NGC 6822 & NIR TRGB &  $23.64\pm0.10$& \textbf{This study}\\
NGC 6822 & Leavitt law & $23.38\pm0.04$ & CHP \citep{2014ApJ...794..107R}\\
NGC 6822 & JAGB method & $23.44\pm0.02$ & \citetalias{2020arXiv200510793F} \\
NGC 6822 & JAGB method & $23.24\pm0.04$  & \citetalias{2021arXiv210502120Z} \\
NGC 6822 & JAGB method & $23.52\pm0.03$&\citetalias{2023MNRAS.tmp..926P} \\
NGC 6822  & JAGB method & $23.57\pm0.11$ & \textbf{This study}\\
\hline
IC 1613 & I-band TRGB & $24.40\pm0.04$& \cite{2017ApJ...845..146H} & $m_{F814W}=20.35\pm0.01$~mag, assumed zero Galactic extinction\\
IC 1613  & NIR TRGB & $24.32\pm0.05$& \cite{2018ApJ...858...11M}\\
IC 1613 & NIR TRGB  & $24.45\pm0.03$ & \citetalias{2023MNRAS.tmp..926P} & K-band TRGB\\
IC 1613 & NIR TRGB & $24.36\pm0.06$ & \textbf{This study}\\
IC 1613 & Leavitt law & $24.29\pm0.04$ & CHP \citep{2013ApJ...773..106S}\\
IC 1613 & JAGB method & $24.36\pm0.05$ &\citetalias{2020arXiv200510793F} & $m_J=-6.20\pm0.04$~mag \\ 
IC 1613 & JAGB method & $24.46\pm0.05$& 
\citetalias{2023MNRAS.tmp..926P}\\
IC 1613 & JAGB method & $24.45\pm0.11$& \textbf{This study}\\
\hline
NGC 3109 & I-band TRGB & $25.64\pm0.10$ & \citetalias{2009AJ....138..332J} & $m_{F814W}=21.69\pm0.09$~mag, $A_{F814W}=0.10$~mag\\ 
NGC 3109 & I-band TRGB  & $25.60\pm0.04$& \citetalias{2009ApJS..183...67D} & 2 fields, $m_{F814W}=21.65\pm0.02$~mag, $A_{F814W}=0.10$~mag \\
NGC 3109 & NIR TRGB & $25.58\pm0.04$ & \citetalias{2023MNRAS.tmp..926P}&K-band TRGB\\
NGC 3109 & NIR TRGB & $25.47\pm0.08$ & \textbf{This study}\\
NGC 3109 & JAGB method & $25.52\pm0.05$ &  \citetalias{2021arXiv210502120Z} \\
NGC 3109 & JAGB method & $25.59\pm0.03$ & \citetalias{2023MNRAS.tmp..926P}\\
NGC 3109 & JAGB method & $25.56\pm0.05$ &\citetalias{2020arXiv200510793F} \\
NGC 3109 & JAGB method & $25.46\pm0.06$ & \textbf{This study}\\
\hline
Sextans B & I-band TRGB  & $25.80\pm0.05$ & \citetalias{2009AJ....138..332J} & $m_{F814W}=21.80\pm0.03$~mag, $A_{F814W}=0.05$~mag\\
Sextans B & I-band TRGB & $25.82\pm0.04$ & \citetalias{2009ApJS..183...67D} & $m_{F814W}=21.82\pm0.02$~mag, $A_{F814W}=0.05$~mag\\ 
Sextans B & NIR TRGB & $25.62\pm0.07$ & \textbf{This study}\\
Sextans B & JAGB method & $25.50\pm0.10$ & \textbf{This study}\\
\hline
Sextans A & I-band TRGB & $25.82\pm0.05$ & \citetalias{2009ApJS..183...67D} & $m_{F814W}=21.84\pm0.03$~mag, $A_{F814W}=0.07$~mag \\
Sextans A & NIR TRGB & $25.76\pm0.07$ & \textbf{This study}\\
Sextans A & JAGB method & $25.71\pm0.15$ & \textbf{This study}\\
\hline
NGC 0300& I-band TRGB & $26.51\pm0.04$& \citetalias{2009ApJS..183...67D} & 6 fields, $m_{F814W}=22.48\pm0.02$~mag, $A_{F814W}=0.02$~mag\\
NGC 0300 & NIR TRGB & $26.29\pm0.06$ & \textbf{This study}\\
NGC 0300 &JAGB method& $26.30\pm0.02$ & \cite{2022ApJ...926..153M} & Median JAGB magnitude \\
NGC 0300 & JAGB method & $26.47\pm0.06$ &  \citetalias{2021arXiv210502120Z} \\
NGC 0300 & JAGB method & $26.27\pm0.05$ & \textbf{This study}\\
\hline
NGC 0055 & I-band TRGB & $26.66\pm0.04$& \citetalias{2009ApJS..183...67D} & 4 fields, $m_{F814W}=22.63\pm0.01$~mag, $A_{F814W}=0.02$~mag\\
NGC 0055 & I-band TRGB & $26.69\pm0.05$&\citetalias{2009AJ....138..332J} & $m_{F814W}=22.66\pm0.03$~mag, $A_{F814W}=0.02$~mag \\
NGC 0055 & NIR TRGB & $26.42\pm0.06$ & \textbf{This study}\\
NGC 0055 & JAGB method & $26.36\pm0.02$&\citetalias{2021arXiv210502120Z}   \\ 
NGC 0055  & JAGB method & $26.35\pm0.05$ & \textbf{This study}\\
\hline
NGC 7793 & I-band TRGB &$27.78\pm0.07$ & \citetalias{2009AJ....138..332J} & $m_{F814W}=23.76\pm0.06$~mag, $A_{F814W}=0.03$~mag\\
NGC 7793 & NIR TRGB & $27.44\pm0.07$ & \textbf{This study}\\
NGC 7793 & JAGB method & $27.70\pm0.03$&\citetalias{2021arXiv210502120Z} \\
NGC 7793 & JAGB method &  $27.46\pm0.11$& \textbf{This study}\\
\hline
NGC 0247& I-band TRGB & $27.76\pm0.06$&\citetalias{2009ApJS..183...67D} & 2 fields, $m_{F814W}=23.74\pm0.04$~mag, $A_{F814W}=0.03$~mag \\
% NGC 0247 & NIR TRGB & $27.61\pm0.12$ & \textbf{This study}\\
NGC 0247 &JAGB method& $27.41\pm0.02$ & \citetalias{2021arXiv210502120Z} \\ 
NGC 0247 & JAGB method & $27.54\pm0.05$ & \textbf{This study}\\
\hline
NGC 5253 & I-band TRGB & $27.65\pm0.04$& \citetalias{2009AJ....138..332J} & $m_{F814W}=23.69\pm0.02$~mag, $A_{F814W}=0.09$~mag\\
NGC 5253 & NIR TRGB & $27.71\pm0.07$ & \textbf{This study}\\
NGC 5253 & JAGB method & $27.60\pm0.06$ &  \textbf{This study}\\
\hline
NGC 1313 & I-band TRGB &$28.17\pm0.05$ & \citetalias{2009AJ....138..332J} & $m_{F814W}=24.29\pm0.03$~mag, $A_{F814W}=0.17$~mag \\
NGC 1313 & JAGB method & $27.95\pm0.06$ & \textbf{This study}\\
\enddata
\tablecomments{All I-band TRGB distance moduli have been standardized to use the \cite{2021ApJ...919...16F} zero point of $M_{F814W}=-4.05\pm0.04$~mag. }
\tablecomments{If there were multiple TRGB measurements from different fields in the same paper, we took the average.}
\end{deluxetable*}
\clearpage

\section{Conclusion}\label{sec:summary}
In this paper we have created a repository of near-infrared stellar photometry for nearby galaxies ($d<4$~Mpc). This photometry was temporally averaged from multiple-epoch observations, designed to decrease the dispersion in the observed JAGB star luminosity function. 
The resulting catalogs will be useful for comprehensive studies of nearby stellar populations in the NIR. 

The method used in this paper for measuring the apparent JAGB magnitude via the modal magnitude of the GLOESS-smoothed LF (after selecting JAGB stars in the color region $1.5<(J-K)<2.0$~mag) was shown to be robust. 
The residuals obtained from subtracting the JAGB distance moduli from the TRGB distance moduli yielded an RMS scatter of $\sigma=0.07$~mag. For this sample of galaxies this scatter puts upper limits ($<2\%$) on the impact of metallicity differences, internal reddening, and star formation history differences between galaxies on the JAGB method distances. 
Furthermore, we showed the composite JAGB star luminosity function formed from the diverse sample of galaxies is closely Gaussian in form. This demonstrates that the underlying JAGB star distribution of a complete star formation history is Gaussian and symmetric.

Further work should quantify directly how distances measuring using the JAGB method are affected by astrophysical systematics. For example, \cite{2023arXiv230502453L} showed that metallicity and age were not significantly correlated with the mode of the JAGB star LF in M31. However, M31 is a relatively metal-rich galaxy. Extending this analysis to a wider range of environments will help further quantify effects of metallicity and star formation history on the JAGB star LF. Furthermore, we plan to simulate artificial star tests for several galaxies in this sample, which will allow us to more accurately quantify effects of the choice of smoothing parameter on the JAGB star LF.

With this further development and testing, the JAGB method has significant potential to provide an independent calibration of Type Ia supernovae with JWST and therefore a measurement of the Hubble constant.
\acknowledgments
We gratefully acknowledge the efforts
and dedication of the Las Campanas Observatory staff for support, particularly in remote observing during the COVID-19 pandemic. In particular, we thank Jorge Araya, Carlos Contreras, Matías Díaz, and Carla Fuentes. We also want to especially thank Eric Persson who created, built and supported FourStar over many years, and without whom this project never would have come to fruition. AJL thanks Saurabh Jha for helpful comments on error budgets. We thank Peter Stetson for providing us with a copy of DAOPHOT and continually helping us troubleshoot problems. We thank Taylor Hoyt for his TRGB measurement code. Finally, we thank the anonymous referee for their constructive and helpful suggestions that improved this work.

AJL was supported by the Future Investigators in NASA Earth and Space Science
and Technology (FINESST) award number 80NSSC22K1602 during the completion of
this work.
AJL thanks the LSSTC Data Science Fellowship Program, which is funded by LSSTC, NSF Cybertraining Grant \#1829740, the Brinson Foundation, and the Moore Foundation; her participation in the program has benefited this work. Finally, we thank the {\it Observatories of the Carnegie Institution for
Science} and the {\it University of Chicago} for their support of our long-term research into the calibration and determination of the expansion rate of the Universe. 

This paper is based on data obtained using the FourStar infrared imager on the 6.5 meter Magellan Telescopes located at Las Campanas Observatory, Chile. 
This research has made use of the NASA/IPAC infrared Science Archive (IRSA), which is operated by the Jet Propulsion Laboratory, California Institute of Technology, under contract with the National Aeronautics and Space Administration.
This publication makes use of data products from the Two Micron All Sky Survey, which is a joint project of the University of Massachusetts and the Infrared Processing and Analysis Center/California Institute of Technology, funded by the National Aeronautics and Space Administration and the National Science Foundation.
This research has made use of NASA's Astrophysics Data System Bibliographic Services.

The data are available on Zenodo under an open-source 
Creative Commons Attribution license: 
\dataset[doi:10.5281/zenodo.10606945]{https://doi.org/10.5281/zenodo.10606945}.

\facility{Magellan:Baade (FourStar)}

\software{\textsc{daophot} \citep{1987PASP...99..191S}, Astropy \citep{2013A&A...558A..33A, 2018AJ....156..123A, 2022ApJ...935..167A}, Matplotlib \citep{2007CSE.....9...90H}, NumPy \citep{2020Natur.585..357H}, Pandas \citep{pandas}, scipy \citep{2020NatMe..17..261V}}

\clearpage
\appendix \label{appendix}
\restartappendixnumbering

\section{JAGB \& TRGB CMDs}\label{sec:jagb_cmd_cont}
In this section, we display the JAGB and TRGB CMDs for the galaxies studied in this paper. 

\begin{figure*}\figurenum{3}
\centering
\includegraphics[width=\textwidth]{"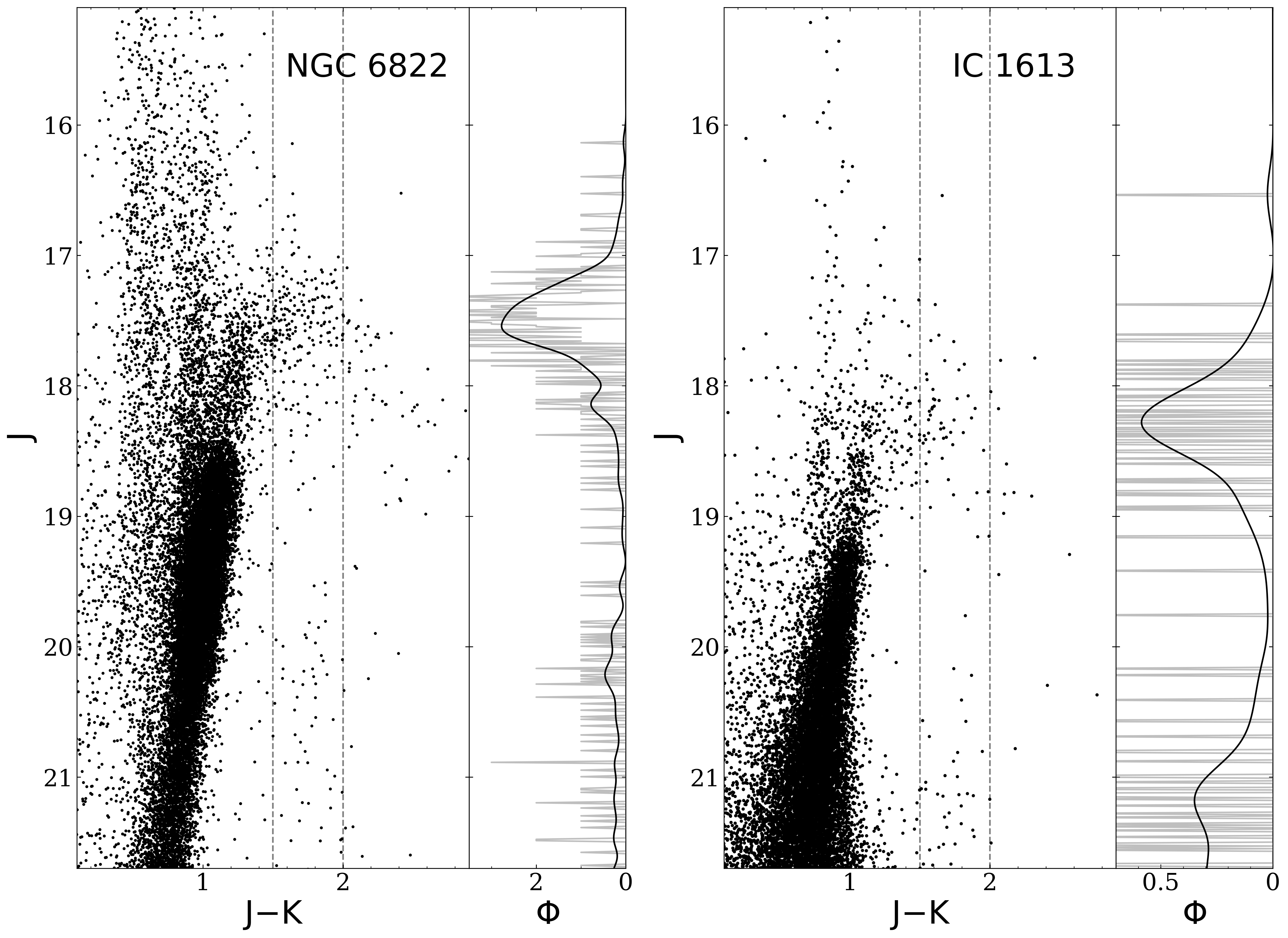"}
\caption{(Continued)}
\end{figure*}

\begin{figure*}\label{fig:cmd2}\figurenum{3}
\centering
\includegraphics[width=\textwidth]{"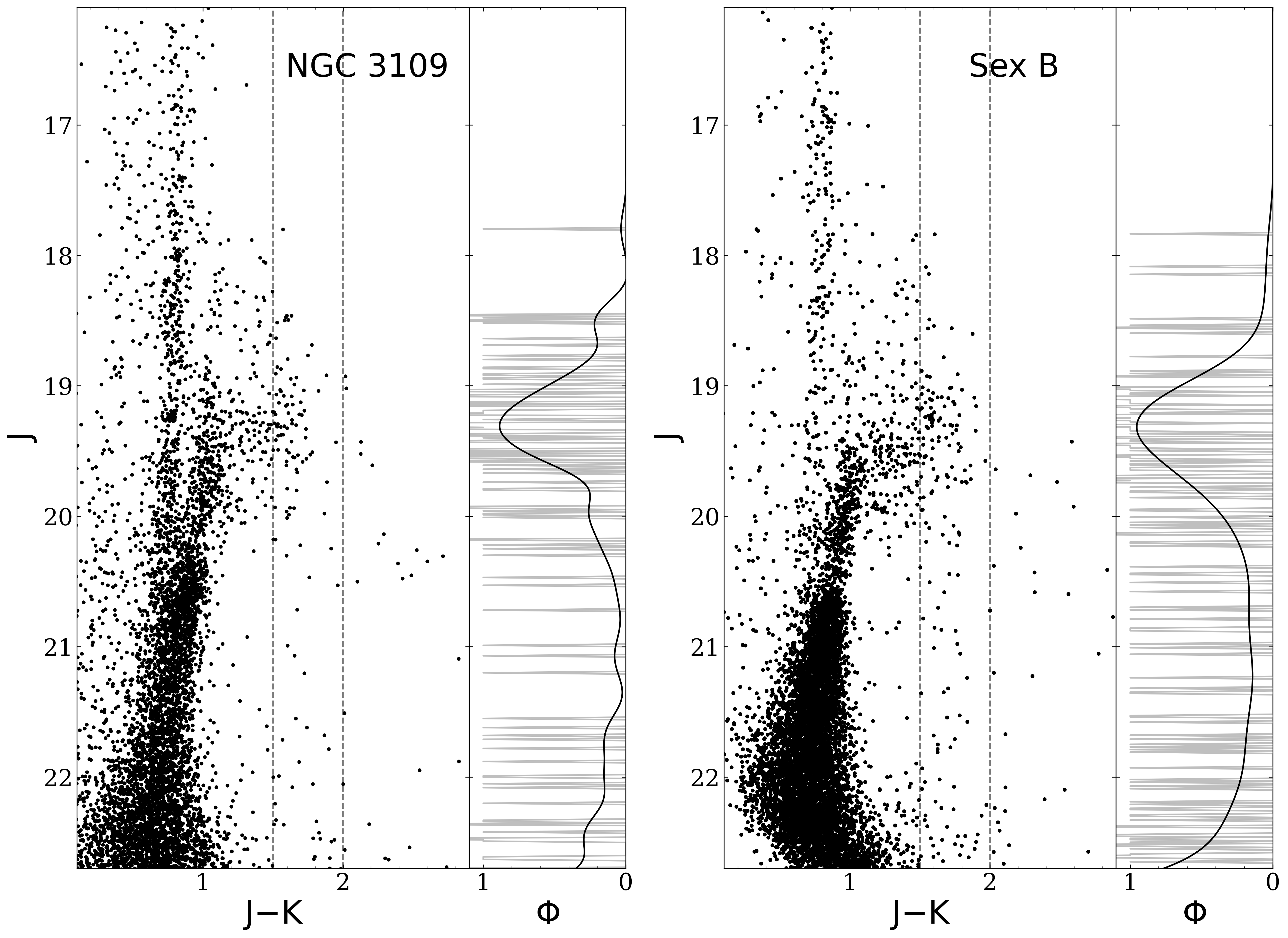"}
\caption{(Continued.) }
\end{figure*}

\begin{figure*}\figurenum{3}
\centering
\includegraphics[width=\textwidth]{"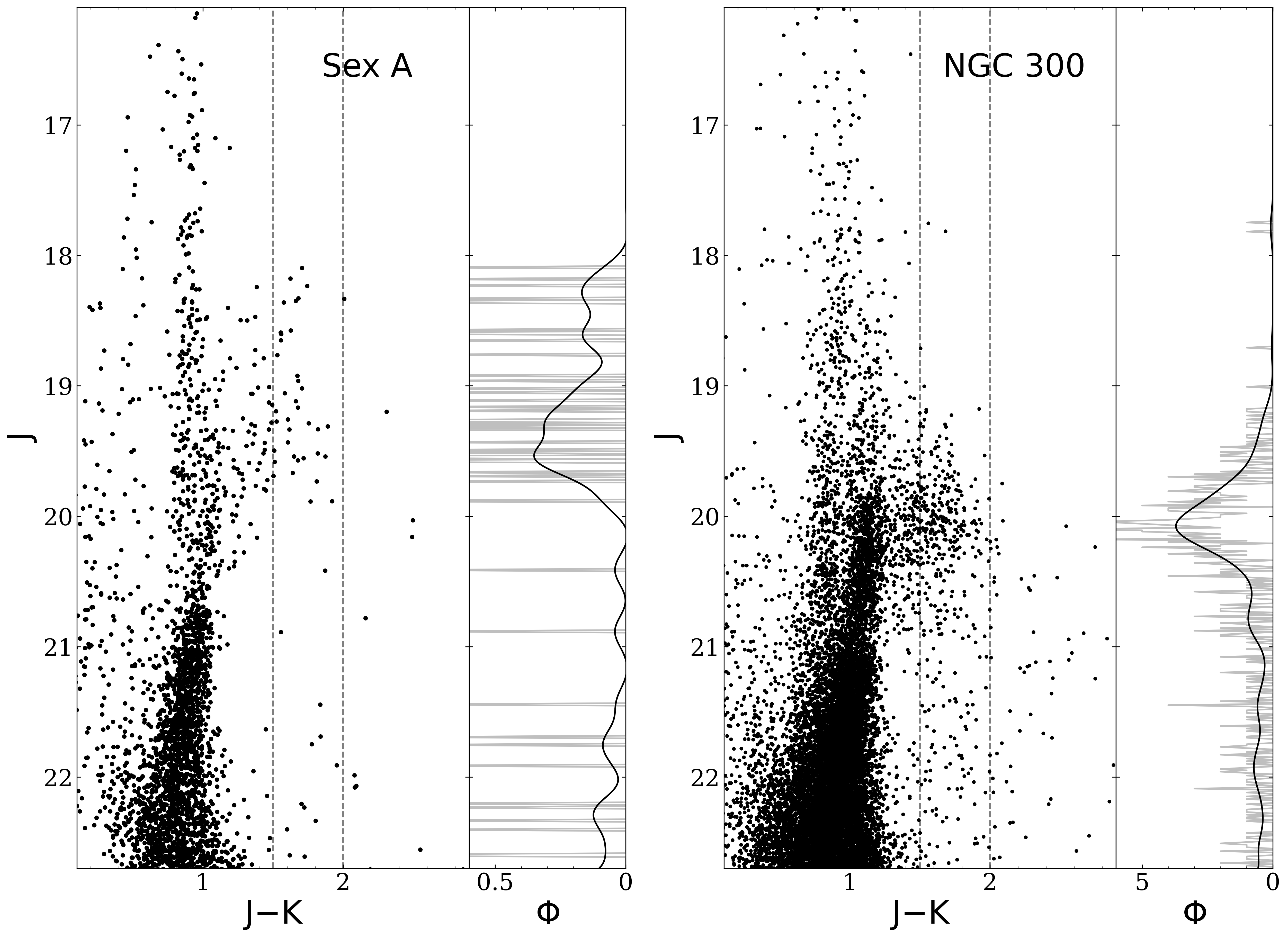"}
\caption{(Continued.) }
\end{figure*}

\begin{figure*}\figurenum{3}
\centering
\includegraphics[width=\textwidth]{"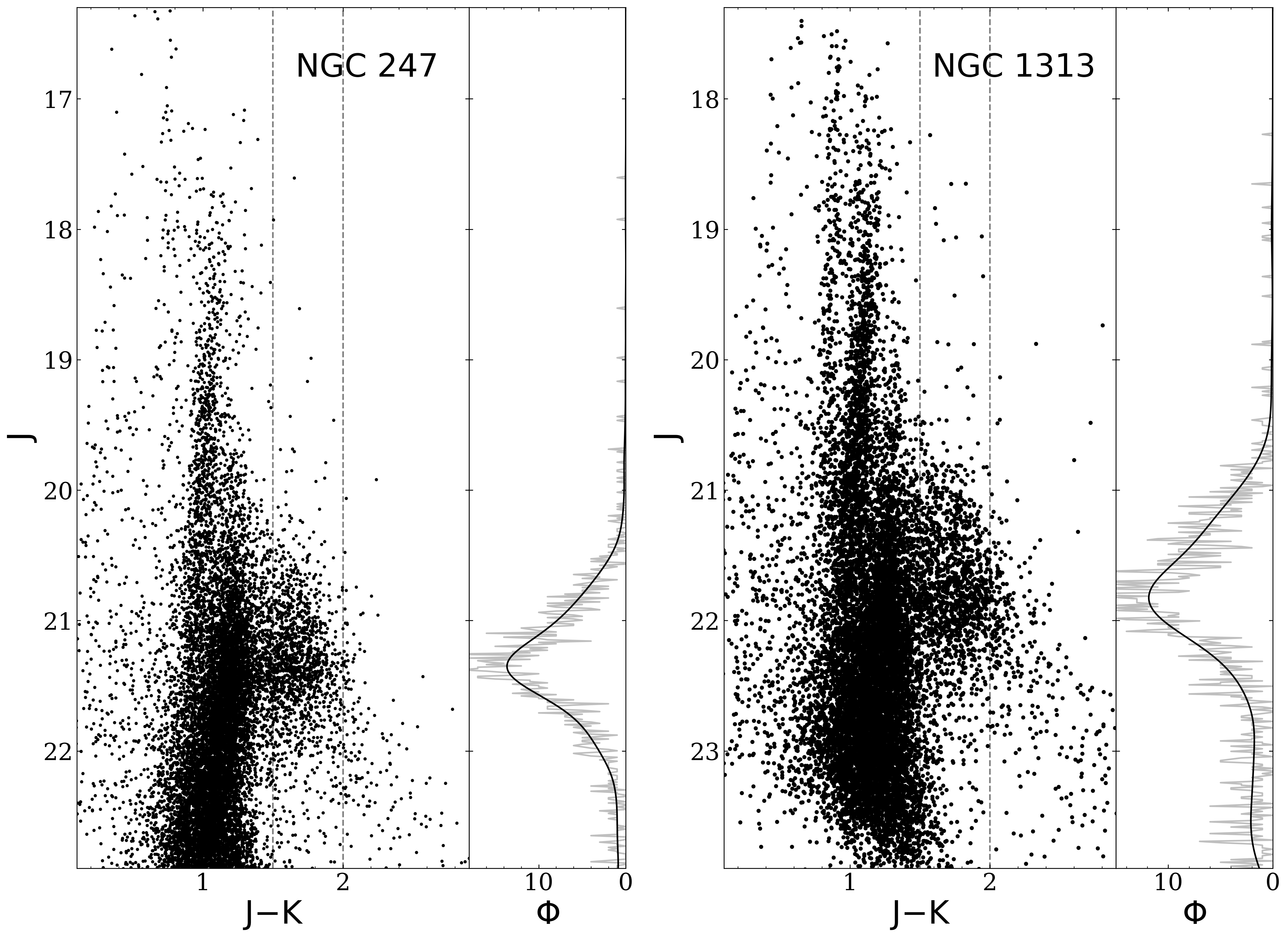"}
\caption{(Continued.) }
\end{figure*}

\begin{figure*}\figurenum{3}
\centering
\includegraphics[width=.5\textwidth]{"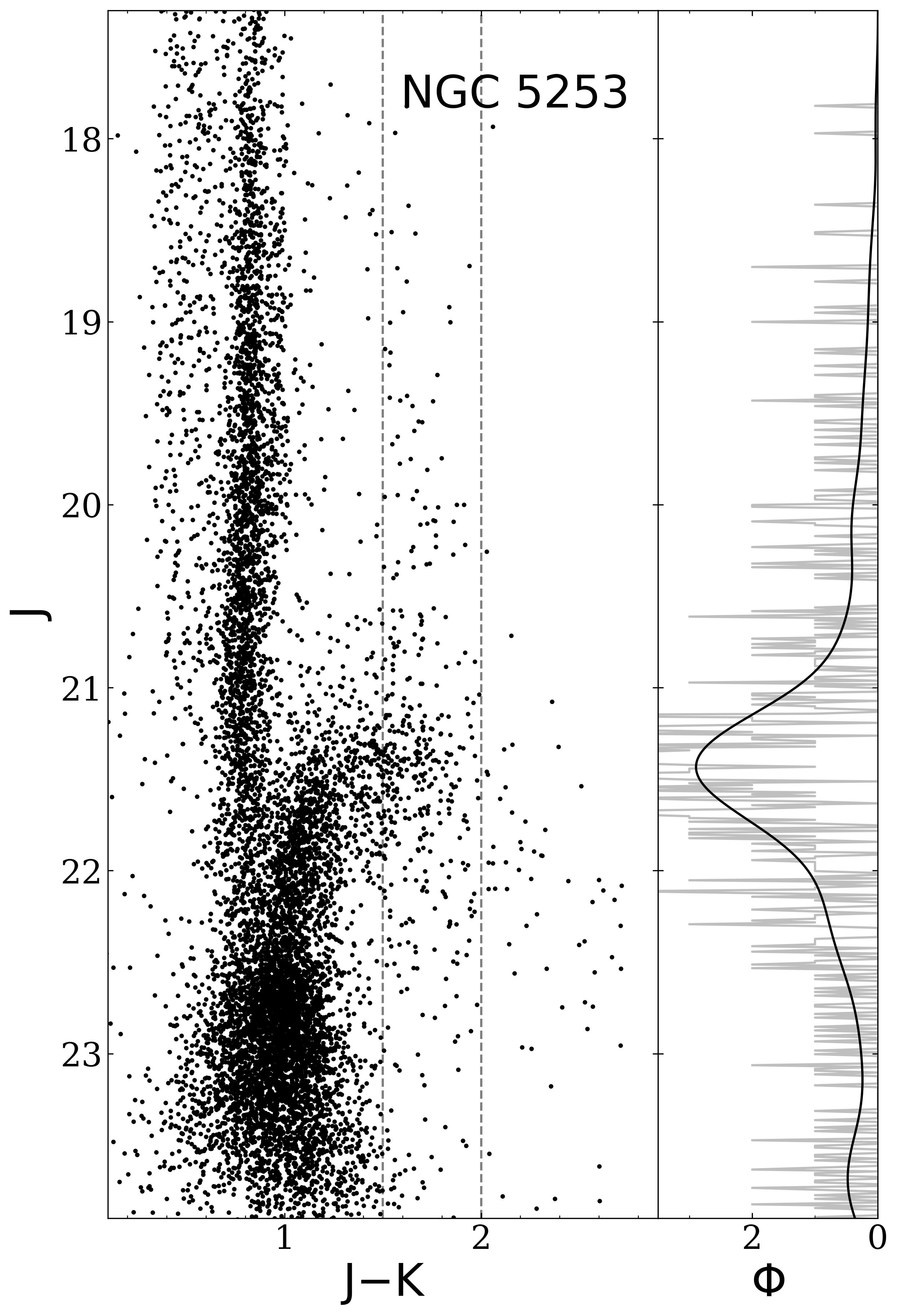"}
\caption{(Continued.) }
\end{figure*}

\begin{figure*}\figurenum{5}
\centering
\includegraphics[width=\textwidth]{"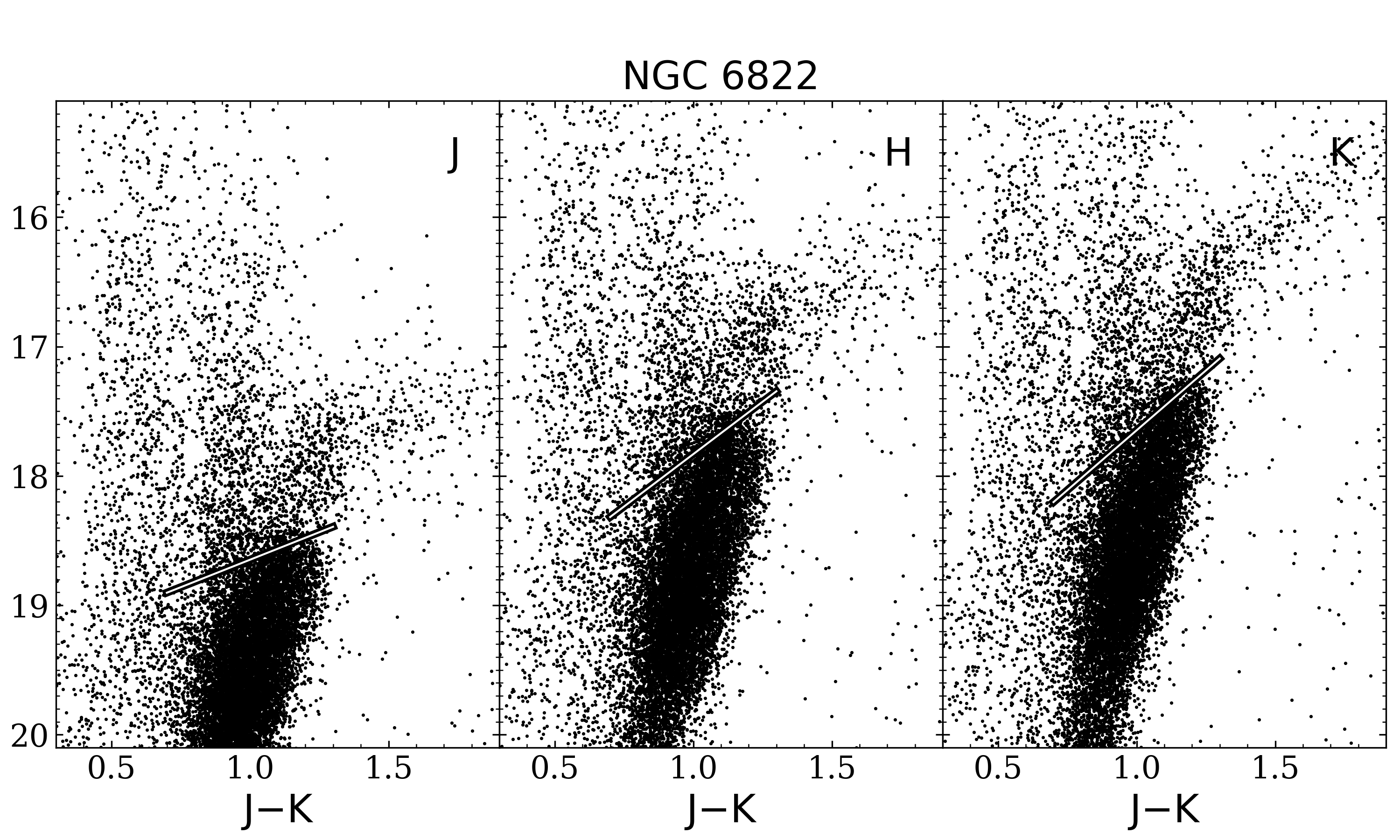"}
\caption{(Continued.)} 
\end{figure*}

\begin{figure*}\figurenum{5}
\centering
\includegraphics[width=.95\textwidth]{"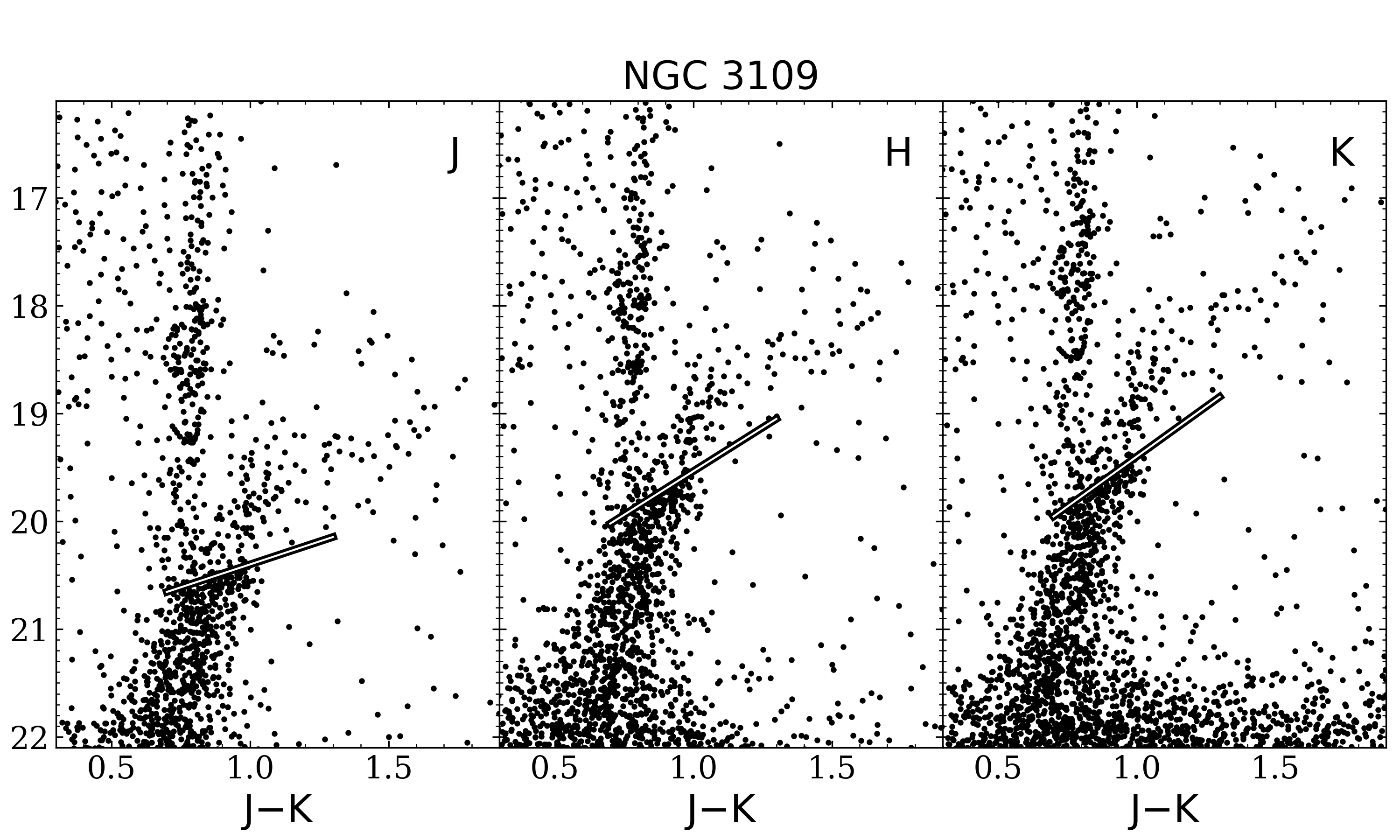"}
\caption{(Continued.)} 
\end{figure*}

\begin{figure*}\figurenum{5}
\centering
\includegraphics[width=.95\textwidth]{"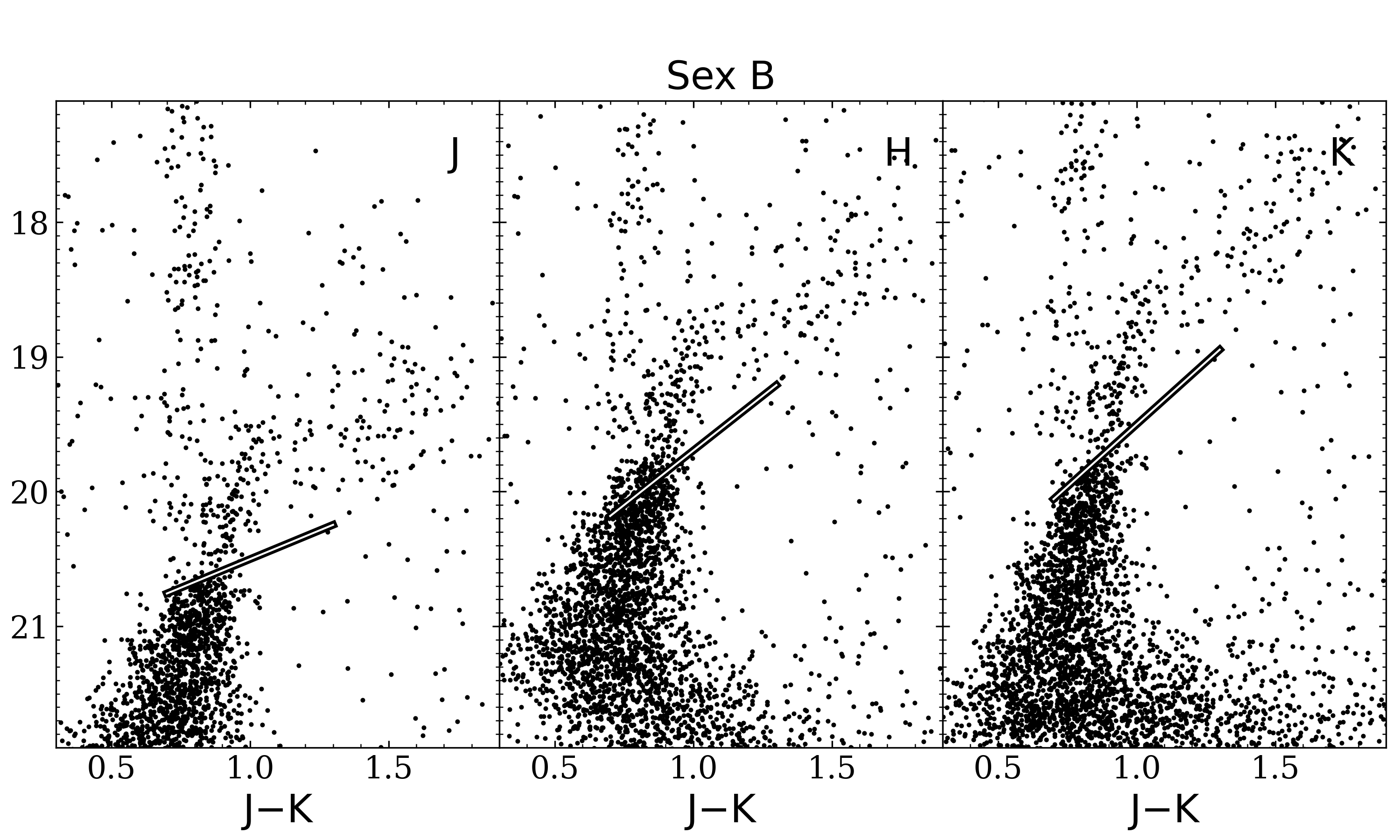"}
\caption{(Continued.)} 
\end{figure*}

\begin{figure*}\figurenum{5}
\centering
\includegraphics[width=.95\textwidth]{"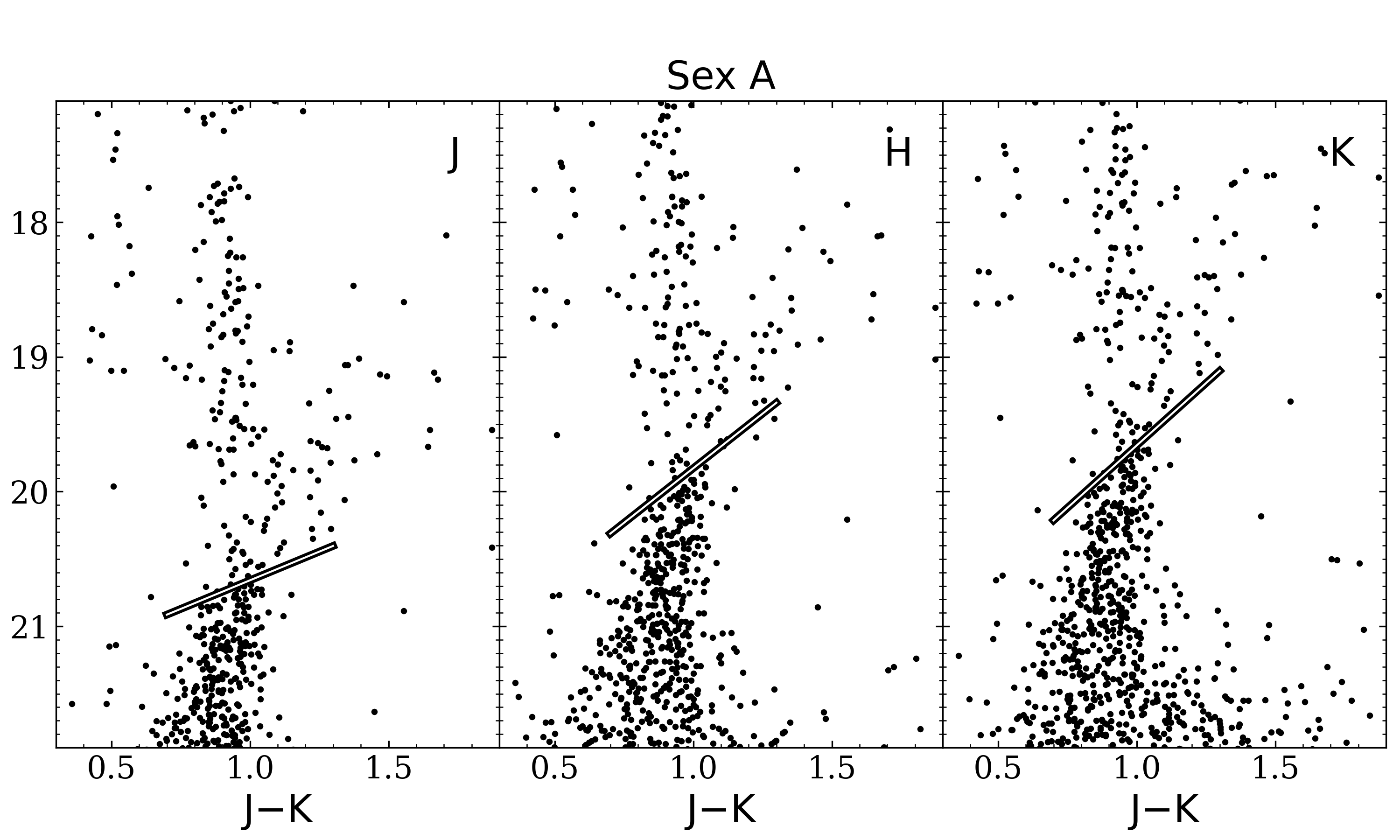"}
\caption{(Continued.)} 
\end{figure*}

\begin{figure*}\figurenum{5}
\centering
\includegraphics[width=.95\textwidth]{"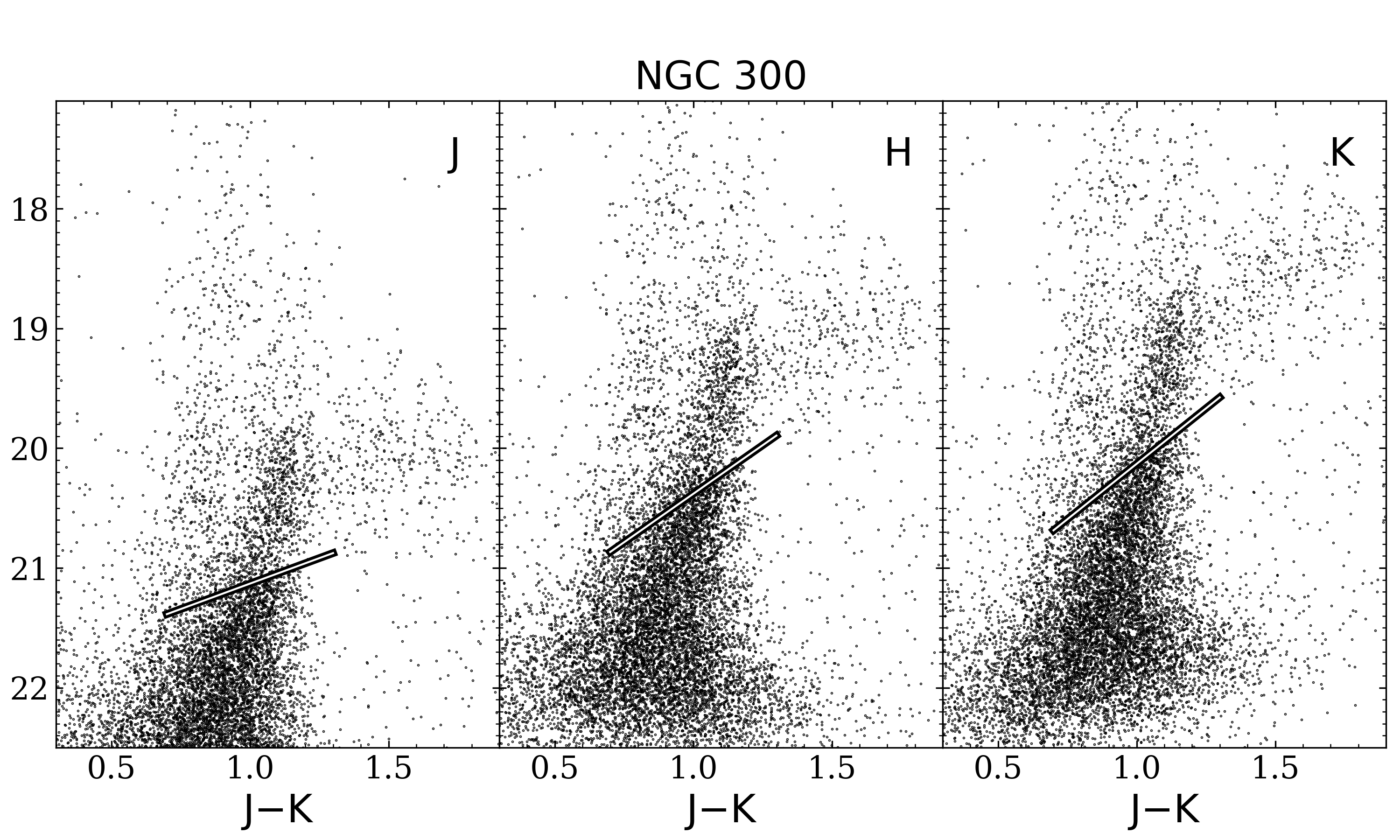"}
\caption{(Continued.)} 
\end{figure*}

\begin{figure*}\figurenum{5}
\centering
\includegraphics[width=.95\textwidth]{"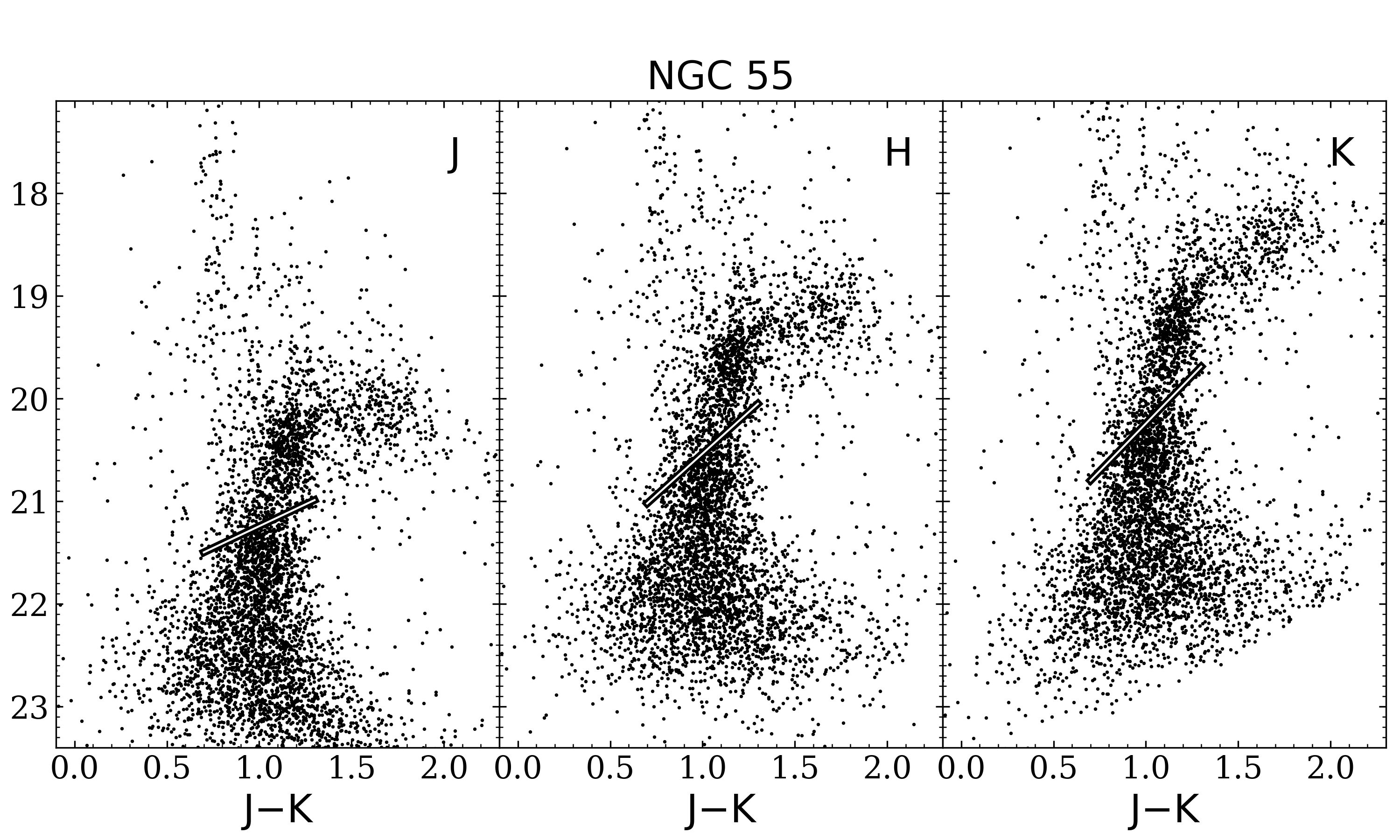"}
\caption{(Continued.)} 
\end{figure*}

% \begin{figure*}\figurenum{3}
% \centering
% \includegraphics[width=.95\textwidth]{"cmd_trgb_1h.png"}
% \caption{(Continued.)  \al{removed because i think NGC 247 doesnt have enough RGB stars for a quality TRGB measurement.}} 
% \end{figure*}

\begin{figure*}\label{fig:cmd2_trgb}\figurenum{5}
\centering
\includegraphics[width=\textwidth]{"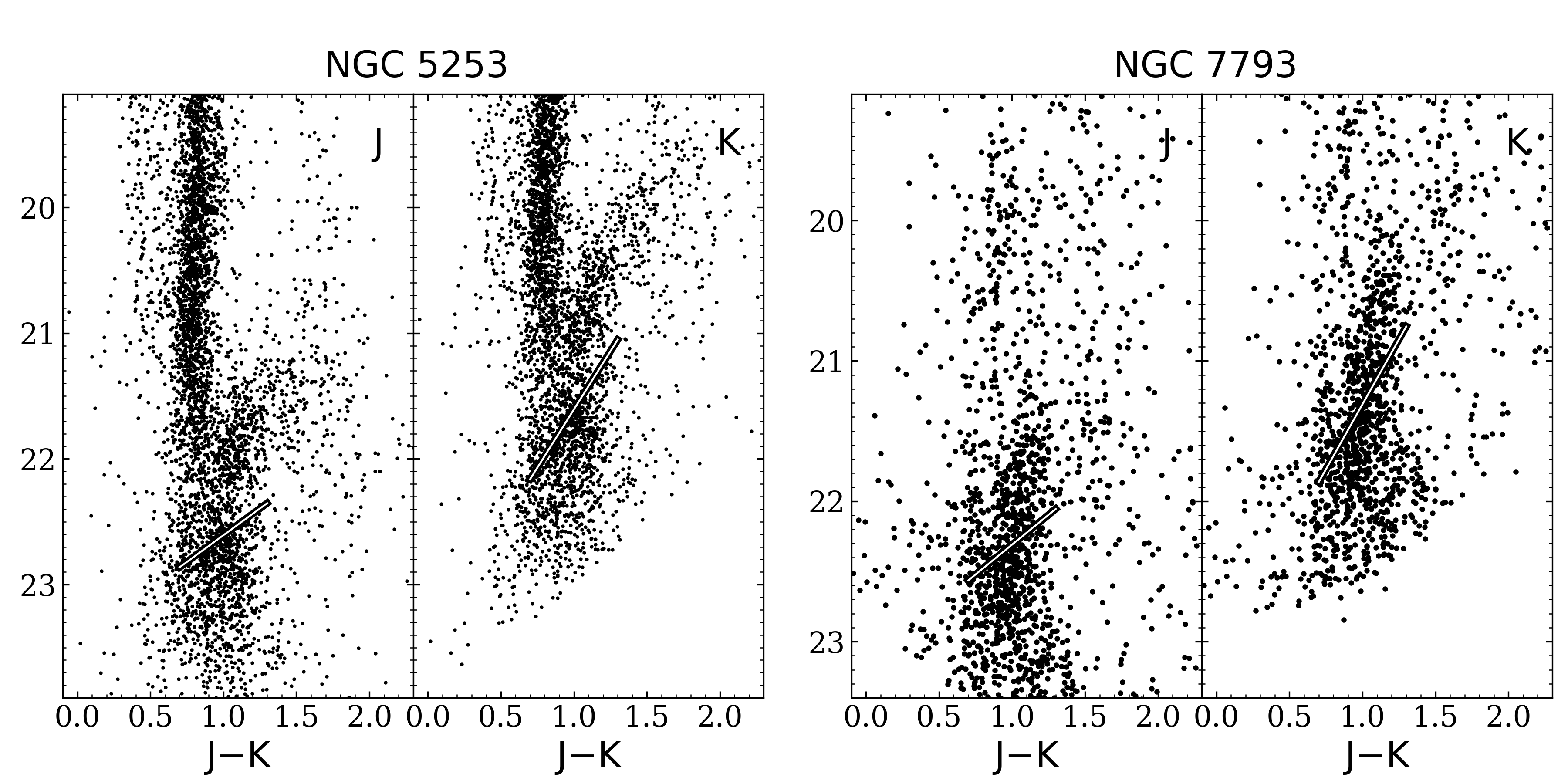"}
\caption{Near-infrared CMDs for the two galaxies without deep H-band data available. The solid line shows the measured TRGB detection in the J band. The TRGB was measured for stars with $0.7<(J-K)<1.3$~mag color, which is shown by the width of the solid line. The J-band TRGB was then projected into the K band and shown here in the right panel for each galaxy.}
\end{figure*}

\section{Anomalous galaxies}

\begin{figure}\figurenum{B2}
\centering
\includegraphics[width=\columnwidth]{"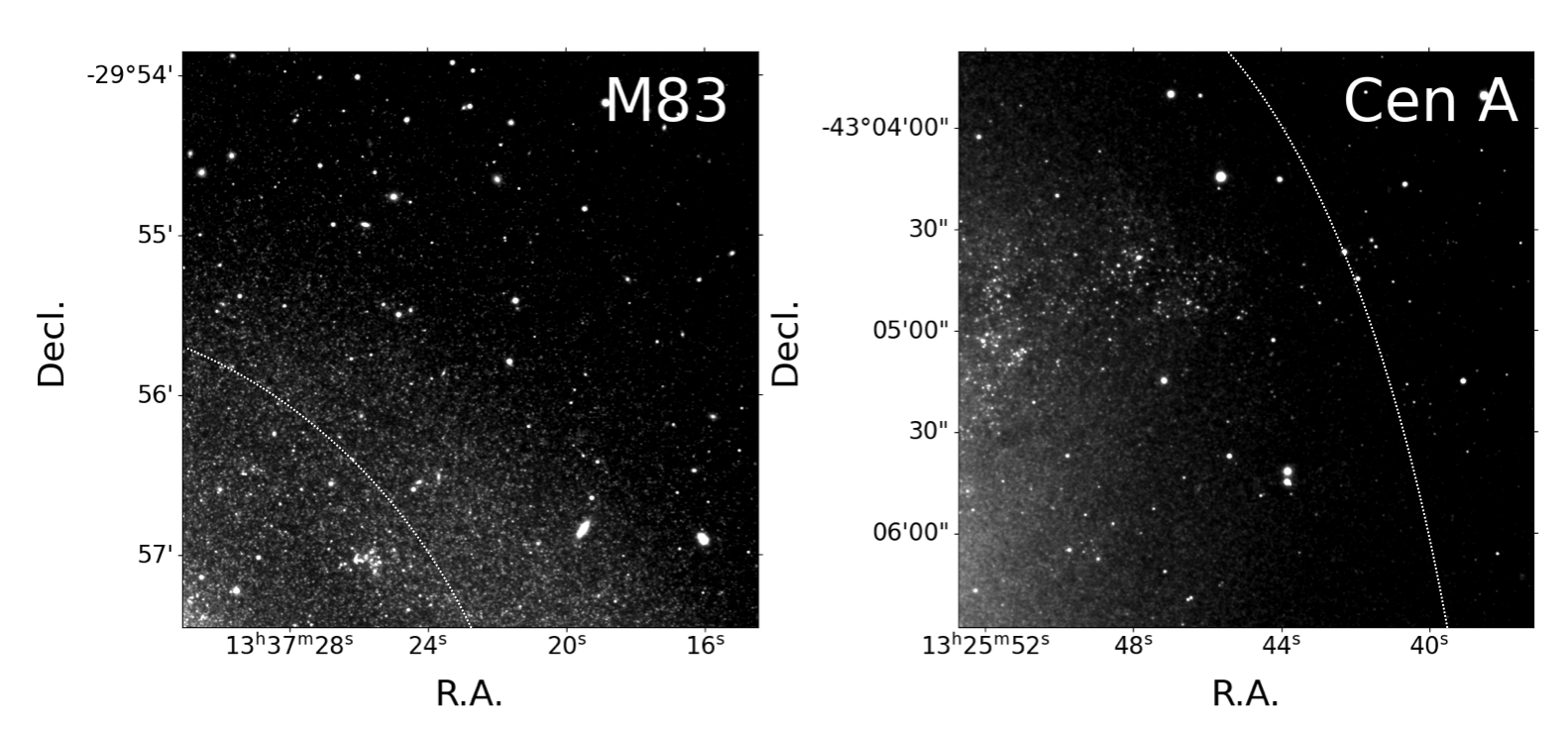"}
\caption{Images of M83 and Cen A observed with the FourStar camera on the 6.5 Magellan-Baade telescope. The JAGB measurements were measured outside the dotted white ellipse. In these two galaxies we only observed the northwest quadrants of the galaxy. }
\label{fig:appendix2}
\end{figure}

We were unable to measure precise distances to two galaxies in our sample, M83 and Cen A. As shown in Figure \ref{fig:appendix1}, neither of the JAGB star LFs for these galaxies has a clearly defined peak location. 
In appendices \ref{subsec:cena} and \ref{subsec:m83}, we speculate on potential explanations for our difficulties in measuring precise distances to these galaxies. The images of both galaxies are shown in Figure \ref{fig:appendix2}.

\subsection{Cen A}\label{subsec:cena}

\begin{figure}\figurenum{B1}
\centering
\includegraphics[width=\columnwidth]{"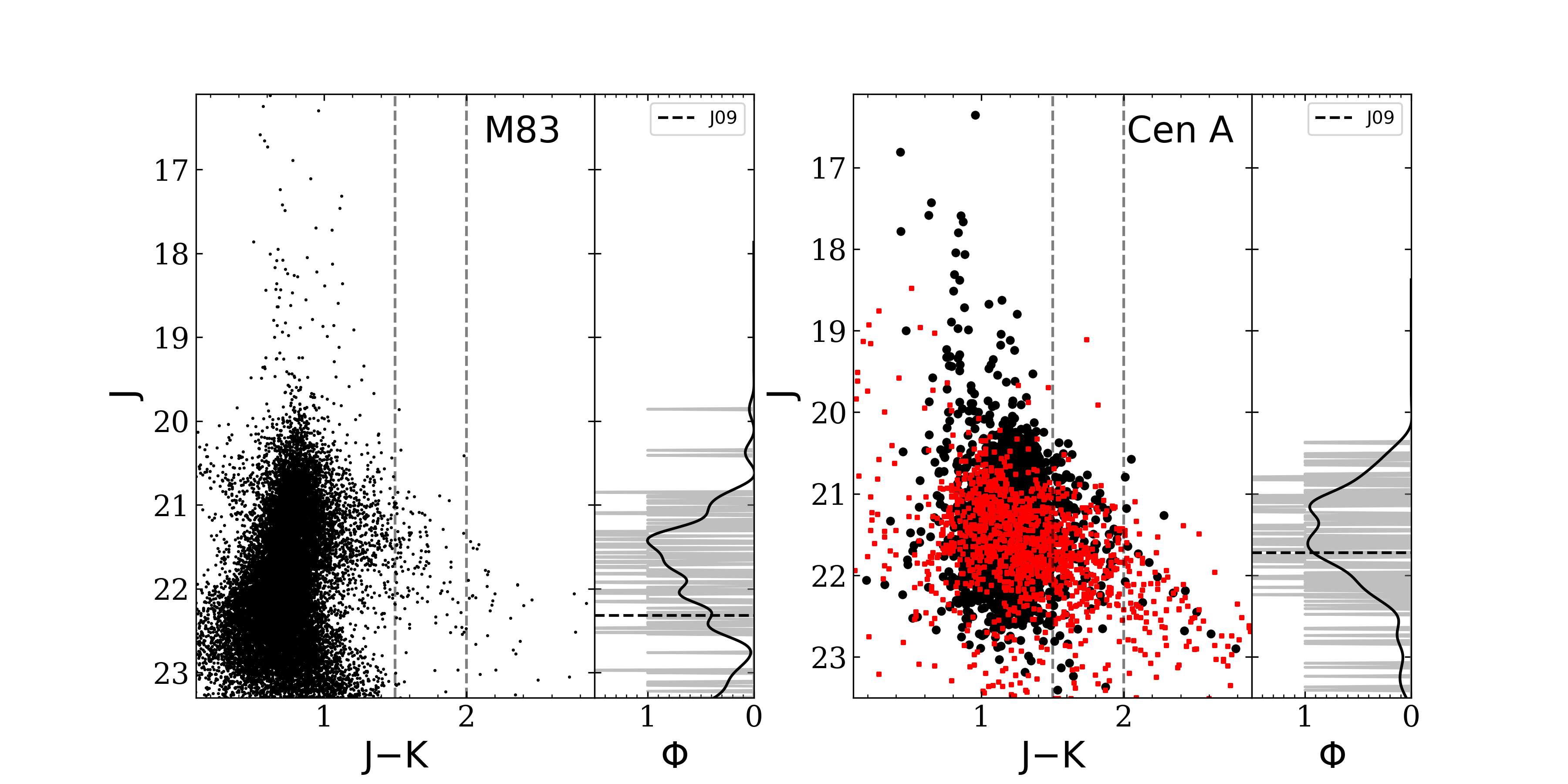"}
\caption{J vs. (J-K) color-magnitude diagrams (left panels) and GLOESS-smoothed luminosity functions in black overplotted on the binned LF in grey (right panels) for M83 and Cen A. There is no clearly defined peak in either luminosity function. The dotted line represents the predicted JAGB luminosity from the EDD I-band TRGB distance modulus from \citetalias{2009AJ....138..332J}, with a foreground extinction correction of $A_J=0.05$~mag added for M83 and $A_J=0.08$~mag for Cen A. Photometry of long-period variable stars from \cite{2003A&A...406...75R} are shown as red points in the CMD of Cen, showing the location of the JAGB stars in our CMD are consistent with the location of the JAGB stars observed in \cite{2003A&A...406...75R}.}
\label{fig:appendix1}
\end{figure}

\begin{deluxetable*}{ccccc}
\tablecaption{Literature Distances}\label{tab:compare2}
\tablehead{
\colhead{Galaxy} & 
\colhead{Method} & 
\colhead{$\mu_0$ (mag.)} & 
\colhead{Reference} & 
\colhead{Notes}
}
\startdata
Cen A & JAGB method & $27.95\pm0.02$ & \citetalias{2020arXiv200510793F}  \\
Cen A & I-band TRGB & $27.84\pm0.04$ & \citetalias{2009AJ....138..332J} & $m_{F814W}=23.97\pm0.03$~mag, $A_{F814W}=0.18$~mag\\
Cen A & Mira variables & $27.96\pm0.11$ & \cite{miras} \\
Cen A & Cepheids & $27.67\pm0.20$ & \cite{2007ApJ...654..186F} \\
\hline
M83 & I-band TRGB & $28.47\pm0.05$& \citetalias{2009AJ....138..332J} & $m_{F814W}=24.52\pm0.03$~mag, $A_{F814W}=0.10$~mag\\
M83 & Cepheids & $28.25\pm0.15$ & \cite{2003ApJ...590..256T}\\
\enddata
\end{deluxetable*}

Cen A (NGC 5128) has a notoriously complex star formation history, as well as radio jets, dust lanes, and X-ray sources. Its morphological type, either an elliptical galaxy with prominent dust lane or a peculiar S0 has been source of debate in the literature  \citep{2010PASA...27..475H}. As shown in Table \ref{tab:compare2}, recent distances in the literature to Cen A range from the Cepheid-based distance of $\mu_0=27.67\pm0.20$~mag \citep{2007ApJ...654..186F} to the Mira-based distance of $\mu_0=27.96\pm0.11$~mag \citep{miras} (a difference of 0.5 Mpc!). 

To perform an additional cross-check on our photometry of Cen A, we compared our photometry to the photometry of \cite{2003A&A...406...75R}, who observed 1504 long-period variable stars using the ISAAC NIR imaging spectrometer at the ESO Paranal UT1 Antu 8.2 m telescope between 1999 and 2002 (332 of these stars have colors of $1.5<(J-K)<2.0$~mag and are therefore JAGB stars). We overplot these stars in Figure \ref{fig:appendix1}, showing that the location of our JAGB stars in the CMD are consistent with the location of the JAGB stars observed in \cite{2003A&A...406...75R}. Our photometry is likely not the source of the problem.

The magnitudes of the JAGB stars in Cen A range from $20.5<J<23$~mag, a significantly larger range of JAGB star magnitudes than what was seen for the 11 nearby galaxies in Figure \ref{fig:jagb_cmd}, which was typically less than $1$~mag. The large scatter of JAGB star magnitudes could therefore be due to Cen A's far distance (resulting in relatively larger photometry errors for the JAGB stars) or its complex star formation history.

\subsection{M83}\label{subsec:m83}
M83 is the farthest galaxy in our sample at around 4.9~Mpc \citep{2009AJ....138..332J}, with the second farthest being NGC 1313 at about 4.0~Mpc away. We speculate this is now the limit of where accurate JAGB distances can be measured with ground-based telescopes. Only two distances have recently been measured to M83, which range from 4.5 Mpc via Cepheid variables stars \citep{2003ApJ...590..256T} to 4.9 Mpc via the I-band TRGB \citep{2009AJ....138..332J}. The lack of a clearly defined peak in the JAGB star LF indicates that the photometric scatter of the stars in this galaxy is significantly large, and that JAGB stars in galaxies at this distance may not be able to be accurately used as standard candles with ground-based telescopes.

\section{T-band Luminosity Functions and Edge Response Functions}\label{sec:tbandlf}

In this section, the T[J, (J, K)]-band and T[H, (J, K)]-band luminosity functions and edge detector response functions are shown in Figure \ref{fig:tband}3.

\begin{figure*}\label{fig:tband}\figurenum{C1}
\gridline{\fig{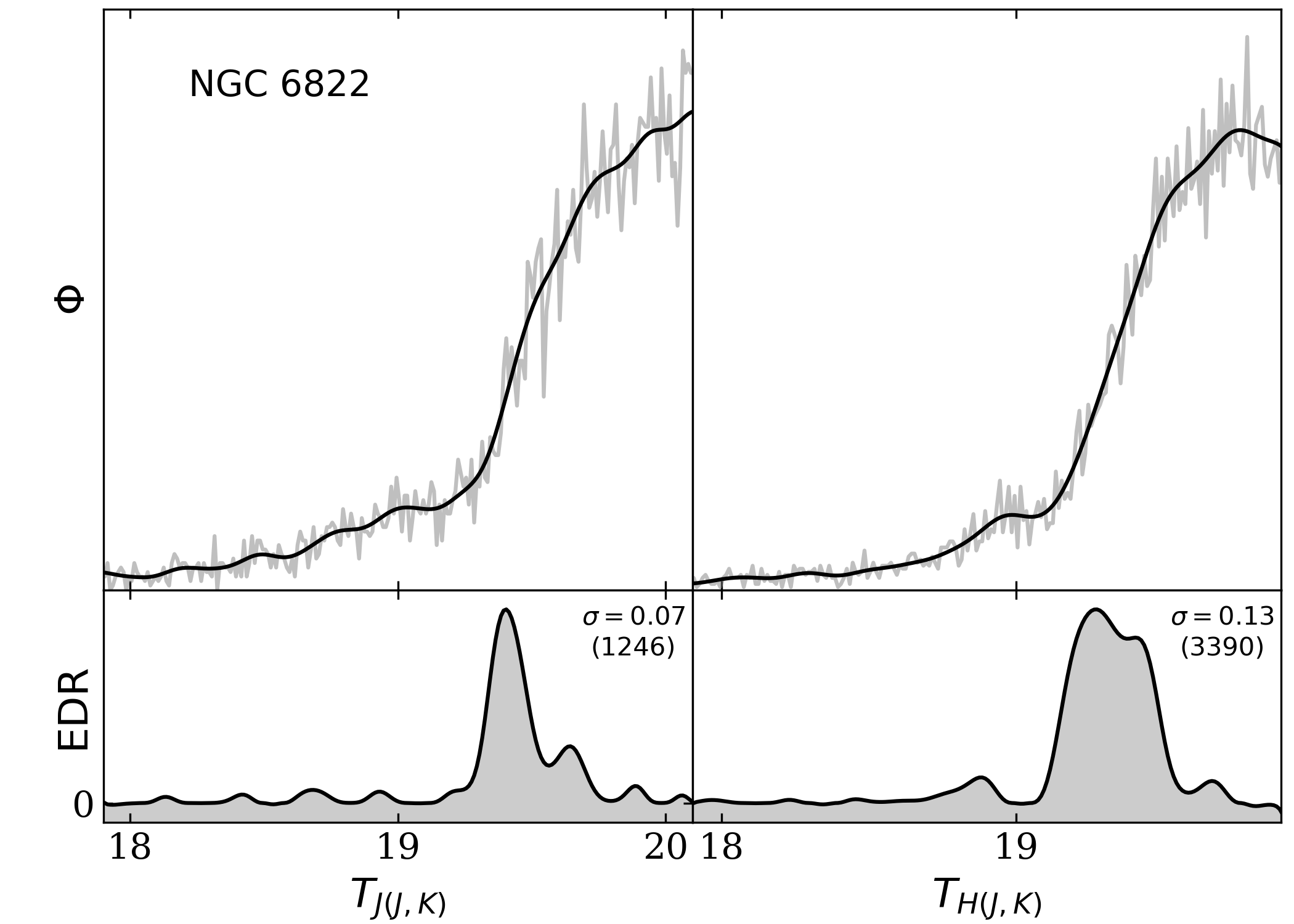}{0.5\textwidth}{}
          \fig{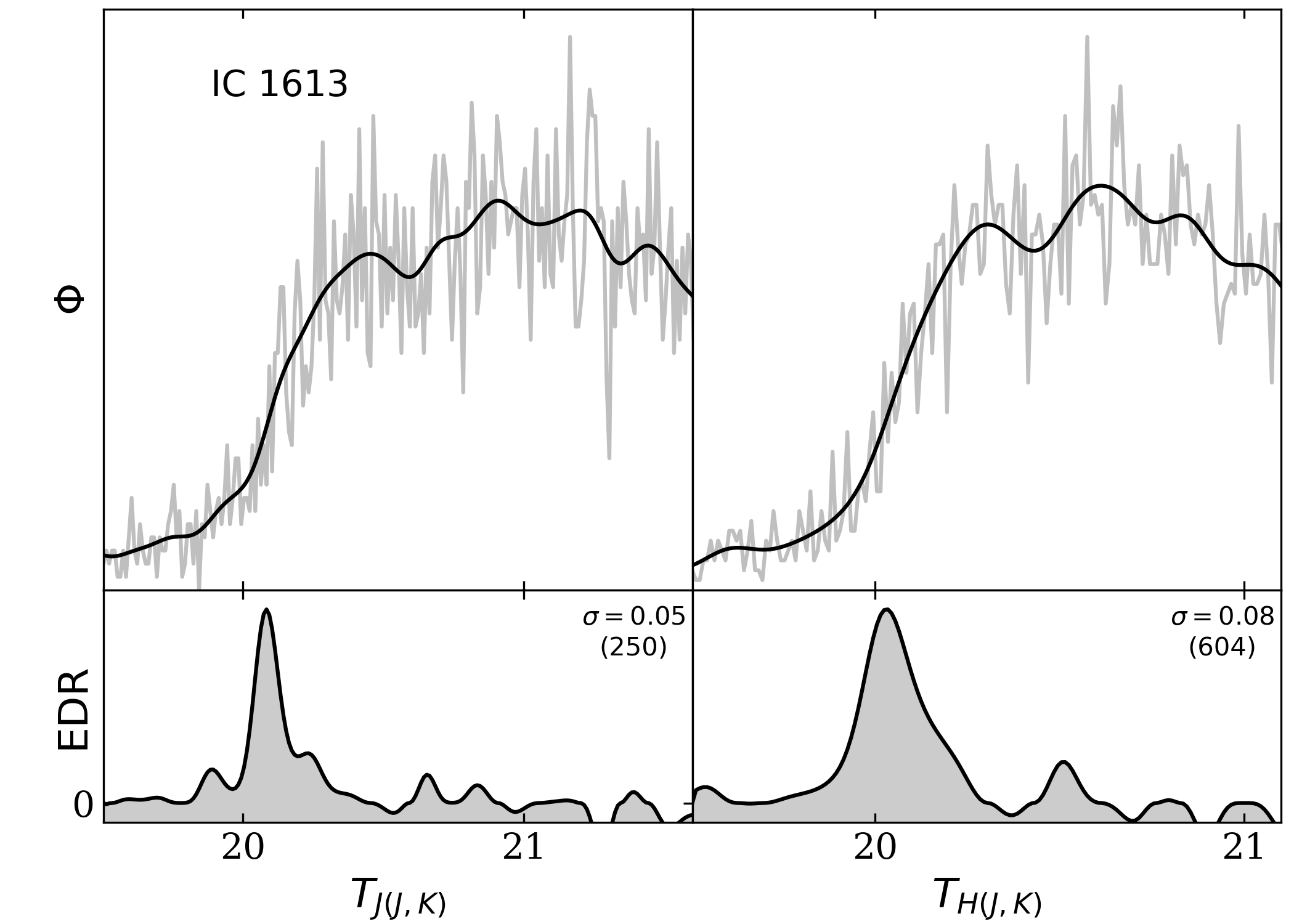}{0.5\textwidth}{}}
\gridline{\fig{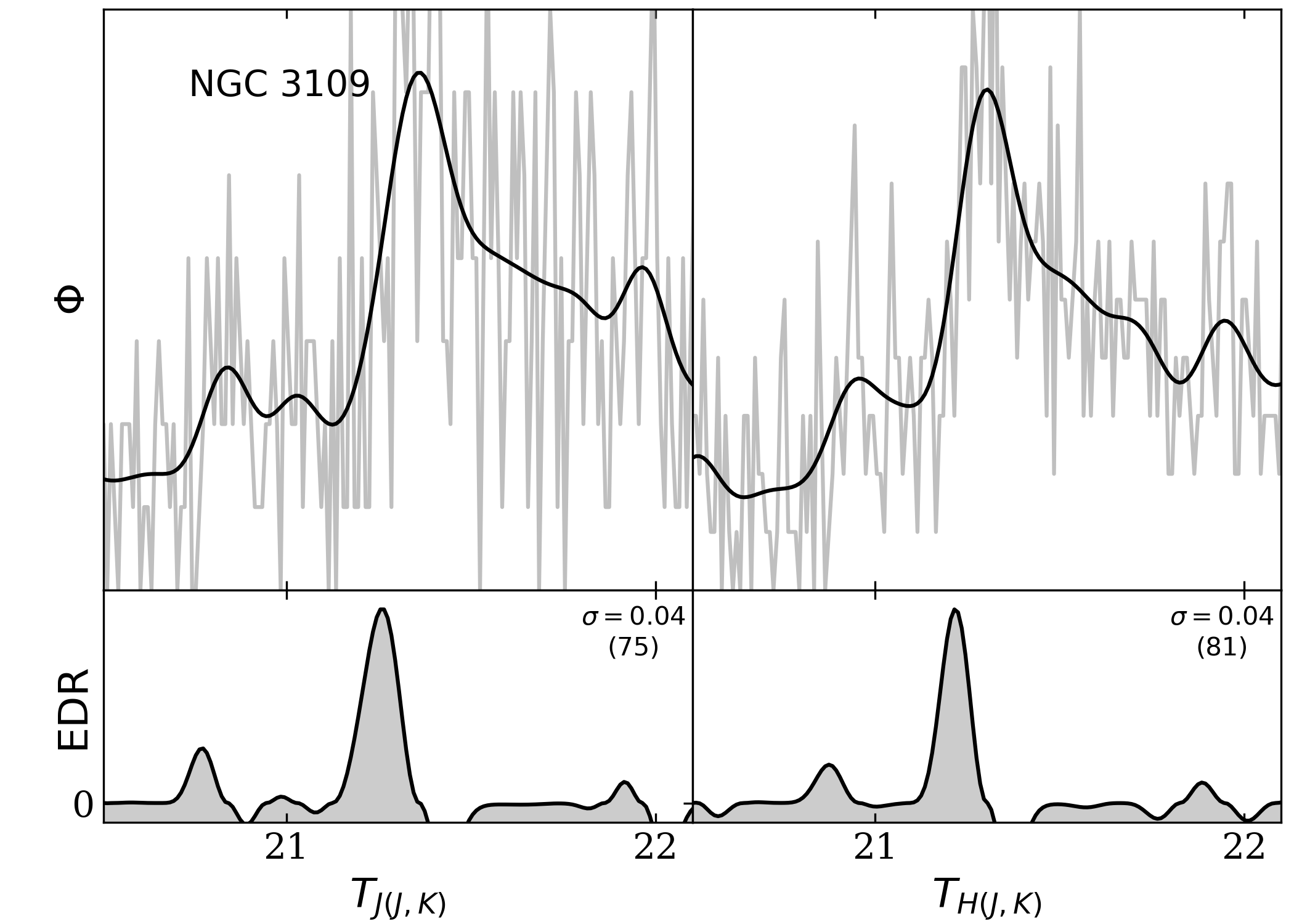}{0.5\textwidth}{}
          \fig{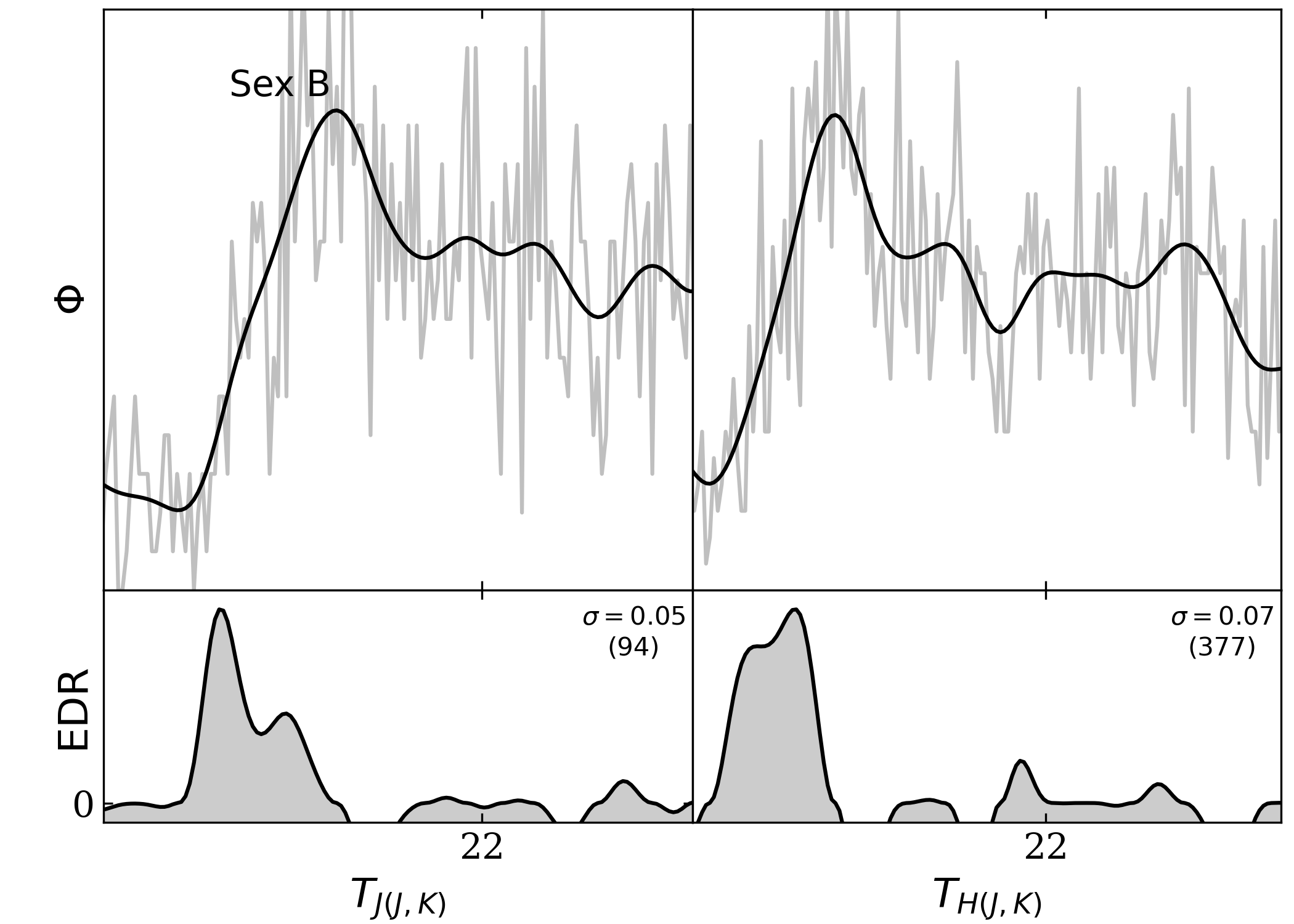}{0.5\textwidth}{}}
\gridline{\fig{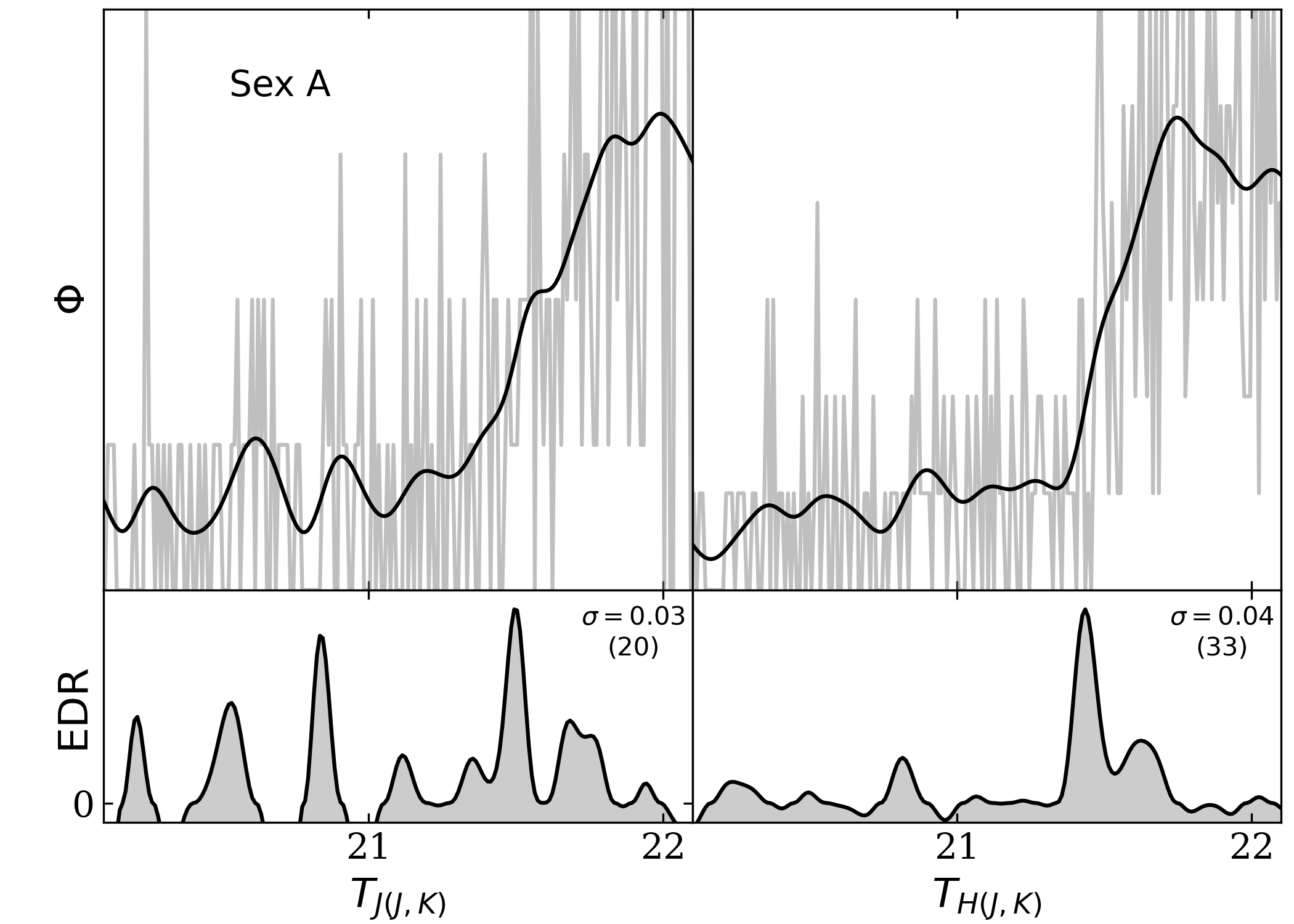}{0.5\textwidth}{}
          \fig{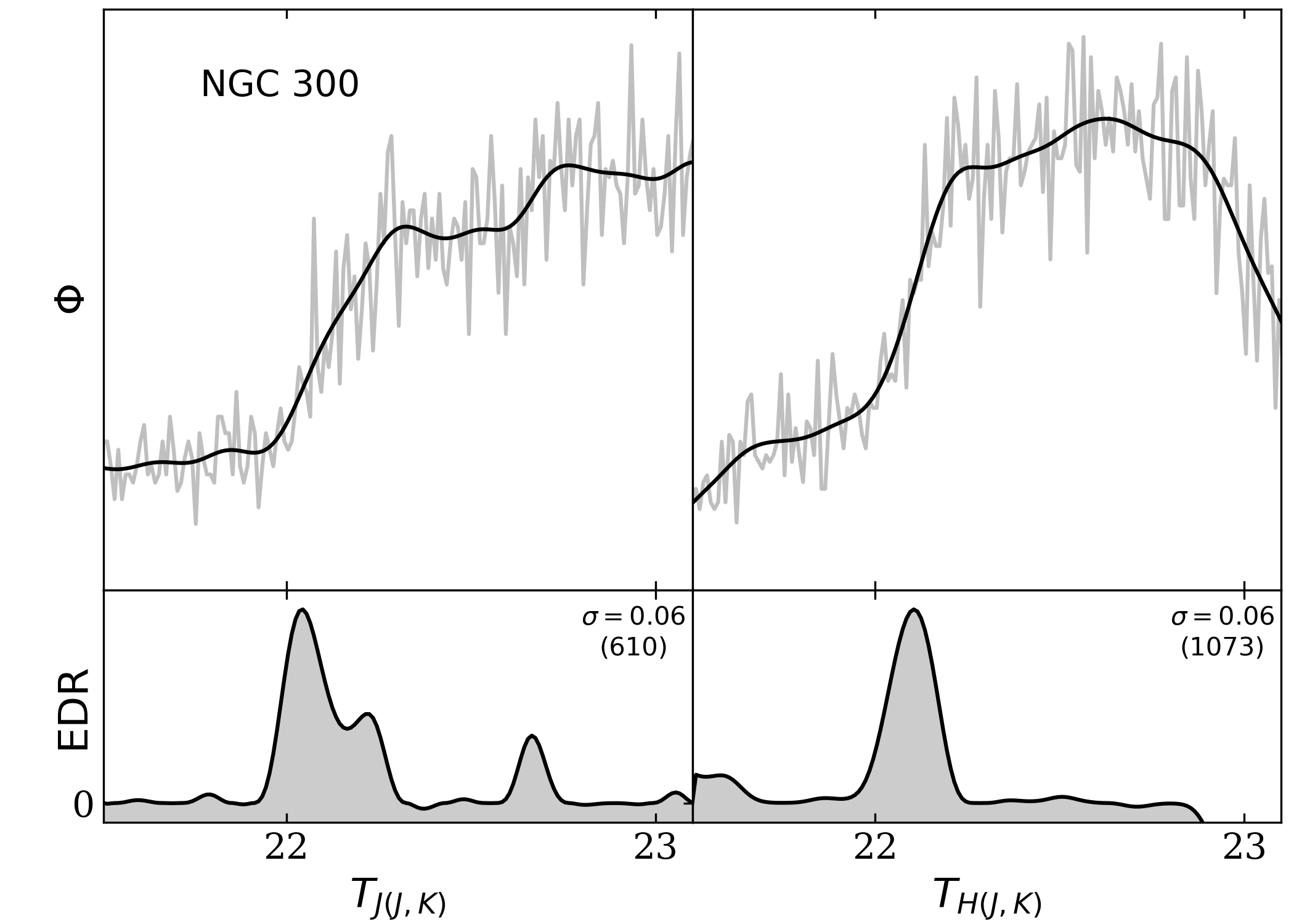}{0.5\textwidth}{}}
\caption{Rectified T[J(J,K)] and T[H(J,K)] luminosity functions (top panels) and edge detector response (EDR) functions (bottom panels) for the 10 galaxies with TRGB distances. All the data were transformed so the TRGB is now insensitive to color. T-band magnitudes were binned using bins of 0.01~mag and then smoothed using a smoothing parameter of $\sigma_s=0.10$~mag. The TRGB marks the point where the EDR is the largest, i.e., where the the first derivative of the luminosity function is the greatest.  The measured width of the EDR, $\sigma$, and number of TRGB stars contributing to the EDR are shown in the bottom panel of every figure.}
\end{figure*}

\begin{figure*}\figurenum{C3}
\gridline{\fig{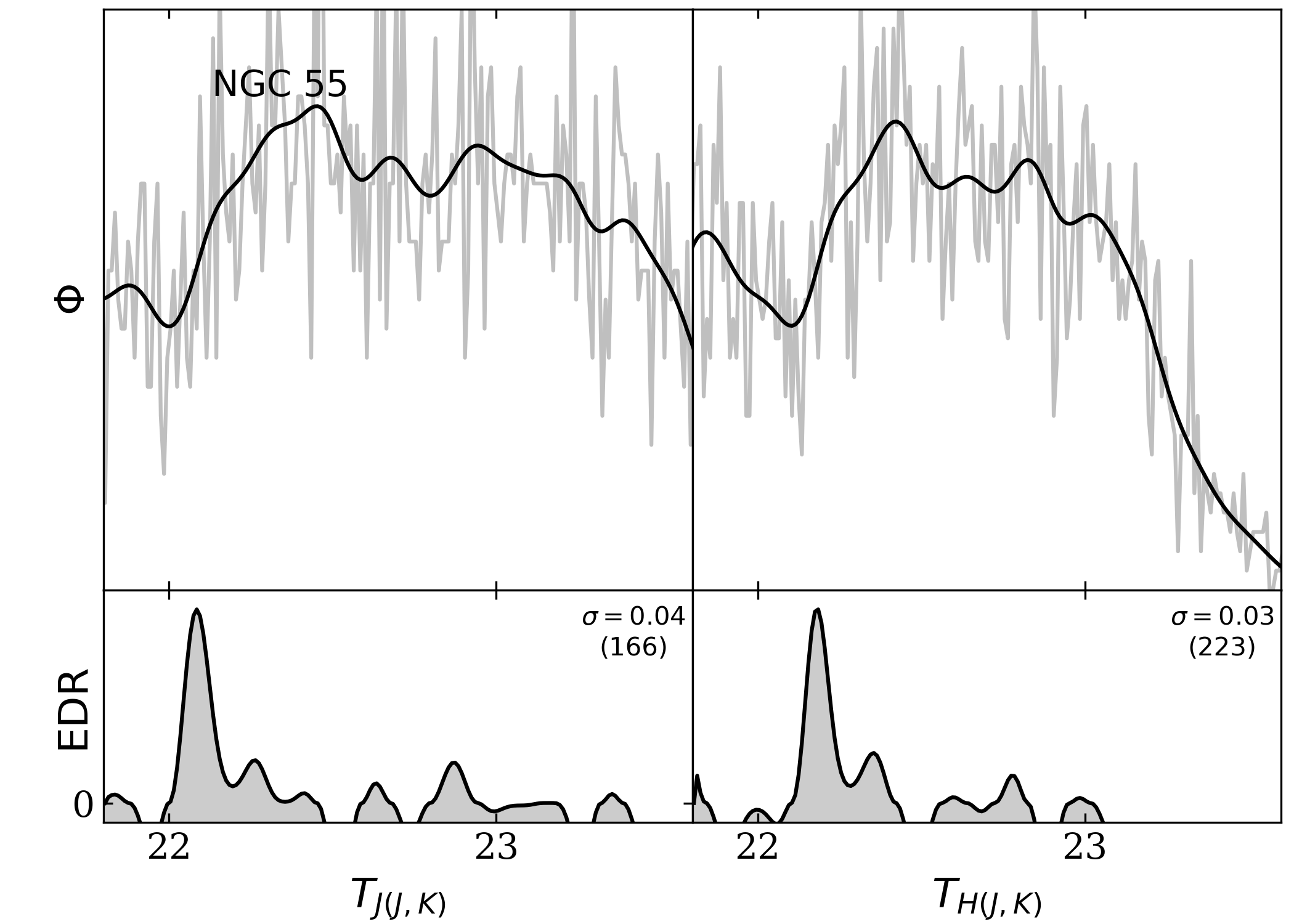}{0.5\textwidth}{}
        \fig{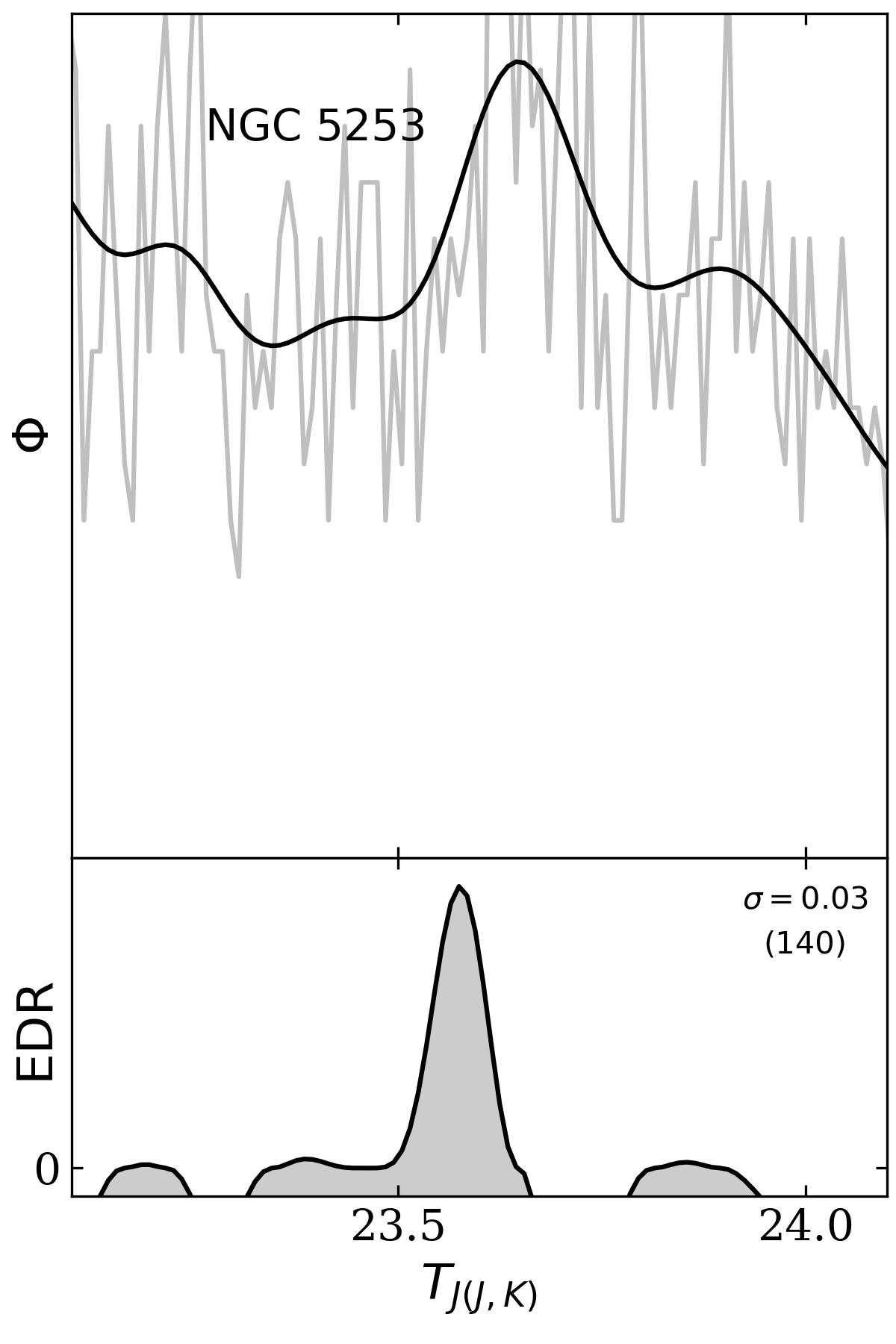}{0.24\textwidth}{}
          \fig{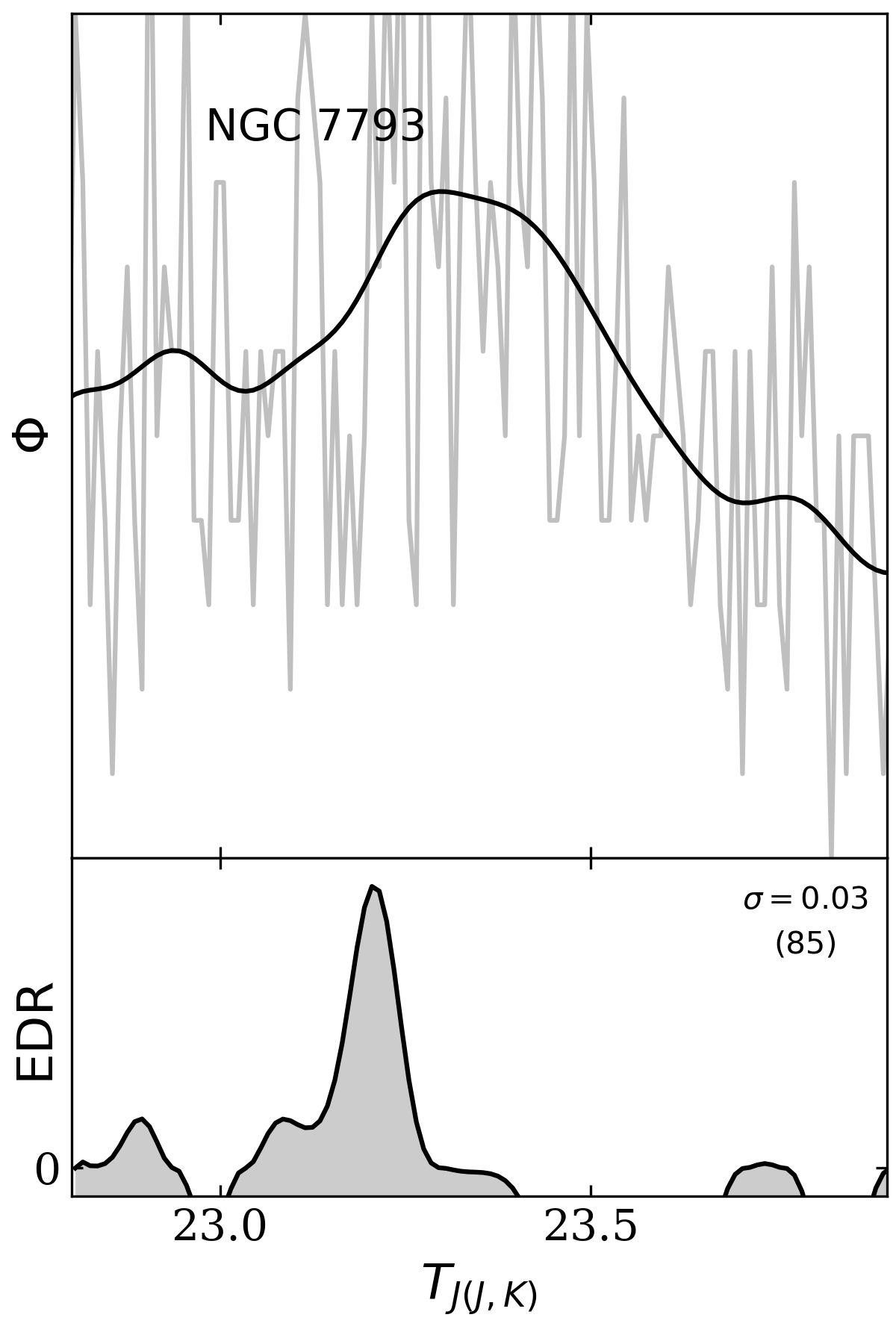}{0.24\textwidth}{}}
                    % \fig{ngc247_t.png}{0.5\textwidth}{}}
\caption{(Continued.)}
\end{figure*}

\end{document}